\documentclass[12pt]{iopart}

\usepackage{iopams}

\expandafter\let\csname equation*\endcsname\relax

\expandafter\let\csname endequation*\endcsname\relax

\usepackage{lineno}
\usepackage{graphicx}

\usepackage{amsmath}
\usepackage{amssymb}
\usepackage{graphicx}
\usepackage{afterpage}
\usepackage{amsfonts}
\usepackage{graphicx}
\usepackage{physics}
\usepackage{textcomp}
\usepackage{bm}
\usepackage{bbm}
\usepackage{color}
\usepackage{bbold}
\usepackage{hyperref}
\usepackage[sort,compress]{cite}
\usepackage{appendix}
\usepackage{etoolbox}
\usepackage{makecell}

\newcommand{\ketL}[1]{\big|#1\big)}
\newcommand{\braL}[1]{\big(#1\big|}
\newcommand{\expL}[1]{\big(#1\big)}

\makeatletter
\def\@mkboth#1#2{}
\newlength\appendixwidth
\preto\appendix{\addtocontents{toc}{\protect\patchl@section}}
\newcommand{\patchl@section}{%
	\settowidth{\appendixwidth}{\textbf{Appendix }}%
	\addtolength{\appendixwidth}{1.5em}%
	\patchcmd{\l@section}{1.5em}{\appendixwidth}{}{}%
}

\newcommand{\mainmatter}{%
	\setcounter{footnote}{0}%
	\patchcmd{\@makefntext}{\fnsymbol}{\arabic}{}{}%
	\patchcmd{\@thefnmark}{\fnsymbol}{\arabic}{}{}%
	\def\@makefnmark{\textsuperscript{\arabic{footnote}}}%
}
\makeatother

\begin{document}
\title{Topical review on acousto-optical Floquet engineering of single-photon emitters}

\author{Daniel~Groll,$^{1}$ Daniel~Wigger,$^{2}$ Matthias Wei{\ss},$^{3}$ Mingyun~Yuan,$^{4}$ Alexander~Kuznetsov,$^{4}$ Alberto~Hern{\'a}ndez-M{\'i}nguez,$^{4}$ Hubert~J.~Krenner,$^{3}$ Tilmann~Kuhn$^1$ \& Pawe\l{}~Machnikowski$^{5}$}

\address{$^{1}$Institute of Solid State Theory, University of M{\"u}nster, 48149 M{\"u}nster, Germany}
\address{$^{2}$Department of Physics, University of M{\"u}nster, 48149 M{\"u}nster, Germany}
\address{$^{3}$Institute of Physics, University of M\"unster, 48149 M\"unster, Germany}
\address{$^{4}$Paul-Drude-Institut für Festk\"{o}rperelektronik, Leibniz-Institut im Forschungsverbund Berlin e.V., 10117 Berlin, Germany}
\address{$^{5}$Institute of Theoretical Physics, Wroc\l{}aw University of Science and Technology, 50-370 Wroc\l{}aw, Poland}

\ead{daniel.groll@uni-muenster.de}

\vspace{10pt}
\begin{indented}
	\item[]February 2026
\end{indented}

\begin{abstract}
The combination of solid state single-photon emitters and mechanical excitations on a common platform is a promising approach for the development of hybrid quantum technologies. In this topical review we discuss state-of-the-art platforms for emitter-based acousto-optics and their feasibility for acousto-optical Floquet engineering. To this aim we investigate theoretically the resonance fluorescence (RF) spectrum of an acoustically modulated single-photon emitter under arbitrarily strong optical driving. In the spectrum, the combination of Mollow triplet physics and phonon sidebands results in a complex structure of crossings, anti-crossings, and line suppressions. We apply Floquet theory to develop an analytical expression for the RF spectrum. Complemented with perturbative and non-perturbative techniques, this allows us to fully understand the underlying acousto-optical double dressing physics of the hybrid quantum system, explaining the observed spectral features. We use these insights to perform an experimental feasibility study of existing emitter-based acousto-optical platforms and come to the conclusion that surface and bulk acoustic waves interfaced with quantum dots as an established Mollow triplet platform represent particularly promising infrastructures for acousto-optical Floquet engineering.
\end{abstract}
\maketitle

\tableofcontents

\clearpage
\mainmatter

\section{Introduction}\label{sec:intro}
As the number of specific tasks in quantum information processing increases with the progress in technology, hybrid approaches are developed that combine the strengths of different physical platforms into single devices~\cite{kurizki2015quantum}. In this context, single photons are key ingredients as carriers for quantum information in long-haul quantum communication~\cite{duan2001long,couteau2023applications} and can also be used for quantum computation~\cite{kok2007linear,couteau2023applications} as well as quantum simulation~\cite{Georgescu2014quantum} employing integrated photonic circuits~\cite{Wang2020integrated}. To exploit the full potential of photons, on the one hand, their spectral properties must be controlled very precisely. On the other hand, they need to be converted into or generated from other (often stationary) quantum degrees of freedom, e.g., charge or spin excitations in quantum emitters~\cite{mi2018coherent,garcia2021semiconductor}. Ideally, all these tasks are performed by single devices enabling quantum transduction~\cite{lauk2020perspectives}. However, such all-optical devices, i.e., interfacing single emitters with single photons, are challenging to miniaturize since the operation wavelength of the light can set a lower bound on feasible structure scales via the diffraction limit~\cite{gramotnev2010plasmonics,gao2024super}.

In this context, manipulation of quantum emitters via elastic oscillations in solids, or phonons in their quantized form, is particularly promising. Due to the much smaller propagation speed of phonons compared with photons, the wavelength of phonons is significantly reduced when operating at comparable frequencies, allowing for advances in miniaturization~\cite{delsing20192019,clerk2020hybrid}. Since these mechanical deformations of the crystal lattice are the fundamental excitation of solid state crystals and couple to virtually any other excitation embedded in the same medium, phonons, e.g., in the form of acoustic waves, are relevant for diverse cross-disciplinary applications from quantum acoustics to the life sciences~\cite{delsing20192019}. To date, acoustic waves are already the backbone of modern classical wireless communication for signal processing at radio- and microwave frequencies~\cite{campbell2012surface,hashimoto2000surface,hagelauer2023microwave}. In the realm of hybrid quantum technologies, phonons have been employed in a broad spectrum of applications. For instance, they have been established as universal quantum transducers~\cite{schuetz2015universal,neuman2021phononic,raniwala2025spin,wigger2021remote,niehues2024excitons} and ultimately as on-chip carriers of quantum information~\cite{bienfait2019phonon,qiao2023splitting,yang2024mechanical}. Furthermore, optically active quantum emitter systems can be interfaced with phonons. Here, applications include the dynamical (time-domain) control of quantum dot lasers~\cite{bruggemann2012laser,czerniuk2014lasing,wigger2017systematic} or single quantum emitters~\cite{gell2008modulation,schulein2015fourier,weiss2018interfacing,lazic2019dynamically,patel2024surface}, as well as acoustic manipulation of exciton-polariton Bose-Einstein condensates~\cite{kuznetsov2023microcavity,kuznetsov2025ground}. The underlying modulation scheme can be used as a key element in (emitter-based) opto-mechanical devices~\cite{fuhrmann2011dynamic,blattmann2014entanglement,weiss2016surface,buhler2022chip,lodde2024strain}, 
to program the spectral properties of the generated single photons by parametric transduction~\cite{metcalfe2010resolved,golter2016optomechanical,weiss2021optomechanical,wigger2021resonance,imany2022quantum}, and to drive coherent dynamics with phonons and photons~\cite{spinnler2024single,decrescent2024coherent}.

A common phenomenon in the above-mentioned hybrid phononic systems is the appearance of time-periodic dynamics, e.g., from modulating transition energies of solid state single-photon emitters with acoustic waves. In the theoretical description of such time-periodic systems, Floquet theory has a long-standing history and can be applied to gain analytical insight into the underlying models, as well as to improve the efficiency of numerical simulations~\cite{shirley1965solution,salzman1974quantum,chu1985recent,ben1993effect,breuer2002theory,ikeda2020general}. In the context of purely optical driving of single emitters, this naturally leads to a description in terms of optically dressed states, where Floquet theory is especially helpful when considering bichromatic driving, i.e., excitation with two optical modes of different frequency~\cite{agarwal1991spectrum,ficek1993resonance,ficek1996fluorescence}. Such bichromatic optical driving leads to intricate resonance patterns observed experimentally~\cite{peiris2014bichromatic,he2015dynamically,gustin2021high}, as well as to suppressions of spectral lines~\cite{zhu1996spectral,rudolph1998shift, ficek1999quantum}. Recently,  these phenomena have also been investigated in the radio frequency domain employing Rydberg atoms~\cite{jayaseelan2023electromagnetically}.

Here, we extend the concept of optical dressing to situations where the semiclassical mechanical or elastic modulation of an emitter, called acoustic modulation throughout this paper for linguistic simplicity, leads to independent acoustic dressing of the already optically dressed states.
This results in acousto-optical double-dressing of the emitter, which is efficiently described in terms of Floquet theory. Manipulating the parameters of the acoustic modulation allows for Floquet engineering of the optical properties of the light-driven emitter, i.e., precise acoustic control over light emission and scattering. Achieving this acousto-optical double dressing experimentally requires sufficient control over the two underlying interfaces between (a) light and emitter and (b) acoustics and emitter. Therefore, in Sec.~\ref{sec:introplatforms} we start with a short review of the existing platforms for the two interfaces. Then we proceed by introducing the theoretical model for an optically driven, acoustically modulated, two-level emitter and discuss its properties in the context of Floquet theory in Sec.~\ref{sec:model}. We give a brief overview of applications of Floquet theory in the realm of optics and solid state physics in Sec.~\ref{sec:floquet_review} before making explicit use of the theory in Sec.~\ref{sec:Floquet-rep} to derive a simple closed form for the emitter's resonance fluorescence spectrum. In Sec.~\ref{sec:discussion} we present numerical simulations of such spectra to discuss spectral features that demonstrate hybrid control of the emitter via the acousto-optical double-dressing. These results are complemented by an analytical perturbative treatment of Floquet theory in Sec.~\ref{sec:floquet_bands} aiming for a deeper understanding of the physical processes leading to the observed spectral features. This theoretical discussion allows us to identify crucial system specifications that would need to be met in an experiment to achieve full acousto-optical Floquet engineering of the emitters' emission properties. This finally leads us to a discussion on the feasibility of different experimental platforms regarding their potential performance in acousto-optical Floquet engineering in Sec.~\ref{sec:platforms}.

\section{Review of experimental platforms}\label{sec:introplatforms}
\begin{figure}[b] 
\centering
\includegraphics[width=0.7\linewidth]{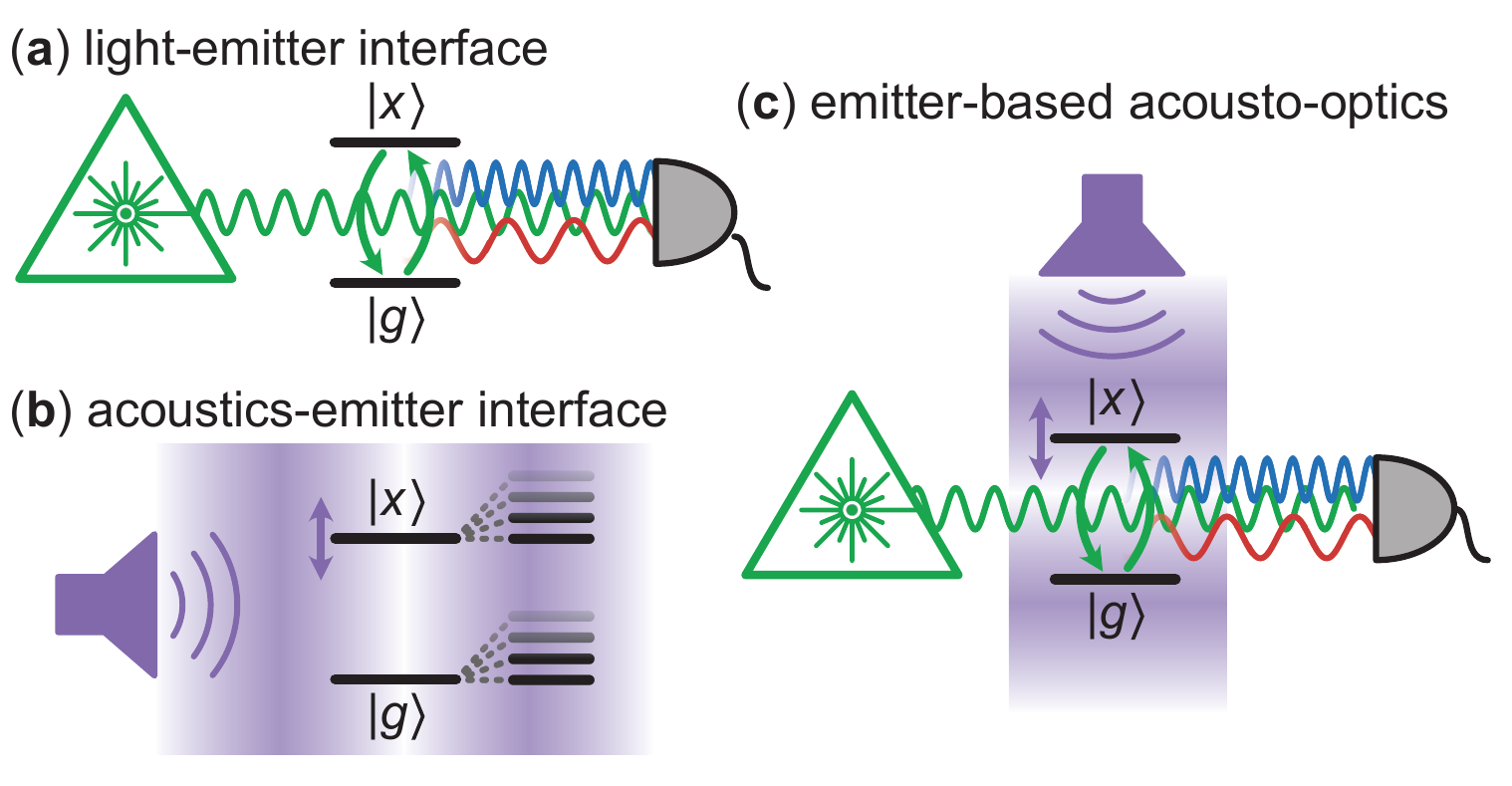}
\caption{Schematic of the considered hybridization interfaces. (a)~The light-emitter interface: A continuous wave laser (green) drives transitions in a two-level system generating light emission. (b)~The acoustics-emitter interface modulates the transition energy of the two-level system which leads to the generation of phonon replica. (c)~Combined acousto-optical driving of a quantum emitter.}
\label{fig:1}
\end{figure}

Hybrid acousto-optical schemes natively require versatile platforms that combine both elastic wave excitations and high-quality optically addressable and interfaced single-photon emitters. The fundamental constituents of such a platform are depicted schematically in Fig.~\ref{fig:1}. It comprises (a) the light-emitter interface, sketched here as an optically driven two-level system (TLS), and (b) the acoustically driven emitter. The combination of both driving schemes gives rise to (c) the fully-fledged acousto-optical dressing of the quantum emitter. 
To successfully implement the full scheme, each of the two interfaces sketched in Fig.~\ref{fig:1}(a, b) needs to be carefully optimized. Next we briefly review typical state-of-the-art solutions for both types of interfaces.

(a) \textit{Light-emitter interface}: The efficiency of the coupling between the quantum emitter, behaving as a driven TLS, and the light field is evaluated by the observation of the Mollow triplet~\cite{mollow1969power} appearing under strong optical driving. The Mollow triplet is therefore one of the hallmarks of efficient light-matter coupling. In order to reach a sufficient optical field amplitude at the location of the single-photon emitter, different cavity structures can be used~\cite{vahala2003optical}. In this regard, semiconductor quantum dots (QDs) have commonly been integrated into distributed Bragg reflector (DBR) cavities to resolve the Mollow triplet~\cite{flagg2009resonantly,ates2009post,ulhaq2012cascaded}. Furthermore, photonic crystal cavities have been used to bring silicon vacancy (SiV) color centers in nano-diamonds~\cite{zhou2017coherent}, as well as QDs~\cite{fischer2016self} into the strong driving regime. However, the Mollow triplet may also be observed without full cavity integration, i.e., without a cavity or only a DBR at the bottom of the sample for efficient light collection~\cite{vamivakas2009spin,unsleber2015observation,gustin2021high,nawrath2021resonance}, which naturally requires stronger optical pumping.

(b) \textit{Acoustics-emitter interface:} Here, among the plethora of solid-state light emitters, the interaction between single QDs and elastic excitations is arguably one of the most extensively studied combinations. QDs have been interfaced with all five broader classes of mechanical modulation platforms: bulk acoustic waves (BAWs)~\cite{bruggemann2012laser,czerniuk2017picosecond,wigger2017systematic}, clamped and free-standing nanomechanical resonators~\cite{yeo2014strain,carter2017sensing,carter2019tunable,yuan2019frequency,kettler2021inducing,finazzer2023chip,tanos2024high,spinnler2024quantum}, propagating surface acoustic waves (SAWs)~\cite{metcalfe2010resolved,schulein2015fourier,weiss2021optomechanical,wigger2021resonance,yuan2021remotely,imany2022quantum}, monolithic phononic waveguides and resonators~\cite{vogele2020quantum,imany2022quantum,lodde2024strain,decrescent2024coherent,spinnler2024single}, and heterogeneously integrated platforms combining QDs and strong piezoelectrics~\cite{weiss2014dynamic,weiss2014radio,pustiowski2015independent,nysten2017multi,nysten2020hybrid,lienhart2023heterogeneous,descamps2024acoustic}. Another widely used class of single-photon emitters considered in quantum technology are color centers in diamond, most prominently the nitrogen vacancy (NV) center~\cite{awschalom2018quantum}. The latter has been interfaced with mechanical resonators~\cite{ovartchaiyapong2014dynamic,lee2016strain,li2023mechanical} to modulate the optical emission. The connected spin degrees of freedom of color centers have been addressed resonantly by SAWs in diamond~\cite{golter2016optomechanical,maity2020coherent} and silicon carbide (SiC)~\cite{whiteley2019spin,hernandez2021acoustically}. NV centers in diamond have also been coupled to BAW resonators~\cite{mccullian2024coherent}, as well as nanomechanical oscillators~\cite{pigeau2015observation}. More recently, quantum emitters in layered van der Waals materials have shifted into focus. Single- or multilayers of these materials can be elegantly transferred onto arbitrary surfaces and substrates. This strength allows for direct interfacing with SAWs on piezoelectric materials. This approach has been demonstrated for color centers in hexagonal Boron Nitride (hBN)~\cite{lazic2019dynamically,iikawa2019acoustically} and strain-induced emitters in transition metal dichalcogenides (TMDCs)~\cite{patel2024surface,mohajerani2025acoustic}.

{\it (c) Emitter-based acousto-optics:} Finally, the integration of both the (a) light-emitter and (b) acoustics-emitter interface into emitter-based acousto-optics has been performed in a variety of different combinations of optically driven emitters and types of acoustic modulation, typically for either weak acoustic driving or weak optical driving. We will provide an overview of these platforms as well as a discussion on their feasibility in the context of acousto-optical double dressing, i.e., both strong acoustic and strong optical driving, in Sec.~\ref{sec:platforms}.

Note, that no unified terminology exists in this broad and diverse field and the identical underlying physical phenomena are commonly referred to as acoustic, elastic, (nano)mechanical, or phononic excitation which are often linked to specific experimental platforms. Thus, for simplicity, we will use the term {\it acoustic} in the remainder of this article. This term encompasses SAWs, BAWs, Lamb modes in suspended systems, and the tuning fork character of (nano)mechanical resonators.

\section{Model and Floquet formalism}\label{sec:model}
In this section we present the theoretical basics required for the discussion of hybrid acousto-optical Floquet engineering of a single-photon emitter.
\subsection{System, model and equation of motion}\label{sec:model-eom}
	We consider a solid state single-photon emitter, e.g., a QD, as schematically depicted in Fig.~\ref{fig:1}(c), whose transition frequency is modulated acoustically, e.g., via SAWs, (violet) and which is optically driven by an external continuous-wave (cw) laser (green). The emitter is modeled as a TLS with a ground state $\ket{g}$ and a bright excited state $\ket{x}$. In the frame rotating with the laser frequency, the Hamiltonian of the system is given by~\cite{weiss2021optomechanical,wigger2021resonance}
	\begin{align}\label{eq:Hamilton}
		H(t)&=\hbar\frac{\Delta(t)}{2}\qty(\ket{g}\bra{g}-\ket{x}\bra{x})\notag\\
        &\quad+\hbar\frac{\Omega_{\rm  R}}{2}\qty(\ket{x}\bra{g}+\ket{g}\bra{x})\,,
	\end{align}
	where $\Delta(t)=\omega_l-\omega_{\rm  TLS}(t)$ is the acoustically modulated detuning between laser frequency $\omega_l$ and transition frequency $\omega_{\rm  TLS}$, $\Omega_{\rm  R}$ is the Rabi frequency that describes the strength of the cw-laser driving, and we employ the dipole and rotating-wave approximations for the interaction between TLS and cw-laser~\cite{cohen2024photons,cohen2024atom}. It is crucial for the following discussion that we assume the acoustically modulated detuning to be periodic with period $T$, such that $\Delta(t+T)=\Delta(t)$, which renders the Hamiltonian itself periodic with $H(t+T)=H(t)$. 
	
    In addition to the unitary dynamics described by the Hamiltonian in Eq.~\eqref{eq:Hamilton}, we consider excited state decay with the rate $\gamma_{\rm  xd}$ and pure dephasing with the rate $\gamma_{\rm  pd}$ as described by the Lindblad dissipators~\cite{breuer2002theory,roy2011influence,groll2020four,preuss2022resonant}
	\begin{subequations}\label{eq:dissipators}
		\begin{align}
			\mathcal{D}_{\rm  xd}(\rho)&=\gamma_{\rm  xd}\qty(\ket{g}\bra{x}\rho\ket{x}\bra{g}-\frac{1}{2}\acomm{\ket{x}\bra{x}}{\rho})\,,\\
			\mathcal{D}_{\rm  pd}(\rho)&=\gamma_{\rm  pd}\qty(\ket{x}\bra{x}\rho\ket{x}\bra{x}-\frac{1}{2}\acomm{\ket{x}\bra{x}}{\rho})\,,
		\end{align}
	\end{subequations}
	acting on the system's density matrix $\rho$. A typical source for excited state decay of the emitter is spontaneous emission of a photon which can then be detected~\cite{cohen2024atom,breuer2002theory}. The origin of pure dephasing, however, may vary strongly, depending on the kind of emitter that is investigated. Possible sources are spectral jitter induced by the presence of additional fluctuating charges in the vicinity of the TLS~\cite{spokoyny2020effect,preuss2022resonant}, coupling to anharmonic phonon modes~\cite{machnikowski2006change,groll2021controlling} or the presence of additional energetically higher lying states that are virtually excited via phonons~\cite{muljarov2004dephasing}. The dynamics of the systems' density matrix $\rho(t)$ are finally described by the Lindblad equation~\cite{lindblad1976generators,breuer2002theory}
	\begin{align}\label{eq:lindblad}
		\dv{t}\rho(t)&=\mathcal{L}(t)\qty[\rho(t)]\notag\\
        &=-\frac{i}{\hbar}\comm{H(t)}{\rho(t)}+\mathcal{D}_{\rm  xd}\qty[\rho(t)]+\mathcal{D}_{\rm  pd}\qty[\rho(t)]
	\end{align}
	with $\mathcal{L}(t)=\mathcal{L}(t+T)$ denoting the Lindblad super-operator, which is defined by the right-hand side of the Lindblad equation and which is periodic due to $H(t)$ being periodic.

    Solid state single-photon emitters are inevitably coupled to phonons of the surrounding bulk medium, in addition to the coherently driven semi-classical phonon mode leading to the detuning modulation in Eq.~\eqref{eq:Hamilton}. This coupling is usually modeled in terms of the independent boson model and the coupling strength to phonon modes with frequency $\Omega$ is then quantified in terms of the phonon spectral density $J(\Omega)$. When considering the coupling to long wavelength acoustic phonons, e.g., via the deformation potential coupling, the interaction can be characterized by a super-ohmic spectral density of the form~\cite{ramsay2010phonon,roy2011influence, nazir2016modelling,groll2020four, preuss2022resonant}
    \begin{equation}
    J(\Omega)=\alpha\Omega^3 f\qty(\frac{\Omega}{\Omega_{\mathrm{c}}})
\end{equation}
with a cutoff function obeying
\begin{equation}
    f(0)=1\,,\qquad f(x\ll 1)\approx1\,,\qquad f(x\gg 1)\approx0\,.
\end{equation}
Here, $\alpha$ is a measure for the overall coupling strength and $\Omega_{\mathrm{c}}$ is the cutoff frequency, i.e., modes with higher frequencies couple inefficiently to the emitter.

    As discussed in detail in Sec.~\ref{sec:platforms} on potential experimental platforms, for the acoustic modulation of the TLS we consider frequencies on the order of $1/T\sim 5~$GHz. As will become clear in the following sections, the acousto-optical resonance $2\pi/T=\Omega_{\mathrm{R}}$ is especially interesting, implying a similar order of magnitude for the Rabi frequencies. Typical cutoff frequencies for the coupling of solid state emitters to acoustic phonons however are on the order of $\hbar\Omega_{\mathrm{c}}\sim1~$meV or higher~\cite{ramsay2010phonon,roy2011influence, alkauskas2014first,goldman2015phonon, nazir2016modelling,groll2020four, norambuena2020quantifying, preuss2022resonant}, i.e., $\Omega_{\mathrm{c}}/(2\pi)\sim250~$GHz$\gg 1/T, \Omega_{\mathrm{R}}/(2\pi)$. As discussed in detail in App.~\ref{app:phonons}, this implies that the TLS is modulated adiabatically from the point of view of the acoustic phonon modes~\cite{machnikowski2007quantum,wigger2014energy}. By making a unitary transformation to the polaron frame, we find that the polaron, i.e., the phonon-dressed TLS, and the phonon modes approximately decouple and that the polaron's dynamics are described by Eq.~\eqref{eq:lindblad}, when accounting for phonon-induced renormalization of parameters. Phonon-induced dissipation of the optically driven polaron is found to be negligible for cryogenic temperatures $\sim4$~K in the considered adiabatic driving regime for a typical coupling strength of $\alpha\sim 10^{-2}~$ps$^2$ or smaller~\cite{ramsay2010phonon, alkauskas2014first,goldman2015phonon, nazir2016modelling,groll2020four, norambuena2020quantifying, preuss2022resonant}. We furthermore discuss that the phonons lead to a constant but negligible background in the resonance fluorescence spectrum presented in the following sections, since we are interested in the form of the spectra in a frequency range determined by the acoustic modulation frequency $1/T\sim 5$~GHz, where the bulk acoustic phonon sidebands are approximately constant at $\sim4$~K. The height of this background as well as the dissipative impact of the acoustic phonons increase at least linearly with temperature, restricting the direct applicability of the polaronic two-level model in Eq.~\eqref{eq:lindblad} in an experimental context to cryogenic temperatures.

\subsection{Liouville space and formal solution of the Lindblad equation}\label{sec:liouville}
	In the following discussion it is very convenient to work in Liouville space, in which Hilbert space operators such as the density matrix $\rho$ are represented by vectors $\ketL{\rho}$ and super-operators such as $\mathcal{L}$ are represented by operators acting on these vectors~\cite{mukamel1995principles,breuer2002theory}. Since the underlying Hilbert space of the TLS is two-dimensional, Liouville space vectors such as $\ketL{\rho(t)}$ can be represented as four-component column vectors, while Liouville space operators such as $\mathcal{L}(t)$ can be represented as $4\times 4$-matrices acting on these column vectors. For any 'ket'-vector $\ketL{A}$ we can furthermore define a 'bra'-vector $\braL{A}$ by identifying it with the hermitian conjugate (complex conjugate and transpose) of the column vector corresponding to $\ketL{A}$, i.e., $\braL{A}$ is represented as a row vector. Such Liouville space representations are explicitly used in App.~\ref{app:liouville_rep}. Finally, we will make use of the Frobenius inner product in the operator space, which corresponds to the standard scalar product in Liouville space~\cite{breuer2002theory}
    \begin{equation}\label{eq:scalar_prod}
		\expL{A\big|B}=\Tr(A^{\dagger}B)\,.
	\end{equation}
    We can now formally solve the Lindblad equation with the time-evolution (super-)operator defined by~\cite{mukamel1995principles,breuer2002theory}
    \begin{equation}
        \ketL{\rho(t)}=\mathcal{V}(t,t_0)\ketL{\rho(t_0)}\,,\label{eq:action_V}
    \end{equation}
    which obeys
	\begin{equation}\label{eq:V_eom}
		\dv{t}\mathcal{V}(t,t_0)=\mathcal{L}(t)\mathcal{V}(t,t_0)\,,\qquad \mathcal{V}(t_0,t_0)=1\,,
	\end{equation}
    due to Eq.~\eqref{eq:lindblad},	as well as the semigroup property
	\begin{equation}\label{eq:semigroup}
		\mathcal{V}(t_2,t_1)\mathcal{V}(t_1,t_0)=\mathcal{V}(t_2,t_0)\,.
	\end{equation}
    Note that here and in the following we write compositions of super-operators, i.e., of Liouville space operators, as simple products. Since the solution to the ordinary differential equation \eqref{eq:V_eom} is uniquely determined by the initial condition, we find that the time-evolution super-operator obeys the discrete time translation symmetry~\cite{salzman1974quantum,chu1985recent,dai2016floquet,chen2024periodically}
	\begin{equation}\label{eq:V_symm}
		\mathcal{V}(t+T,t_0+T)=\mathcal{V}(t,t_0)
	\end{equation}
	since $\mathcal{V}(t+T,t_0+T)$ is subject to the same initial condition for $t=t_0$ as $\mathcal{V}(t,t_0)$ and obeys the same differential equation~\eqref{eq:V_eom}. This last property is due to the periodicity of the Lindblad super-operator $\mathcal{L}(t+T)=\mathcal{L}(t)$. Using Eq.~\eqref{eq:V_symm} together with the semigroup property in Eq.~\eqref{eq:semigroup} we can separate the full time evolution into the following products of operators 
	\begin{align}
		\mathcal{V}(t+nT,0)&=\mathcal{V}(t+nT,nT)\mathcal{V}(nT,nT-T)...\mathcal{V}(T,0)\notag\\&=\mathcal{V}(t+nT,nT)\mathcal{V}(T,0)^n=\mathcal{V}(t,0)\mathcal{T}^n\,,\label{eq:propagation_rewritten_floquet}
	\end{align}
    where $\mathcal{T}\equiv\mathcal{V}(T,0)$ denotes the single-period time translation. Without loss of generality, we may assume $0\leq t\leq T$ such that the dynamical properties of the system are fully encoded in the super-operators $\mathcal{V}(t,0)$ for $t\in[0,T]$. It is therefore sufficient to investigate the dynamics over a single period. 
    The property \eqref{eq:propagation_rewritten_floquet}, known as the Floquet theorem, underlies the Floquet theory of periodic dynamical systems~\cite{shirley1965solution,salzman1974quantum,chu1985recent,breuer2002theory,ikeda2020general,chen2024periodically}, which will be exploited in the following.

\subsection{Floquet state properties}
    The Floquet states of the periodically modulated system, i.e., eigenstates of the single-period time translation, obey~\cite{shirley1965solution,salzman1974quantum,chu1985recent,breuer2002theory,ikeda2020general,chen2024periodically}
	\begin{equation}\label{eq:eigen_floquet}
		\mathcal{T}\ketL{M_r}=\mu_r\ketL{M_r}\,.
	\end{equation}
	These eigenstates provide a complete basis of Liouville space, apart from possible countably many exceptional points, which have measure zero in the parameter space of our model and will therefore be ignored in the following derivation~\cite{muller2008exceptional,minganti2019quantum}. Exceptional points in non-unitary dynamical systems may lead to non-trivial behavior when varying the parameters of the system adiabatically around such points in parameter space~\cite{nasari2022observation}, which is however not relevant to the work presented here. From the eigenvalue equation~\eqref{eq:eigen_floquet} and the fact that the Lindblad equation~\eqref{eq:lindblad} conserves the trace of the density matrix~\cite{lindblad1976generators,breuer2002theory,chen2024periodically}, it follows that
	\begin{equation}\label{eq:trace_eigen_floquet}
		\Tr(M_r)=\Tr[\mathcal{T}(M_r)]=\mu_r \Tr(M_r)\,,
	\end{equation}
	i.e., the Floquet eigenvalues have to be $\mu_r=1$ or otherwise the eigenstates have to be traceless $\Tr(M_r)=0$. This shows that while the eigenstates $\ketL{M_r}$ provide a complete basis for Liouville space (within the constraints discussed above) and thus also for the density matrix, they do not all possess the property $\Tr(\rho)=1$ of a density matrix themselves. It is convenient to reparametrize the eigenvalues $\mu_r$ via amplitude and phase as
	\begin{equation}\label{eq:parametrization}
		\mu_r=e^{-i\delta_r T}e^{-\gamma_r T}
	\end{equation}
	with $\delta_r$ and $\gamma_r$ real. The values of the Floquet frequencies $\delta_r$ are obviously only defined mod\,$2\pi/T$, i.e., we may choose $\delta_r\in\left[-\pi/T,\pi/T\right)$.

    As shown in App.~\ref{app:floquet_eigen_constraints} there is exactly one solution $\ketL{M_0}$ of Eq.~\eqref{eq:eigen_floquet} with non-vanishing trace and thus $\mu_0=1$, $\delta_0=\gamma_0=0$ and there are three traceless solutions $M_r$, $r=1,2,3$ with
	\begin{equation}\label{eq:floquet_rates}
		\min(\gamma,\gamma_{\rm xd})\leq\gamma_r\leq\max(\gamma,\gamma_{\rm xd})\,,\qquad\gamma=\frac{\gamma_{\rm xd}+\gamma_{\rm pd}}{2}\,,
	\end{equation}
	i.e., they decay over time with the period-averaged Floquet decay rates $\gamma_r\geq\gamma_{\rm xd}/2>0$ if we have non-vanishing excited state decay $\gamma_{\rm xd}>0$, since [see Eq.~\eqref{eq:propagation_rewritten_floquet}]
	\begin{equation}
		\mathcal{V}(t+nT,0)\ketL{M_r}=\mathcal{V}(t+nT,nT)\mathcal{T}^n\ketL{M_r}\sim e^{-\gamma_rnT}\,.
	\end{equation}
	Due to this, the state of the system at $t=0$, assumed to have been driven for an infinite amount of time starting from an arbitrary initial state $\ketL{\rho(-\infty)}$, can formally be written via the non-decaying solution $\ketL{M_0}$ as [see also Eq.~\eqref{eq:stationary_rho}]
	\begin{equation}
		\ketL{\rho(0)}=\lim\limits_{n\rightarrow\infty}\mathcal{V}(0,-nT)\ketL{\rho(-nT)}=\frac{\ketL{M_0}}{\Tr(M_0)}\,.
        \label{eq:rho0-M0}
	\end{equation}
    
    Apart from determining the (periodic) state of the system at $t=0$, we can use Floquet theory to rewrite the time-evolution super-operator defined in Eq.~\eqref{eq:V_eom}. To do so we first define time-periodic Floquet states analog to Bloch states in spatially periodic systems via~\cite{chu1985recent,breuer2002theory,chen2024periodically}
	\begin{equation}\label{eq:def_V_r}
		\ketL{V_r(t)}=\mu_r^{-t/T}\mathcal{V}(t,0)\ketL{M_r}=e^{i\delta_r t+\gamma_rt}\mathcal{V}(t,0)\ketL{M_r}\,,
	\end{equation}
	whose periodicity can be confirmed using Eqs.~\eqref{eq:semigroup}, \eqref{eq:V_symm} and \eqref{eq:eigen_floquet} via
	\begin{align}
		\ketL{V_r(t+T)}&=\mu_r^{-1}\mu_r^{-t/T}\mathcal{V}(t+T,0)\ketL{M_r}\notag\\&=\mu_r^{-1}\mu_r^{-t/T}\mathcal{V}(t+T,T)\mathcal{T}\ketL{M_r}=\ketL{V_r(t)}\,.
	\end{align}
	With the help of these periodic states we can write
	\begin{align}
		\mathcal{V}(t,t_0)\ketL{V_r(t_0)}&=\mathcal{V}(t,t_0)e^{i\delta_r t_0+\gamma_r t_0}\mathcal{V}(t_0,0)\ketL{M_r}\notag\\
        &=e^{i\delta_r t_0+\gamma_r t_0}\mathcal{V}(t,0)\ketL{M_r}\notag\\
		&=e^{-i\delta_r(t-t_0)-\gamma_r(t-t_0)}\ketL{V_r(t)}\,.\label{eq:action_time_evol_V_r}
	\end{align}

    As discussed in App.~\ref{app:floquet_eigen_constraints}, the relation between $\ketL{M_r}$ and $\ketL{V_r(t)}$ is invertible, such that also the periodic Floquet states $\ketL{V_r(t)}$ constitute a complete basis for Liouville space. Both of these sets of basis states however are not necessarily orthonormal, due to dissipation in the system leading to non-unitary dynamics in Eq.~\eqref{eq:lindblad}. Therefore, it is convenient to use the reciprocal (dual) basis $\ketL{\overline{V}_r(t)}$ to $\ketL{V_r(t)}$~\cite{lang2012algebra}, which fulfills the joint orthonormality
    \begin{subequations}\label{eq:reciprocal_basis}
	\begin{equation}
		\expL{\overline{V}_r(t)\big|V_{r'}(t)}=\delta_{rr'}\,
	\end{equation}
   	and completeness relation
   	\begin{equation}
   		1=\sum_r \ketL{V_r(t)}\braL{\overline{V}_r(t)}\,,
        \label{eq:completeness}
   	\end{equation}
    \end{subequations}
    such that the time evolution super-operator can finally be written as
    \begin{equation}\label{eq:V_t_t_0_floquet}
        \mathcal{V}(t,t_0)=\sum_re^{-i\delta_r(t-t_0)-\gamma_r(t-t_0)}\ketL{V_r(t)}\braL{\overline{V}_r(t_0)}\,.
    \end{equation}

\section{Short overview on applications of Floquet theory}\label{sec:floquet_review}
Before using Floquet theory to calculate the resonance fluorescence spectrum of an acousto-optically modulated TLS, we briefly review some important fields in the context of quantum physics, where Floquet theory has been applied historically or is applied currently. Due to the ubiquitous nature of periodically driven systems the spectrum of potential applications is very broad and we focus mainly on topics related to this article, i.e., optics and solid state physics. 

Floquet theory has been a cornerstone in the study of atoms and molecules interacting with intense electromagnetic radiation, i.e., when the rotating wave approximation is not applicable~\cite{autler1955stark,shirley1965solution,chu1977intense,chu1985recent,chu1985threshold,potvliege1989multiphoton}. Especially the widespread availability of intense laser sources has led to an increased interest in non-linear optical phenomena in these systems. If the radiation is monochromatic, this defines the periodic driving necessary for the application of Floquet theory. One can even extend the application to cases of bichromatic driving, if the rotating wave approximation is viable~\cite{agarwal1991spectrum,ficek1993resonance,ficek1996fluorescence,peiris2014bichromatic,he2015dynamically,gustin2021high}. In such semiclassical light-matter systems Floquet theory can be used to understand the impact of non-linear optical phenomena like multi-photon transitions between the atomic or molecular levels.

While all quantum systems are open to some degree, especially in the context of solid state physics dissipation plays a central role for understanding decoherence in and thermalization behavior of different materials. When introducing an additional strong periodic driving, Floquet theory is a useful tool to analyze such driven dissipative systems~\cite{kohler1997floquet, breuer2002theory,hone2009statistical,ikeda2020general, mori2023floquet}. The external driving then leads to the presence of non-equilibrium steady states, whose properties can be inferred directly from the Floquet eigenstates. Their properties can however also be modified by tuning the parameters of the external driving, leading naturally to the concept of Floquet engineering~\cite{oka2019floquet}.

Especially in recent years, Floquet engineering of condensed matter systems has seen a significant rise in popularity, allowing for the manipulation of the Floquet (pseudo-)band structure of materials, as well as the control of topological properties and non-equilibrium phases. Many-body Floquet systems can host numerous of these non-equilibrium Floquet phases~\cite{harper2020topology} including discrete time crystals~\cite{else2016floquet,yao2017discrete,zhang2017observation, choi2017observation,rovny2018observation,else2020discrete}.

Originally proposed for graphene~\cite{oka2009photovoltaic}, periodic Floquet driving can be used to engineer a topological band structure creating Floquet topological insulators~\cite{lindner2011floquet,cayssol2013floquet,rudner2020band}. This effect was confirmed experimentally for graphene~\cite{mciver2020light,merboldt2025observation}, for surface states of a topological insulator~\cite{wang2013observation,mahmood2016selective}, and can furthermore be demonstrated in systems with acoustic~\cite{peng2016experimental} or photonic band structures~\cite{rechtsman2013photonic}, as well as in ultracold quantum gases~\cite{jotzu2014experimental}. The latter system is an excellent testbed for predictions of Floquet theory~\cite{bordia2017periodically,weitenberg2021tailoring}. Here, Floquet engineering furthermore allows for an additional avenue of controlling the interaction strength between the atoms~\cite{guthmann2025floquet}.

The band structure of materials can in principle be tailored at will, when employing optimal control theory~\cite{castro2022floquet,castro2023floquet}. Apart from graphene, TMDCs have shifted into focus in recent years, where Floquet engineering of the band structure via optical driving~\cite{aeschlimann2021survival,uchida2022diabatic}, as well as via excitonic driving, i.e., exciton-induced modulation of the band structure~\cite{pareek2026driving}, has been demonstrated. Furthermore, direct Floquet engineering of excitons in these materials can be achieved via intense lasers~\cite{kobayashi2023floquet} and the effective field can be drastically enhanced via cavity integration~\cite{zhou2024cavity}.

This short overview demonstrates the large variety of potential applications of Floquet engineering. In the following we will continue by focusing on the optically driven and acoustically modulated solid state single-photon emitter system presented in the previous section.

\section{Floquet analysis of the resonance fluorescence spectrum}
In the following Sec.~\ref{sec:Floquet-rep} we present calculations on resonance fluorescence (RF) spectra of periodically modulated, optically driven TLSs. We find a closed form for the spectrum in terms of Floquet eigenstates and eigenvalues, which allows for a particularly efficient numerical calculation. We then proceed in Sec.~\ref{sec:discussion} by presenting numerical simulations on a sinusoidally modulated TLS, whose RF spectrum exhibits resonance structures in the form of anticrossings and line suppressions. These results are analyzed in detail in Sec.~\ref{sec:floquet_bands} by employing perturbation theory.

\subsection{Floquet representation of the spectrum}\label{sec:Floquet-rep}
    Here we investigate the RF spectrum of the periodically modulated, optically driven TLS, discussed in the previous Sec.~\ref{sec:model}, in the framework of Floquet theory. The time-integrated RF spectrum is given by~\cite{mollow1969power,wigger2021resonance,groll2025read}
	\begin{align}\label{eq:I_RF}
		I_{\rm RF}(\omega_l+&\omega;\Gamma)=\notag\\
        &2\text{Re}\left[\int\limits_0^{\infty}\dd\tau\int\limits_0^T\frac{\dd t}{T}\,C(t+\tau,t)e^{-i\omega\tau-\Gamma\tau}\right]\,,
	\end{align}
	where $\Gamma$ describes the resolution of the spectrometer and $\omega$ is the frequency shift of the detected light relative to the laser frequency $\omega_l$. The argument $\omega_l+\omega$ on the left-hand side of Eq.~\eqref{eq:I_RF} accounts for the fact that we perform the calculations on the right-hand side in the frame rotating with the laser frequency $\omega_l$. The two-time correlation function $C(t+\tau,t)=\expval{\sigma_+(t+\tau)\sigma_-(t)}$, with $\sigma_+=\ket{x}\bra{g}$ and $\sigma_-=\ket{g}\bra{x}$ describing transitions in the TLS, can be calculated using the quantum regression theorem for the case of Markovian dynamics described by Eq.~\eqref{eq:lindblad}~\cite{breuer2002theory}
	\begin{equation}
		C(t+\tau,t)=\Tr\qty{\sigma_+\mathcal{V}(t+\tau,t)\qty[\sigma_-\rho(t)]}\,.\label{eq:def_C}
	\end{equation}
    Starting from an initial ground state $\rho(-\infty)=\ket{g}\bra{g}$ the system reaches the state in Eq.~\eqref{eq:rho0-M0}. Its form implies
    \begin{equation}\label{eq:rho_t}
        \ketL{\rho(t)}=\frac{\ketL{V_0(t)}}{\Tr(M_0)}\,,\qquad t\geq 0\,
    \end{equation}
    due to $\mu_0=1$, i.e., the system reaches the periodic state $\ketL{V_0(t)}$ as a consequence of dissipation and periodic driving. Thus, the $t$-integration over a single period $T$ in Eq.~\eqref{eq:I_RF} is sufficient.  In Liouville space the correlation function can be written as
	\begin{equation}\label{eq:C_liou}
		C(t+\tau,t)=\braL{\sigma_-}\mathcal{V}(t+\tau,t)\ketL{\sigma_-\rho(t)}\,,
	\end{equation}
	where we used the standard scalar product from Eq.~\eqref{eq:scalar_prod}.	
     Using $\mathcal{V}(t,t_0)$ from Eq.~\eqref{eq:V_t_t_0_floquet}, we can now write the correlation function in Eq.~\eqref{eq:C_liou} as
	\begin{align}
		&C(t+\tau,t)=\braL{\sigma_-}\mathcal{V}(t+\tau,t)\sigma_-\ketL{\rho(t)}\notag\\
		&=\frac{1}{\Tr(M_0)}\braL{\sigma_-}\sum_re^{-i\delta_r\tau-\gamma_r\tau}\ketL{V_r(t+\tau)}\braL{\overline{V}_r(t)}\sigma_-\ketL{V_0(t)}\notag\\
		&=\sum_r e^{-i\delta_r\tau-\gamma_r\tau}\frac{\braL{\sigma_0}\sigma_+\ketL{V_r(t+\tau)}\braL{\overline{V}_r(t)}\sigma_-\ketL{V_0(t)}}{\expL{\sigma_0\big|M_0}}\,,
        \label{eq:C-V}
	\end{align}
	where we used the Liouville representation of ordinary Hilbert space operators $\ketL{\sigma_-\rho}=\sigma_-\ketL{\rho}$ and the Hilbert space identity operator $\sigma_0=\ket{g}\bra{g}+\ket{x}\bra{x}$. 
    
    The virtue of Eq.~\eqref{eq:C-V} is that we can now use the periodicity of the $V_r(t)$ and $\overline{V}_r(t)$ to provide a rather simple form for the RF spectrum in Eq.~\eqref{eq:I_RF} using a representation of the matrix elements appearing in the previous result via their Fourier components. For a general time-periodic function $f(t)=f(t+T)$ we use the Fourier series representation
	\begin{equation}
		f(t)=\sum_m e^{-im\Omega t}f^{(m)}
	\end{equation}
	with $\Omega=2\pi/T$ and using a superscript $(m)$ to denote the Fourier components of $f(t)$
	\begin{equation}
		f^{(m)}=\frac{1}{T}\int\limits_0^T \dd t\,f(t)e^{im\Omega t}\,.
	\end{equation}
	The RF spectrum in Eq.~\eqref{eq:I_RF} is finally given by~\cite{ho1986floquet,yan2016exotic,felicetti2018ultrastrong}
	\begin{align}\label{eq:I_RF_full}
		&I_{\rm RF}(\omega_l+\omega;\Gamma)\\
		&=\frac{2}{\expL{\sigma_0\big|M_0}}\sum_m\sum_r \text{Re}\left[\frac{\expL{V_r\big|\sigma_-\big|\sigma_0}^{(m)*}\expL{\overline{V}_r\big|\sigma_-\big|V_0}^{(m)}}{i(\omega+\delta_r-m\Omega)+(\Gamma+\gamma_r)}\right]\notag\,.
	\end{align}
	The spectrum is a sum of Lorentzians with positions given by the Floquet frequencies $\delta_r$, plus integer multiples of the periodicity frequency $\Omega$. The widths are determined by the period-averaged Floquet decay rates $\gamma_r$ with an additional broadening due to the spectrometer resolution $\Gamma$. As discussed previously, there is one Floquet state with $\gamma_0=0$, whose corresponding peaks in the spectrum are broadened only due to the spectrometer resolution. This is the coherent component of the RF spectrum. The other three Floquet states have widths bound by $\gamma_r\geq \gamma_{\rm xd}/2$, constituting the incoherent part of the RF spectrum~\cite{konthasinghe2012coherent,phillips2020photon,hanschke2020origin}.
    
    Note that the form of the RF spectrum in Eq.~\eqref{eq:I_RF_full} is valid not only for the specific Hamiltonian in Eq.~\eqref{eq:Hamilton}, but for any TLS with a time-periodic Hamiltonian subject to excited state decay $\gamma_{\rm xd}>0$, as this ensures $\gamma_r>0$ for three traceless solutions, while there is only one non-decaying solution with periodic Floquet state $\ketL{V_0(t)}$ as discussed in App.~\ref{app:floquet_eigen_diss}.

\subsection{Resonances and line suppressions in resonance fluorescence under sinusoidal acoustic modulation}\label{sec:discussion}

\begin{figure*}[t] 
\centering
\includegraphics[width=\textwidth]{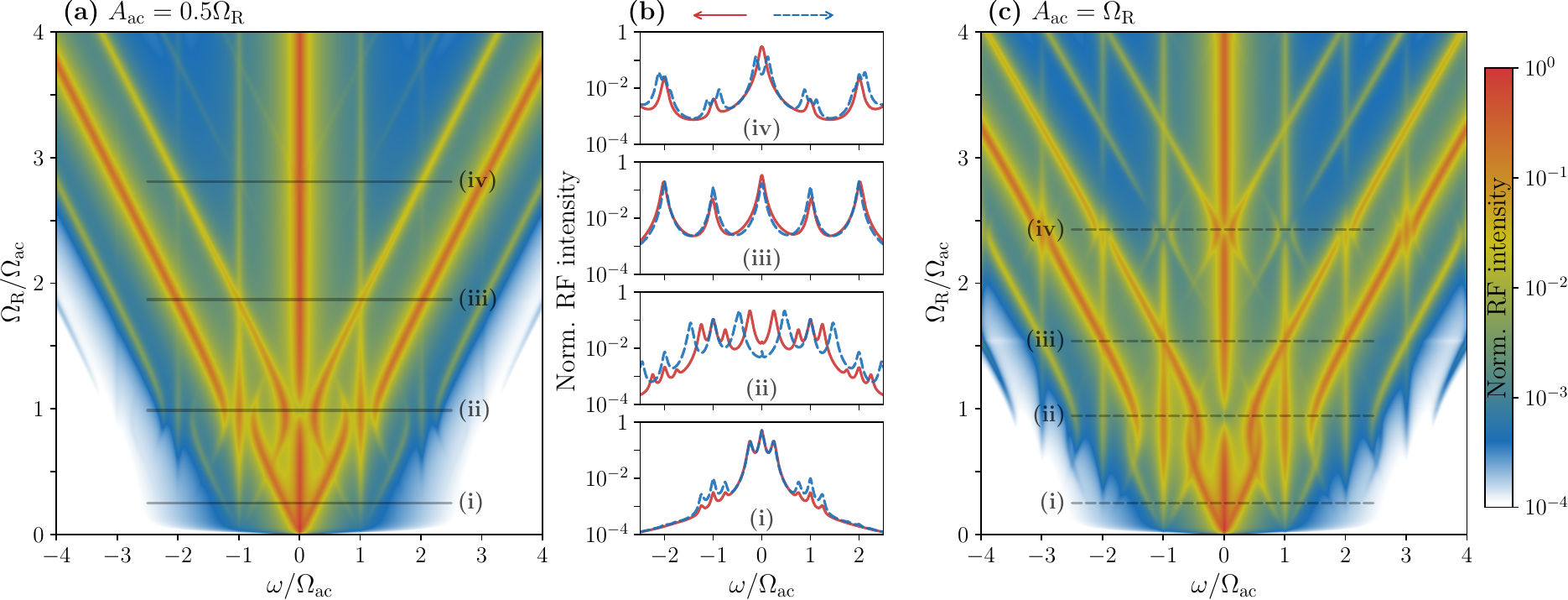}
\caption{Numerical simulations of RF spectra using Eq.~\eqref{eq:I_RF_full}. We consider sinusoidal acoustic modulation and resonant optical excitation of the TLS, i.e., the detuning given in Eq.~\eqref{eq:detuning_sinus}, implying $\Omega=\Omega_{\rm ac}$ in Eq.~\eqref{eq:I_RF_full}. (a) and (c): RF spectra as a function of the emission frequency shift $\omega$ and the Rabi frequency $\Omega_{\rm R}$ in units of the acoustic modulation frequency $\Omega_{\rm ac}$. We consider a fixed ratio between acoustic modulation amplitude $A_{\rm ac}$ and Rabi frequency $\Omega_{\rm R}$ for better visibility of the resonance structure in the spectrum, where in (a) we have $A_{\rm ac}/\Omega_{\rm R}=0.5$ and in (c) we have $A_{\rm ac}/\Omega_{\rm R}=1$. In addition (b) shows cuts in (a) and (c) at fixed values of the Rabi frequency $\Omega_{\rm R}$. Solid red lines correspond to solid horizontal cuts in (a), while dashed blue lines correspond to dashed horizontal cuts in (c), as indicated by the labels (i)-(iv), respectively. The finite line broadening is due to the spectrometer resolution $\Gamma/\Omega_{\rm ac}=10^{-2}$ and the dissipation rates $\gamma_{\rm xd}/\Omega_{\rm ac}=4\times10^{-2}$, $\gamma_{\rm pd}/\Omega_{\rm ac}=2\times10^{-2}$.}
\label{fig:2}
\end{figure*}  
In the following we will present numerical simulations for the RF spectrum employing Eq.~\eqref{eq:I_RF_full}, which in this form is a much less demanding numerical calculation, compared to a more direct simulation of Eq.~\eqref{eq:I_RF}~\cite{gustin2021high}. The reason is of course that all properties of the Floquet states, Floquet frequencies and decay rates appearing in Eq.~\eqref{eq:I_RF_full} can be determined from a simulation of the dynamics on the single-period time interval $[0,T]$, which then leads to the solution of the Floquet eigenvalue equation~\eqref{eq:eigen_floquet}. We will focus on the case of a sinusoidal acoustic modulation and resonant cw-laser driving, leading to the time-dependent detuning~\cite{weiss2021optomechanical,wigger2021resonance}
	\begin{equation}\label{eq:detuning_sinus}
		\Delta(t)=A_{\rm ac}\sin\qty(\Omega_{\rm ac} t)
	\end{equation}
	with $A_{\rm ac}$ and $\Omega_{\rm ac}$ denoting the amplitude and frequency of the modulation, respectively. The period of the dynamics is therefore $T=2\pi/\Omega_{\rm ac}$ and with respect to Eq.~\eqref{eq:I_RF_full} this implies $\Omega=\Omega_{\rm ac}$. The Hamiltonian in Eq.~\eqref{eq:Hamilton} thus contains the three independent parameters $\Omega_{\rm R}$, $\Omega_{\rm ac}$ and $A_{\rm ac}$.

	Figure~\ref{fig:2}(a) shows numerical results for the RF spectrum as a function of the emission frequency shift $\omega$ and Rabi frequency $\Omega_{\rm R}$, both in units of the acoustic frequency $\Omega_{\rm ac}$. We keep the ratio of acoustic modulation amplitude to Rabi frequency fixed at $A_{\rm ac}=0.5\Omega_{\rm R}$, i.e., the amplitude increases with the Rabi frequency, to provide better visibility for the resonance structure that is discussed in the following. The dissipation rates are chosen as $\gamma_{\rm xd}=4\times10^{-2}\Omega_{\rm ac}$ and $\gamma_{\rm pd}=2\times10^{-2}\Omega_{\rm ac}$, while the spectrometer resolution is set to $\Gamma=10^{-2}\Omega_{\rm ac}$. Note that in the displayed spectra $\omega=0$ means that the detected light actually has the frequency $\omega_l$ of the driving cw-laser, since we perform the calculations in the frame rotating with this laser frequency. In addition to the false color plot in Fig.~\ref{fig:2}(a), we provide spectra at fixed Rabi frequency $\Omega_{\rm R}$ in (b) (solid red lines). These correspond to horizontal cuts in (a), marked by the solid lines.
    
    Starting at small Rabi frequencies $\Omega_{\rm R}\ll\Omega_{\rm ac}$ and acoustic modulation amplitudes $A_{\rm ac}\ll\Omega_{\rm ac}$ in Fig.~\ref{fig:2}(a), i.e., below line (i), we observe the well-known Mollow triplet structure around $\omega=0$~\cite{mollow1969power}. To understand its origin, we consider the Hamiltonian from Eq.~\eqref{eq:Hamilton}. In absence of any acoustic modulation and for resonant cw-laser driving, i.e., with $\Delta(t)=0$, the eigenstates of the system are the following optically dressed states~\cite{cohen2024atom}
	\begin{equation}\label{eq:def_dressed_states}
		\ket{\pm}=\frac{1}{\sqrt{2}}\qty(\ket{g}\pm\ket{x})\,.
	\end{equation}
    Written in this basis, the Hamiltonian from Eq.~\eqref{eq:Hamilton} reads
	\begin{align}\label{eq:H_dressed_states}
		H(t)&=\hbar\frac{\Omega_{\rm  R}}{2}\qty(\ket{+}\bra{+}-\ket{-}\bra{-})\notag\\
        &\quad+\hbar\frac{\Delta(t)}{2}\qty(\ket{+}\bra{-}+\ket{-}\bra{+})\,.
	\end{align}
	  In the limit $\Delta(t)\rightarrow 0$, which approximately describes the situation below line (i) in Fig.~\ref{fig:2}(a), the optically dressed states $\ket{\pm}$ have the energies $\pm\hbar\Omega_{\rm{R}}/2$. The three peaks stem from light-induced scattering processes leading to transitions between these optically dressed states $\ket{\pm}$. There are thus four possible transitions, i.e., $\ket{\pm}\rightarrow\ket{\mp}$ with transition frequencies $\omega=\pm\Omega_{\rm R}$ and $\ket{\pm}\rightarrow\ket{\pm}$ with transition frequency $\omega=0$, which together constitute the Mollow triplet around $\omega=0$~\cite{mollow1969power,scully1997quantum}.

    In addition to this triplet at $\omega=0$ we find its replicas around $\omega=\pm\Omega_{\rm ac}$. These replicas stem from the periodic acoustic modulation and can be interpreted in terms of light-scattering processes where a phonon is absorbed or emitted, leading to energy gain $(\omega=+\Omega_{\rm ac})$ or loss $(\omega=-\Omega_{\rm ac})$, respectively~\cite{metcalfe2010resolved,wigger2017systematic,weiss2021optomechanical,wigger2021resonance}. They are therefore usually named phonon sidebands (PSBs), while the center triplet structure at $\omega=0$ corresponds to the zero-phonon line (ZPL). With respect to the spectrum in Eq.~\eqref{eq:I_RF_full} we will thus call the contribution with $m=0$ the ZPL and the contributions with $m\neq 0$ the PSBs. Since we are in the semiclassical regime, both phonon emission and absorption are equally likely, leading here to equal intensities for both PSBs at $\omega=\pm\Omega_{\rm ac}$~\cite{wigger2021resonance,groll2025read}. The center Mollow triplet and its two PSB replicas at $\omega=\pm\Omega_{\rm ac}$ are also clearly visible in the cut shown in Fig.~\ref{fig:2}(b, i) (red, solid). 

    Moving to larger Rabi frequencies around $\Omega_{\rm R}=\Omega_{\rm ac}$ in Fig.~\ref{fig:2}(a), i.e., around the position of cut (ii), we find anti-crossings of spectral lines together with a suppression of the center line at $\omega=0$. As displayed in the cut in (b, ii) (red, solid), this leads to a doublet peak structure around the center with a minimum at $\omega=0$, apart from a barely visible narrow peak sitting on top of this minimum, whose origin is discussed later in the context of Fig.~\ref{fig:2}(c). This behavior is a clear indication of resonant interactions between the optically dressed states $\ket{\pm}$~\cite{agarwal1991spectrum,ficek1993resonance,ficek1996fluorescence,peiris2014bichromatic,he2015dynamically,gustin2021high}, induced here by the acoustic modulation. Looking at the Hamiltonian in Eq.~\eqref{eq:H_dressed_states} together with the form of the detuning in Eq.~\eqref{eq:detuning_sinus}, it is clear that the acoustic modulation can drive transitions between the dressed states effectively, when their transition frequency $\Omega_{\rm R}/2-(-\Omega_{\rm R}/2)=\Omega_{\rm R}$ is resonant to the modulation frequency $\Omega_{\rm ac}$. The details of this process and the suppression of the center line are discussed in the following Sec.~\ref{sec:floquet_bands}.

    Increasing the Rabi frequency further to $\Omega_{\rm R}\approx 2\Omega_{\rm ac}$ in Fig.~\ref{fig:2}(a), i.e., around the position of cut (iii), we find that the spectral lines simply cross instead of showing anti-crossing behavior. This line crossing is here only well visible for the first PSBs, but will be more transparent when discussing larger acoustic modulation amplitudes in the following. The corresponding cut in (b, iii) exhibits a single peak at $\omega=0$ together with PSBs at $\omega=\pm n\Omega_{\rm ac}$ for $n=1,2$ (red line). At $\Omega_{\rm R}\approx 2\Omega_{\rm ac}$ the spectrum is therefore structurally similar to RF spectra at weak optical driving $\Omega_{\rm R}\ll\Omega_{\rm ac}$ but sufficiently strong acoustic driving $A_{\rm ac}\sim \Omega_{\rm ac}$, which are known to exhibit a single ZPL and additional PSBs~\cite{metcalfe2010resolved,weiss2021optomechanical,wigger2021resonance,imany2022quantum,groll2025read}.
    
    The line-crossings in Fig.~\ref{fig:2}(a) around cut (iii) indicate that in contrast to the anti-crossings around (ii), there is no efficient interaction between the optically dressed states induced by the acoustic modulation. If we again look at the Hamiltonian in the dressed state basis in Eq.~\eqref{eq:H_dressed_states}, we see that this resembles a simple Rabi model for the detuning from Eq.~\eqref{eq:detuning_sinus}. It is well known that in the Rabi model odd harmonic generation, i.e., transitions for $\Omega_{\rm R}=\Omega_{\rm ac}, 3\Omega_{\rm ac}, 5\Omega_{\rm ac}, ...$ in our case, is allowed, while even harmonic generation, i.e., transitions for $\Omega_{\rm R}=2\Omega_{\rm ac}, 4\Omega_{\rm ac}, ...$, is parity forbidden~\cite{shirley1965solution, ben1993effect} (see Sec.~\ref{sec:floquet_bands} for a detailed discussion). This explains the anti-crossings at $\Omega_{\rm R}\approx \Omega_{\rm ac}$ (a, ii) and the line crossings at $\Omega_{\rm R}\approx 2\Omega_{\rm ac}$ (a, iii). It also predicts that at even larger Rabi frequencies of $\Omega_{\rm R}\approx 3\Omega_{\rm ac}$ in (a, iv) another anti-crossing structure should appear, which is however not visible here due to line broadening. Consequently there is no doublet structure around a minimum at $\omega=0$ in (b, iv) (red, solid), which was however present in (b, ii). Staying in the language of high-harmonics generation, which is here induced by the acoustic modulation, i.e., the phonons, we can tentatively claim that there is no visible anti-crossing structure at $\Omega_{\rm R}\approx 3\Omega_{\rm ac}$, since we are in a regime where phonon-induced third harmonic generation is much less efficient than phonon-induced resonant excitation such that it cannot overcome the line broadening. 

    To test this claim we increase the influence of this phonon-induced high harmonics generation on the RF spectrum in Fig.~\ref{fig:2}(c) by considering the same situation as in (a), but with a doubled acoustic modulation amplitude $A_{\rm ac}=\Omega_{\rm R}$. As before we show cuts of the spectra for certain Rabi frequencies in (b) as blue dashed lines, corresponding to dashed horizontal lines in (c), marked by (i-iv). We can identify a few overall trends due to the doubling of the acoustic amplitude: 
    
    (A) PSBs are more pronounced. This can be seen well in (b, i), where the intensity of the first PSBs at $\omega=\pm\Omega_{\rm ac}$ is increased between the cut in (a) (red, solid) and the cut in (c) (blue, dashed). Furthermore also second PSBs at $\omega=\pm2\Omega_{\rm ac}$ start to appear now (blue, dashed).

    (B) The positions of the anti-crossings and crossings shift to smaller Rabi frequencies, which is accounted for by shifting the positions of the cuts (ii-iv) in (c). This corresponds to a renormalization of the Rabi frequency, i.e., the transition frequency between the optically dressed states, that is induced by the stronger acoustic modulation. This effect is well-known from spins in magnetic fields as Bloch-Siegert shift~\cite{bloch1940magnetic,shirley1965solution} and is discussed further in App.~\ref{app:bloch_siegert}.

    (C) The line crossing behavior for even harmonic generation in (c, iii) is now visible for the center line as well as for more PSBs compared with (a, iii). The two lines crossing the center line are suppressed at crossing. As discussed in App.~\ref{app:crossing_suppression}, this is a general feature of the spectrum, valid for any acoustic modulation amplitude $A_{\rm ac}$. It is therefore not a new feature compared with (a), but is now visible due to the overall stronger PSBs, discussed in point (A). In simulations with less line broadening this suppression of lines crossing the center line becomes much better visible (see Fig.~\ref{fig:appB} in App.~\ref{app:unitary_limit}).
    
    (D) The anti-crossing in (c, ii) becomes more pronounced, i.e., the peaks in the doublets around $\omega=0$ in (b, ii) are repelled from each other (blue, dashed vs. red, solid line). In addition the predicted anti-crossing for phonon-induced third harmonic generation in (c, iv) appears, leading to a doublet in (b, iv) around $\omega=0$ (blue, dashed). Furthermore the minimum at $\omega=0$, i.e., the suppression of the center line of the RF spectrum is more pronounced at stronger acoustic modulation [(b, ii) and (b, iv); blue, dashed vs. red, solid lines]. We can now even see such a suppression on the first PSBs at $\omega=\pm\Omega_{\rm ac}$ in (c, iv). This line suppression however is at a slightly different Rabi frequency compared to the center line, which can be seen from the cut in (b, iv) showing triplets and not doublets of peaks at $\omega=\pm\Omega_{\rm ac}$ (blue, dashed).
    
    (E) We finally discuss the narrow peak at exactly $\omega=0$, that is present at anti-crossing in (b, ii). The more pronounced suppression of the center line discussed in (D) makes this feature more visible for stronger acoustic modulation (blue, dashed vs. red, solid line). Its physical origin can be understood when considering the form of the RF spectrum in Eq.~\eqref{eq:I_RF_full}. As discussed in the corresponding derivation, there are three Floquet states decaying with the period-averaged rates $\gamma_r>\gamma_{\rm xd}/2$, while there is one stationary state with vanishing decay rate $\gamma_0=0$ and vanishing Floquet frequency $\delta_0=0$. Thus, the narrow peak at $\omega=0$ stems from the stationary solution, i.e., the coherent component of the RF spectrum broadened only by the spectrometer resolution $\Gamma$. As discussed in App.~\ref{app:unitary_limit}, the corresponding periodic Floquet state is of the form
    \begin{equation}
        \ketL{V_0(t)}=\frac{1}{2}\ketL{\sigma_0}+\mathcal{O}\qty(\frac{\gamma_{\rm xd}}{\Omega_{\rm ac}})\,,
    \end{equation}
    i.e., it describes the optically transparent 50/50 mixture of ground and excited state ($\sigma_0=\ket{g}\bra{g}+\ket{x}\bra{x}$), apart from corrections due to the excited state decay. To ensure normalization of $\ketL{V_0(t)}$, these corrections have to be trace-less, i.e., superpositions of $\ketL{\sigma_+}$, $\ketL{\sigma_-}$, and $\ketL{\sigma_3}$ with $\sigma_3=\ket{g}\bra{g}-\ket{x}\bra{x}$ being the third Pauli matrix, i.e., the inversion of the TLS. This form of the non-decaying periodic Floquet state implies
    \begin{equation}
     \ketL{\overline{V}_0(t)}=\ketL{\sigma_0}+\mathcal{O}\qty(\frac{\gamma_{\rm xd}}{\Omega_{\rm ac}})\,,
    \end{equation}
    such that the coherent component with $r=0$ vanishes in the RF spectrum in Eq.~\eqref{eq:I_RF_full} in zeroth order with respect to the excited state decay rate $\gamma_{\rm xd}$, since, e.g.,
    \begin{equation}
        \expL{\overline{V}_{r=0}\big|\sigma_-\big|V_0}=\frac{1}{2}\expL{\sigma_0\big|\sigma_-\big|\sigma_0}+\mathcal{O}\qty(\frac{\gamma_{\rm xd}}{\Omega_{\rm ac}})=\mathcal{O}\qty(\frac{\gamma_{\rm xd}}{\Omega_{\rm ac}})\,.
    \end{equation}
    Including higher orders in $\gamma_{\rm xd}/\Omega_{\rm ac}$ leads to the appearance of the coherent component, i.e., a peak whose linewidth is determined only by the spectrometer broadening $\Gamma$ and whose position is at $\omega=0$. This is due to the fact that the excited state decay together with optical driving generates inversion ($\sigma_3$) and polarization $(\sigma_\pm)$ contributions in the stationary state. For the parameters considered in Fig.~\ref{fig:2} with $\gamma_{\rm xd}/\Omega_{\rm ac}\sim 10^{-2}$ we thus get a narrow peak at $\omega=0$ with a correspondingly small RF intensity.

    \begin{figure}[t] 
    	\centering
\includegraphics[width=0.7\linewidth]{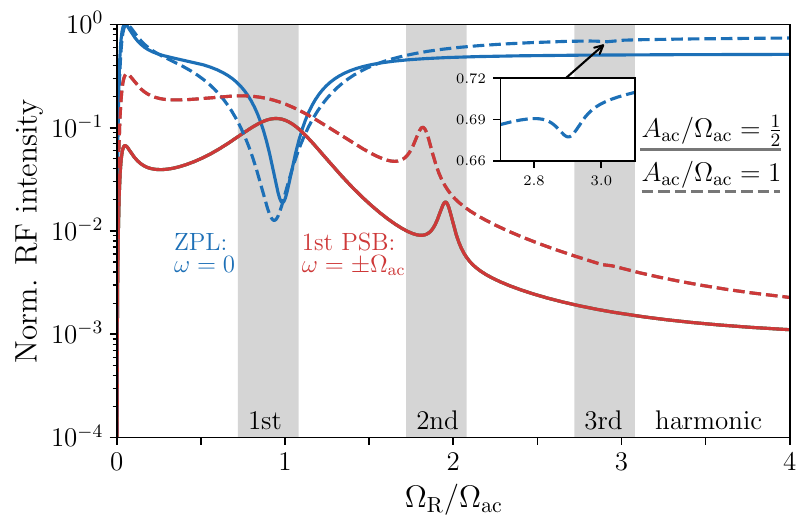}
\caption{Numerical simulations of RF spectra using Eq.~\eqref{eq:I_RF_full} for fixed emission frequency shift $\omega$ as a function of the Rabi frequency $\Omega_{\rm R}$. ZPL ($\omega=0$) in blue, first PSBs $(\omega=\pm\Omega_{\rm ac})$ in red. The acoustic modulation amplitude is fixed at $A_{\rm ac}/\Omega_{\rm ac}=\frac{1}{2}$ (solid lines) and $A_{\rm ac}/\Omega_{\rm ac}=1$ (dashed lines). The parameters $\Gamma$, $\gamma_{\rm xd}$ and $\gamma_{\rm pd}$ constituting the line broadening are kept as in Fig.~\ref{fig:2}. The gray shaded areas denote the regions of possible high harmonic resonances with $\Omega_{\rm R}\approx n\Omega_{\rm ac}$ and $n=1,2,3,...$.}
\label{fig:3}
\end{figure}   

    To get a more quantitative picture of the line suppression of the center line at odd harmonic resonances [feature (D) discussed above], in Fig.~\ref{fig:3} we show the ZPL part of the RF spectrum ($\omega=0$, blue lines) and the first PSBs ($\omega=\pm\Omega_{\rm ac}$, red lines) as a function of the Rabi frequency $\Omega_{\rm R}$ for different strengths of the acoustic modulation. This would correspond to vertical cuts in Figs.~\ref{fig:2}(a) and (c) with the difference being that now instead of $A_{\rm ac}/\Omega_{\rm R}$ we fix $A_{\rm ac}/\Omega_{\rm ac}$ to a value of $\frac{1}{2}$ (solid lines) and $1$ (dashed lines). The parameters $\Gamma$, $\gamma_{\rm xd}$, and $\gamma_{\rm pd}$ constituting the line broadening are kept as in Fig.~\ref{fig:2}. The gray shaded areas denote the regions of possible high harmonic resonances with $\Omega_{\rm R}\approx n\Omega_{\rm ac}$ and $n=1,2,3,...$, which are also the regions where the lines exhibit several extrema. 
    
    Focusing first on the ZPL (blue lines), we see a dominant line suppression for resonant driving, i.e., the first harmonic, $\Omega_{\rm R}\approx\Omega_{\rm ac}$. When increasing the acoustic modulation amplitude (blue, dashed line), the line suppression becomes more pronounced [see (D) from the previous discussion] and its position moves to smaller Rabi frequencies due to the Bloch-Siegert shift [see (B) from the previous discussion]. Furthermore another line suppression just barely starts to emerge on the third harmonic at increased acoustic modulation amplitudes (blue, dashed line;  see also the enlargement in the inset).

    On the first PSBs (red lines), we find a strong line enhancement for the second harmonic $\Omega_{\rm R}\approx 2\Omega_{\rm ac}$, which is due to a line-crossing [see Fig.~\ref{fig:2}(c, iii)]. The absence of such a clear enhancement on the ZPL (blue lines) is consistent with point (C) from the previous discussion, i.e., that lines get dim when crossing the center line for even harmonics (see also Fig.~\ref{fig:appB} in App.~\ref{app:unitary_limit}, as well as App.~\ref{app:crossing_suppression}). We generally find that the intensity of the first PSBs is increased when increasing the acoustic modulation amplitude $A_{\rm ac}$ [dashed vs. solid red lines, see also (A) from the previous discussion].

    To conclude, the resonance structure present in the RF spectra in Figs.~\ref{fig:2} and \ref{fig:3} can be attributed to phonon-induced high-harmonic generation mediated by the semiclassical acoustic modulation in Eq.~\eqref{eq:detuning_sinus}. This origin of the resonance structure will be investigated in more detail in the following section.

\subsection{Floquet spectrum and perturbation theory}\label{sec:floquet_bands}

    In the following we will discuss the origin of the various features of the RF spectrum highlighted in the previous section, i.e., anti-crossings for roughly odd ratios of Rabi frequency and acoustic modulation frequency $\Omega_{\rm R}/\Omega_{\rm ac}$ and resulting suppression of lines, as well as crossings for even ratios. To this aim we investigate the Floquet spectrum, i.e., the quasi-energies, of the system~\cite{shirley1965solution,chu1985recent,breuer2002theory}. To keep the discussion as simple as possible, we consider the unitary limit of the dynamics in Eq.~\eqref{eq:lindblad}, i.e., we want to go to the limit of vanishing pure dephasing $\gamma_{\rm pd}\rightarrow 0$ and vanishing decay $\gamma_{\rm xd}\rightarrow 0$ in a meaningful fashion. This is clearly non-trivial since in the derivation of the RF spectrum in Eq.~\eqref{eq:I_RF_full} we used that three of the Floquet states decay with a rate of at least $\gamma_r\geq \gamma_{\rm xd}/2$, to obtain the correct stationary state. For a detailed discussion of the correct unitary limit we refer to App.~\ref{app:unitary_limit}. 

    In this limit, the Floquet eigenvalue equation~\eqref{eq:eigen_floquet} reads~\cite{salzman1974quantum,chu1985recent,breuer2002theory,chen2024periodically}
    \begin{equation}\label{eq:eigen_floquet_unitary_liouville}
		\mathcal{T}\qty[M_r]=U(T,0)M_rU^{\dagger}(T,0)=e^{-i\delta_r T} M_r\,,\quad \gamma_{\rm xd,pd}=0\,,
	\end{equation}
    with $U(T,0)$ being the unitary time-evolution operator, i.e., the  solution of [analog to Eq.~\eqref{eq:V_eom}]
	\begin{equation}\label{eq:U_eom}
		\dv{t}U(t,t_0)=-\frac{i}{\hbar}H(t)U(t,t_0)\,,\qquad U(t_0,t_0)=1\,.
	\end{equation}
    Note that this operator also fulfills a semigroup property 
    \begin{equation}\label{eq:semigroup_unitary}
        U(t_2,t_1)U(t_1,t_0)=U(t_2,t_0)
    \end{equation}
    analog to Eq.~\eqref{eq:semigroup}. The Floquet eigenvalue equation~\eqref{eq:eigen_floquet_unitary_liouville} can now be solved using its Hilbert space pendant
    \begin{equation}\label{eq:eigen_floquet_unitary}
		U(T,0)\ket{\psi_\alpha}=e^{-i\epsilon_\alpha T}\ket{\psi_\alpha}
	\end{equation}
    leading us to the solutions $M_{r=\alpha\beta}=\ket{\psi_\alpha}\bra{\psi_\beta}$ with $\delta_{r=\alpha\beta}=\epsilon_\alpha-\epsilon_\beta$. From symmetry considerations of the Floquet eigenvalues discussed in Sec.~\ref{app:unitary_nonpert} it follows that the two unitary Floquet frequencies have equal absolute value and opposite sign, $\epsilon_\alpha=\pm|\epsilon|$.
    
    The positions of the peaks in the RF spectrum in Eq.~\eqref{eq:I_RF_full} are therefore determined by the differences between the (quasi)energies $\hbar\epsilon_\alpha$ associated with the Hilbert space Floquet eigenstates $\ket{\psi_\alpha}$ (plus multiples of the acoustic modulation frequency $\Omega=\Omega_{\rm ac}$). Using the periodicity $U(t+T,t_0+T)=U(t,t_0)$ analog to Eq.~\eqref{eq:V_symm}, we can also define periodic Floquet states analog to Eq.~\eqref{eq:def_V_r}~\cite{chu1985recent,breuer2002theory,chen2024periodically}
	\begin{equation}\label{eq:def_u_alpha}
		\ket{u_\alpha(t)}=e^{i\epsilon_\alpha t}U(t,0)\ket{\psi_\alpha}=\ket{u_\alpha(t+T)}
	\end{equation}
	and these can be used to simplify the RF spectrum in the unitary limit, as discussed in App.~\ref{app:unitary_limit}, leading to 
    \begin{equation}\label{eq:I_RF_unitary}
		I_{\rm RF}(\omega_l+\omega;\Gamma)=\sum_{n}\sum_{\alpha\beta}\frac{\Gamma\qty|P_{\alpha\beta}^{(n)}|^2}{\Gamma^2+(\omega-n\Omega+\epsilon_\alpha-\epsilon_\beta)^2}\,.
	\end{equation}
    The peak amplitudes are determined by the Fourier components
	\begin{equation}
		P_{\alpha\beta}^{(n)}=\frac{1}{T}\int\limits_0^T \dd t\,\expval{u_\alpha(t)|\sigma_-|u_\beta(t)}e^{in\Omega t}\,
	\end{equation}
	of matrix elements between the periodic Floquet states
	\begin{equation}\label{eq:def_P_alpha_beta}
		P_{\alpha\beta}(t)=\expval{u_\alpha(t)|\sigma_-|u_\beta(t)}=\expval{u_\alpha(t)|g}\expval{x|u_\beta(t)}\,.
	\end{equation} 
    Note that the cw-laser, interacting with the dipole of the TLS, drives the transitions between $\ket{g}$ and $\ket{x}$ [see Eq.~\eqref{eq:Hamilton}]. Due to this we call these $P_{\alpha\beta}$ the transition dipole matrix elements.
    
    To better understand the resonance structure in Figs.~\ref{fig:2} and \ref{fig:3}, we simply need to investigate the properties of the Floquet frequencies $\epsilon_{\alpha}$ and periodic Floquet states $\ket{u_{\alpha}(t)}$. From the definition of the periodic Floquet states in Eq.~\eqref{eq:def_u_alpha}, using Eq.~\eqref{eq:U_eom}, we obtain the equation
	\begin{equation}\label{eq:floquet_states_eom}
		i\hbar\dv{t}\ket{u_\alpha(t)}=\qty[H(t)-\hbar\epsilon_\alpha]\ket{u_\alpha(t)}\,.
	\end{equation}
    Since the states $\ket{u_\alpha(t)}$, as well as the operator $H(t)$, are periodic with period $T$, it is useful to define a scalar product
    \begin{equation}\label{eq:scalar_prod_floquet_unitary}
        \expval{\expval{u|v}}=\frac{1}{T}\int\limits_0^T\dd t\, \expval{u(t)|v(t)}
    \end{equation}
    for $T$-periodic states in the Hilbert space~\cite{breuer2002theory}. For any orthonormal Hilbert space basis $\qty{\ket{\sigma}}$, the states 
    \begin{equation}\label{eq:floquet_basis}
        \left.\ket{\sigma,n}\right>=\ket{\sigma}e^{in\Omega t}
    \end{equation}
    constitute a complete orthonormal basis with respect to this scalar product. 
    
    Note that by multiplying the periodic Floquet states $\ket{u_\alpha(t)}$ with a time-dependent periodic phase $e^{im\Omega t}$, we obtain a new solution to Eq.~\eqref{eq:floquet_states_eom} with Floquet frequency $\epsilon_{\alpha}+m\Omega$. However such a multiplication by a time-dependent phase does not change any of the physical properties of quantum mechanical states, i.e., all solutions of the form $\ket{u_\alpha(t)}e^{im\Omega t}$ are physically equivalent, as are the corresponding Floquet frequencies $\epsilon_{\alpha}+m\Omega$. This is consistent with the fact that the $\epsilon_\alpha$ in Eq.~\eqref{eq:eigen_floquet_unitary} are only defined modulo~$\Omega$~\cite{breuer2002theory}. Multiplying every basis state in Eq.~\eqref{eq:floquet_basis} by $e^{im\Omega t}$ leads to $\left.\ket{\sigma,n}\right>\rightarrow \left.\ket{\sigma,n+m}\right>$. Since this multiplication with a global phase does not change any of the physical results, only differences between different Floquet quantum numbers $n'-n''$ have physical relevance.    
    
    We can now calculate the matrix elements of both sides of Eq.~\eqref{eq:floquet_states_eom} with respect to the basis states in Eq.~\eqref{eq:floquet_basis}, leading to
    \begin{align}
        &\expval{\expval{\sigma,m\bigg|i\hbar\dv{t}\bigg|\sigma',n}}=-n\hbar\Omega\delta_{mn}\delta_{\sigma\sigma'}\notag\\
        &=-\hbar\epsilon_\alpha \delta_{mn}\delta_{\sigma\sigma'}+\frac{1}{T}\int\limits_0^T\dd t\,e^{i(n-m)\Omega t}\expval{\sigma|H(t)|\sigma'}\,.\label{eq:matrix_elem_unitary}
    \end{align}
    In the following we want to understand the resonance structure in Fig.~\ref{fig:2} from a perturbative standpoint, considering weak acoustic modulation. At vanishing acoustic modulation, $\Delta(t)=0$, the eigenstates of the Hamiltonian in Eq.~\eqref{eq:H_dressed_states} are the optically dressed states $\ket{\pm}$, such that we choose $\sigma=\pm$ for the calculation of the matrix elements in the previous equation. Using the explicit form of the Hamiltonian from Eq.~\eqref{eq:H_dressed_states} with the sinusoidal modulation given by  Eq.~\eqref{eq:detuning_sinus}, this leads us to the eigenvalue equation
    \begin{equation}\label{eq:eigen_floquet_hamiltonian}
        \mathbb{H}\left.\ket{\Psi}\right>=\hbar\epsilon \left.\ket{\Psi}\right>
    \end{equation}
    on the vector space of $T$-periodic Hilbert space states, spanned by the states in Eq.~\eqref{eq:floquet_basis}, where
    \begin{align}\label{eq:Hamiltonian_Floquet}
        &\mathbb{H}=\mathbb{H}_0+\mathbb{V}\\
        &=\sum_{m,\sigma=\pm}\hbar\qty(\sigma\frac{\Omega_{\rm R}}{2}+m\Omega_{\rm ac})\left.\ket{\sigma,m}\right>\left<\bra{\sigma,m}\right.\notag\\
        &+\sum_{m,\sigma=\pm}i\hbar \frac{A_{\rm ac}}{4}\left(\left.\ket{\sigma,m}\right>\left<\bra{-\sigma,m+1}\right.-\left.\ket{\sigma,m+1}\right>\left<\bra{-\sigma,m}\right. \right)\notag
    \end{align}
    is the Floquet Hamiltonian~\cite{shirley1965solution,chu1985recent,breuer2002theory,chen2024periodically}. Note that we used the notation $-\sigma=\mp$ for $\sigma=\pm$ here. The second line constitutes the unperturbed Floquet Hamiltonian $\mathbb{H}_0$, while the third line is the perturbation $\mathbb{V}$ due to the acoustic modulation. This perturbation leads to transitions between the optically dressed states $\sigma\leftrightarrow -\sigma$ while simultaneously changing the acoustic dressing $m\leftrightarrow m+1$. Speaking in terms of phonons, as discussed in the context of sidebands in Fig.~\ref{fig:2}, the perturbation $\mathbb{V}$ describes phonon absorption and emission processes, which lead to transitions between the optically dressed states $\ket{\pm}$.

    We now make a few crucial points for understanding the Floquet spectrum of this system: 
    
    (i) The energies $\epsilon_\alpha$ are only defined modulo $\Omega_{\rm ac}$, as can be seen in Eq.~\eqref{eq:eigen_floquet_unitary}. This implies that we can assume that they lie in a kind of 'first Brillouin zone' (BZ) $-\Omega_{\rm ac}/2\leq \epsilon_\alpha<\Omega_{\rm ac}/2$. While in ordinary band structure calculations the spatial periodicity of the crystal defines the range in which the relevant energies lie, here for the Floquet spectrum it is the temporal periodicity, i.e., $\Omega_{\rm ac}$~\cite{shirley1965solution,chu1985recent,breuer2002theory,rudner2020band}. 
    
    (ii) For each set of parameters $A_{\rm ac}$, $\Omega_{\rm R}$ in the Floquet Hamiltonian in Eq.~\eqref{eq:Hamiltonian_Floquet}, there are exactly two eigenvalues within this first BZ, i.e., in contrast to the electronic band structure in spatially periodic systems, we have two relevant energy values instead of a continuum.

       \begin{figure*}[t] 
       	\centering
\includegraphics[width=0.9\linewidth]{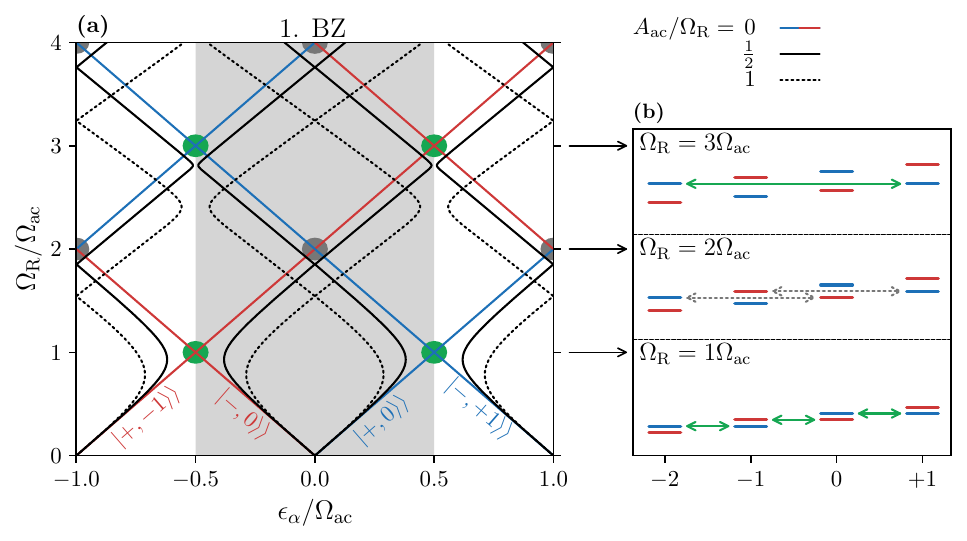}
\caption{Floquet spectrum. (a) Colored lines: Unperturbed eigenvalues of $\mathbb{H}_0$ i.e., $\epsilon_\alpha=\pm\Omega_{\rm R}/2+m\Omega_{\rm ac}$ belonging to the unperturbed states $\left.\ket{\pm,m}\right>$ as a function of the Rabi frequency $\Omega_{\rm R}$. The first BZ is given by the shaded gray region and coloring of the lines denotes the two possible parity values $\pm 1$ (blue/red). Crossing of unperturbed lines is marked by circles for two different cases: same parity (green circles), opposite parity (gray circles). Black lines: The Floquet eigenvalues $\epsilon_\alpha$ for finite acoustic modulation with $A_{\rm ac}=0.5\Omega_{\rm R}$ (solid black lines) and $A_{\rm ac}=\Omega_{\rm R}$ (dotted black lines). (b) Unperturbed level structure for high harmonic resonances $\Omega_{\rm R}=n\Omega_{\rm ac}$. Colors encode parity and degenerate states are connected by horizontal arrows with arrow colors corresponding to the circles in (a).}
\label{fig:4}
\end{figure*} 
    
    (iii) The Floquet Hamiltonian in Eq.~\eqref{eq:Hamiltonian_Floquet} conserves a kind of parity~\cite{ben1993effect,yan2019role}, as it commutes with
    \begin{equation}\label{eq:parity}
        \Pi=\sum_{m,\sigma=\pm}\sigma(-1)^m\left.\ket{\sigma,m}\right>\left<\bra{\sigma,m}\right.
    \end{equation}
    which has the possible eigenvalues $\pm 1$. This means that the perturbation $\mathbb{V}$ in the unitary limit couples only states with the same parity, i.e., only $\left.\ket{+,n}\right>$, $\left.\ket{-,n+1}\right>$, $\left.\ket{+,n+2}\right>$, etc.
    
    In Fig.~\ref{fig:4}(a), with colored lines, we plot the unperturbed eigenvalues of $\mathbb{H}_0$, i.e., $\epsilon_\alpha=\pm\Omega_{\rm R}/2+m\Omega_{\rm ac}$ belonging to the unperturbed states $\left.\ket{\pm,m}\right>$ as a function of the Rabi frequency $\Omega_{\rm R}$. While there are only two values within the first BZ (shaded gray region) for each chosen parameter $\Omega_{\rm R}$, we still obtain a plot that has some analogies to the band structure of quasi-free electrons~\cite{madelung2012introduction} due to this parametric dependence. The coloring of the lines denotes the two possible parity values $\pm 1$ (blue/red). We can identify two different kinds of line crossings: Crossing of lines with the same parity at the edge of the BZs for odd harmonics $\Omega_{\rm R}=\Omega_{\rm ac}, 3\Omega_{\rm ac},...$ (green circles), and  crossing of lines with opposite parity at the center of the BZs for even harmonics $\Omega_{\rm R}=2\Omega_{\rm ac}, 4\Omega_{\rm ac},...$ (gray circles). 
    
    When we now turn on the acoustic modulation [$A_{\rm ac}=0.5\Omega_{\rm R}$, solid black lines in Fig.~\ref{fig:4}(a)], we get anti-crossings of lines for odd harmonics (green circles), while in the case of even harmonics the lines still cross (gray circles), as states with opposite parity cannot be connected via the perturbation~$\mathbb{V}$. Note that the shift of the positions of the anti-crossings and crossings towards smaller Rabi frequencies $\Omega_{\rm R}$ compared to the case of vanishing acoustic modulation is a direct consequence of the Bloch-Siegert shift (see App.~\ref{app:bloch_siegert}) already discussed in the context of Fig.~\ref{fig:2}. Increasing the acoustic modulation further ($A_{\rm ac}=\Omega_{\rm R}$, dotted black lines), leads to stronger level repulsion in the case of the anti-crossing and an even stronger renormalization of the positions of crossings and anti-crossings towards smaller Rabi frequencies. 

    Using perturbation theory we can understand a few features of the spectrum in Fig.~\ref{fig:2} (see App.~\ref{app:pt} for more details), especially the peak positions and the vanishing of certain lines for even and odd harmonic resonances. To help in this discussion, the unperturbed energy levels of the states $\left.\left|\pm,m\right>\right>$ with $m=-2,...,+1$ are sketched in Fig.~\ref{fig:4}(b) for different ratios of Rabi frequency~$\Omega_{\rm R}$ and acoustic modulation frequency~$\Omega_{\rm ac}$. Degenerate states, i.e., high harmonic resonances with respect to the acoustic modulation, are connected by horizontal arrows. 
    
    Starting at the bottom for resonant driving, i.e., the first harmonic $\Omega_{\rm R}=\Omega_{\rm ac}$, the states $\left.\ket{+,m}\right>$ and $\left.\ket{-,m +1}\right>$ are degenerate with respect to the unperturbed Hamiltonian $\mathbb{H}_0$. Furthermore they have the same parity (same color) and can therefore interact via the perturbation $\mathbb{V}$ [green horizontal arrows corresponding to green circles in (a)]. The same is true for the third harmonic $\Omega_{\rm R}=3\Omega_{\rm ac}$  and states $\left.\ket{+,m}\right>$ and $\left.\ket{-,m + 3}\right>$ (top). Considering general odd harmonics with $\Omega_{\rm R}=(2p-1)\Omega_{\rm ac}$ and $p=1,2,3,...$ 
    we can use quasi-degenerate Brillouin-Wigner perturbation theory to determine the energies of the eigenstates formed approximately by the two unperturbed states in resonance~\cite{taylor2002quantum,englert2024lectures} [see App.~\ref{app:pt_odd}, Eqs.~\eqref{eq:app_E1_eps_pm} and \eqref{eq:app_E1_phi_pm} with Eq.~\eqref{eq:pt_phi_alpha}],
    \begin{equation}\label{eq:u_pm_odd}
        \left.\ket{u_{\pm}}\right>\approx\frac{1}{\sqrt{2}}\qty(\left.\ket{+,m}\right>\mp i \left.\ket{-,m+(2p-1)}\right>)
    \end{equation}
    as
    \begin{equation}
        \epsilon_{\pm}=\tilde{\epsilon}^{(0)}\pm\frac{\Delta_{\epsilon}}{2}
    \end{equation}
    with $\tilde{\epsilon}^{(0)}=\Omega_{\rm R}/2+m\Omega+\mathcal{O}\qty(\mathbb{V}^2)$ being the originally unperturbed Floquet frequency, renormalized by the presence of the perturbation $\mathbb{V}$ (Bloch-Siegert shift, discussed in the context of Fig.~\ref{fig:2}, see also App.~\ref{app:bloch_siegert}). Analog to the optically dressed states in Eq.~\eqref{eq:def_dressed_states}, we call these Floquet eigenstates acousto-optically (doubly) dressed. The width of the anti-crossing in the spectrum is then given by~\cite{shirley1965solution} [see Eq.~\eqref{eq:I_RF_unitary} with Eqs.~\eqref{eq:app_v_21} and \eqref{eq:app_E1_eps_pm}]
    \begin{align}\label{eq:line_distance}
        2|\epsilon_+-\epsilon_-|&=2|\Delta_{\epsilon}|=A_{\rm ac}\frac{\qty(\frac{A_{\rm ac}}{8\Omega_{\rm ac}})^{2p-2}}{[(p-1)!]^2}+\mathcal{O}\qty(A_{\rm ac}^{2p})\,.
    \end{align}
    The distance of the peaks at anti-crossings gets larger when increasing the acoustic modulation amplitude, as already discussed in the context of Fig.~\ref{fig:2}. Furthermore, since the dependence on the parameter $p$ in Eq.~\eqref{eq:line_distance} resembles a squared Poisson distribution with a maximum around $p-1\approx A_{\mathrm{ac}}/(8\Omega_{\mathrm{ac}})$, the anti-crossings corresponding to higher odd harmonic resonances, i.e., for $p>1+A_{\mathrm{ac}}/(8\Omega_{\mathrm{ac}})$ ,  are less well visible, as they are suppressed with the factor $\qty(\frac{A_{\rm ac}}{8\Omega_{\rm ac}})^{2p-2}/[(p-1)!]^2$. For normal resonance $\Omega_{\rm R}\approx \Omega_{\rm ac}$ the distance of the peaks at anti-crossing is given by 
    \begin{equation}\label{eq:line_distance_resonant}
        2|\epsilon_+-\epsilon_-|=2|\Delta_{\epsilon}|=A_{\rm ac}+\mathcal{O}\qty(A_{\rm ac}^2)\,,\qquad p=1\,.
    \end{equation}
    
    The intensity of the center line in Fig.~\ref{fig:2} at these anti-crossings can be calculated in the unitary limit according to Eq.~\eqref{eq:I_RF_unitary} via the transition dipole matrix elements
    \begin{align}
        P_{\pm\pm}^{(0)}&=\expval{\expval{u_{\pm}\ket{g}\bra{x}u_{\pm}}}\notag\\&\approx\frac{1}{2}\expval{+|g}\expval{x|+}+\frac{1}{2}\expval{-|g}\expval{x|-}=\frac{1}{4}-\frac{1}{4}=0\,,
    \end{align}
    which vanish due to destructive interference between the two different optically dressed states $\ket{\pm}$ in Eq.~\eqref{eq:u_pm_odd} which are in addition dressed with different acoustic components [Floquet indices $m$ and $m+(2p-1)$]~\cite{ficek1993resonance,ficek1996fluorescence,ficek1999quantum}. As discussed in App.~\ref{app:anticrossing_suppression}, this suppression of the center line at roughly odd harmonic resonances holds also non-perturbatively. From the previous discussion it is clear that this is due to destructive interference of quantum amplitudes for transitions between optically dressed states mediated by the acoustic modulation, i.e., a clear signature of acousto-optical double dressing.

    For even harmonics with $\Omega_{\rm R}=(2p)\Omega_{\rm ac}$ and $p=1,2,3,...$, the states $\left.\ket{+,m-p}\right>$ and $\left.\ket{-,m+ p}\right>$ are degenerate with respect to the unperturbed Hamiltonian $\mathbb{H}_0$, i.e., they are resonant. However in contrast to odd harmonic resonances, they have opposite parity and cannot interact via the perturbation~$\mathbb{V}$, as sketched in Fig.~\ref{fig:4}(b) [center, dashed gray arrows corresponding to gray circles in (a)]. This implies, as discussed in App.~\ref{app:pt_even}, that the states
    \begin{align}
        \left.\ket{u_{\pm}}\right>\approx\left.\ket{\pm,m\mp p}\right>
    \end{align}
    are already approximate eigenstates with identical frequencies at the resonance, whose position is renormalized by the perturbation
    \begin{equation}
        \epsilon_\pm=\tilde{\epsilon}^{(0)}=\Omega_{\rm R}/2 +(m-p)\Omega_{\rm ac}+\mathcal{O}\qty(\mathbb{V}^2)\,.
    \end{equation}
    This explains why the lines that cross the center line in Fig.~\ref{fig:2}(a) and (c) vanish at crossing. According to Eq.~\eqref{eq:I_RF_unitary} the intensity of these lines is given by the transition dipole matrix elements
    \begin{equation}
        P_{\pm\mp}^{(0)}=\expval{\expval{u_\pm|g}\expval{x|u_\mp}} \sim\int\limits_0^T\dd t\, e^{\mp i(2p)\Omega_{\rm{ac}} t}=0\,.
    \end{equation}
    A dipole transition between these states that would contribute to the ZPL is therefore not allowed due to the different phonon content in the eigenstates $\left.\ket{u_\pm}\right>$. Conversely, lines that cross with the PSBs do not necessarily vanish at crossing, as discussed in the context of Fig.~\ref{fig:3} and which can be seen in Fig.~\ref{fig:2}. The vanishing of lines crossing with the center line can be shown completely non-perturbatively, as discussed in App.~\ref{app:crossing_suppression}.

    Note that we discussed here only perturbation theory for the Floquet spectrum in the unitary limit. As presented in App.~\ref{app:pt_nonunitary} for completeness, taking also dissipation into account leads to additional interesting features such as a transition between an over-damped regime for $A_{\rm ac}<|\gamma_{\rm xd}-\gamma_{\rm pd}|/4$, where there is no anti-crossing at resonance $\Omega_{\rm R}\approx \Omega_{\rm ac}$, and an under-damped regime for $A_{\rm ac}>|\gamma_{\rm xd}-\gamma_{\rm pd}|/4$ with a faithful anti-crossing. As discussed previously in Eq.~\eqref{eq:line_distance_resonant}, the peak distance at anti-crossing is in the unitary limit given by $2|\epsilon_+-\epsilon_-|\approx A_{\rm ac}$ for resonant acoustic driving $\Omega_{\rm ac}\approx \Omega_{\rm R}$. The width of the peaks however is determined by $\gamma_r$ according to Eq.~\eqref{eq:I_RF_full}. As discussed in App.~\ref{app:floquet_eigen_diss}, the non-decaying traceless solution is the one with $\gamma_r=0$, $\delta_r=0$, i.e., the anti-crossing lines always belong to the decaying solutions, which are broadened by $\max(\gamma_{\rm xd},\gamma)\geq\gamma_r>\gamma_{\rm xd}/2$. To be able to spectrally resolve such an anti-crossing we would therefore need $A_{\rm ac}>\gamma_{\rm xd}$ and preferably a vanishing pure dephasing rate, i.e., a lifetime-limited two-level emitter. This implies that we want to be in the under-damped regime anyway and the presence of a potential transition between over- and under-damped regimes is not relevant to the following discussion on platforms in which the acoustically induced anti-crossing behavior and suppression of the center line can be made visible.

\section{Feasibility study of acousto-optical Floquet engineering}\label{sec:platforms}

We know from the theoretical analysis in the previous sections that the observation of acousto-optical double-dressing (manifesting through anti-crossings and line-suppressions in the RF spectra under strong optical driving, see Figs.~\ref{fig:2} and \ref{fig:3}), is governed by four parameters: (i) The acoustic sideband splitting $\Omega_{\rm ac}$, (ii) the Rabi frequency which needs to satisfy the resonance condition $\Omega_{\rm R}\approx \Omega_{\rm ac}$ to generate an anti-crossing, (iii) the width of the anti-crossing $A_{\rm ac}$ at this resonance, and (iv) the incoherent linewidths $\gamma_r$, which lie between $\gamma$ and $\gamma_{\rm xd}$, i.e., $\gamma_{\rm xd}/2$ and $\gamma_{\rm xd}$ for lifetime-limited emitters. Thus for simplicity, we consider $\gamma_{\rm xd}$ to be the relevant parameter here, assuming sufficiently weak pure dephasing. We will discuss the impact of pure dephasing for the relevant emitter types separately at the end of the following feasibility study. Since (ii) the Rabi frequency needs to satisfy the resonance condition, we are left with three free parameters $\Omega_{\rm ac}$, $A_{\rm ac}$, and $\gamma_{\rm xd}$ in resonance. From these we can form the two crucial parameter ratios $\Omega_{\rm ac}/\gamma_{\rm xd}$ and $A_{\rm ac}/\gamma_{\rm xd}$, which both need to be $\gtrsim 1$ to be able to observe the anti-crossing at resonance $\Omega_{\rm R}\approx \Omega_{\rm ac}$ (see Fig.~\ref{fig:2}). The feasibility of acousto-optical Floquet engineering in the form of double-dressing thus strongly depends on the performance parameters $\Omega_{\rm ac}$, $\Omega_{\rm R}$, $A_{\rm ac}$, and $\gamma_{\rm xd}$ of the potential light-emitter and acoustics-emitter interface platforms. Before carrying out the actual feasibility study we will shortly review established values of these key system parameters for the different types of interfaces.

{\it (a) Light-emitter interface:}
Table~\ref{tab:Mollow} summarizes the characteristic experimental parameters of observed Mollow triplets, i.e., for the light-emitter interface, for typical emitters embedded in different forms of optical cavities. In lifetime-limited emitters with nearly vanishing pure dephasing, the lifetime $T_1=1/\gamma_{\rm xd}$ is the key limiting factor for observing a Mollow triplet and thus also for potentially observing acousto-optical double dressing via anti-crossings in RF spectra. Therefore, Tab.~\ref{tab:lifetimes} summarizes lifetimes of several single-photon emitters that can be interfaced with acoustic excitations. The additional impact of pure dephasing will be discussed separately at the end of this section.

\begin{table}[t]
    \centering
    \begin{tabular}{c c c c c c}
        \makecell{Emitter\\ type} & \makecell{Cavity\\type} & \makecell{Lifetime\\(ns)} & \makecell{Linewidth \\$\frac{\Delta \omega_{\rm TLS}}{2\pi}$(GHz)} & \makecell{ Rabi splitting \\ $\frac{\Omega_{\rm R}}{2\pi}$ (GHz) }& Ref. \\
         \hline
        QD & none  &  4.4  & 0.34  & 5.5 & \cite{vamivakas2009spin}\\
        QD & DBR & 0.15 & 6.6~ & 4.4 & \cite{flagg2009resonantly}\\
        QD & DBR & 0.63 & 0.28 & 6.5 & \cite{ates2009post}\\
        QD & DBR & 1.0 & 0.73 & 5.4 & \cite{ulhaq2012cascaded}\\
        QD & \makecell{photonic \\ crystal} & $ 0.05$ & $ 10$ & $ 35$ & \cite{fischer2016self}\\
        \makecell{SiV in nano-\\ diamond} & \makecell{photonic \\ crystal} & 1.9 & 0.22 & 3.1 & \cite{zhou2017coherent}
    \end{tabular}
    \caption{Typical (approximate) Mollow triplet parameters.}
    \label{tab:Mollow}
\end{table}

\begin{table}[t]
    \centering
    \begin{tabular}{ccc}
         Emitter type & \makecell{Lifetime\\$T_1$ (ns)} & Ref. \\
         \hline
         Epitaxial QDs & 0.3 &\cite{bacher1999biexciton,mermillod2016dynamics} \\
         QD molecules & 2 - 10 & \cite{nakaoka2006direct}\\
         Quantum posts & $ 10^6$ & \cite{krenner2008semiconductor}\\ 
         SiV centers in diamond & 1.7 &\cite{zuber2023shallow} \\
         NV centers in diamond & 12 - 22 &\cite{aharonovich2016solid,laube2023fluorescence} \\
         Color centers in hBN & 2 - 5 &\cite{jungwirth2016temperature,wigger2019phonon} \\
         Strain-induced emitters in TMDCs & 1 - 10 &\cite{tonndorf2015single,parto2021defect} \\
         Color centers in SiC & 1.2 - 14 &\cite{castelletto2014silicon,falk2014electrically} \\
         Defects in CNTs & 0.1 - 0.5 &\cite{ishii2018enhanced,ma2015room} 
    \end{tabular}
    \caption{Typical (approximate) lifetimes for different single photon emitter types. CNT = carbon nanotube. The decay rates are $\gamma_{\text{xd}}=1/T_1$.}
    \label{tab:lifetimes}
\end{table}

{\it (b) Acoustics-emitter interface:} 
In the following, we will reduce the number of realizations from Sec.~\ref{sec:introplatforms} to three general phononic/mechanical/acoustic platforms to realize acousto-optical Floquet engineering: (A) mechanical resonators, (B) surface or guided  acoustic waves (SAWs), and (C) bulk acoustic waves (BAWs).

One main difference between the platforms is their typical operation frequencies, which roughly increase from (A) to (C)~\cite{wigger2021remote}. From our previous theoretical study, we know that the acoustic frequency $\Omega_{\rm ac}$ is one of the crucial parameters to reach double dressing, since we aim for resonant driving of the optically dressed states $\Omega_{\rm ac}\approx \Omega_{\rm R}$. Table~\ref{tab:acoustics} gives an overview of typical frequency ranges of the different acoustic platforms supporting mechanical modes suitable for coupling to quantum emitters.
\begin{table}[t]
    \centering
    \begin{tabular}{c c c c}
         \makecell{Acoustic\\platform} & Type & \makecell{Operation frequency \\ $\frac{\Omega_{\rm ac}}{2\pi}$ (GHz)} &  Ref. \\[3mm]
         \hline
         GaAs mech. res. & (A) & $5\!\times\!10^{-4}$ - 0.2 & \cite{munsch2017resonant,tanos2024high}\\
         CNT mech. res. & (A) & 0.04 - 3 &\cite{huttel2009carbon,xu2022nanomechanical}\\
         TMDC mech. res. & (A) & 0.06 &\cite{xu2022nanomechanical}\\
         SiC mech. res. & (A) & 0.01 - 1 &\cite{xu2022nanomechanical}\\
         Heterogeneous integr. & (B) & 0.15 - 0.6 & \cite{nysten2017multi,patel2024surface}\\
         SAW on SiC & (B) & 0.5 - 1 &\cite{hernandez2021acoustically,whiteley2019spin}\\
         Suspended GaAs  & (B) & 0.2 - 1.7 & \cite{fuhrmann2011dynamic,vogele2020quantum,spinnler2024single}\\
         SAW on GaAs & (B) & 0.2 - 3.5 & \cite{weiss2018multiharmonic,decrescent2024coherent}\\
         SAW on diamond & (B) & 0.8 - 3.4 &\cite{golter2016optomechanical,maity2020coherent}\\
         Piezogen. BAW & (C) & 1 - 100 &\cite{vorobiev2011correlations,machado2019generation,kramer2025acoustic}\\
         BAW optomech. & (C) & 10 - 1000 &\cite{czerniuk2014lasing,van2015nonlinear}
    \end{tabular}
    \caption{Acoustic platforms with typical (approximate) operation frequency ranges.}
    \label{tab:acoustics}
\end{table}
\begin{table}[t]
    \centering
    \begin{tabular}{c c c c c}
         \makecell{Emitter \& \\ acoustic platform} & Type & \makecell{Modul. ampl.\\ $\hbar A_{\rm ac}$ (meV)} & \makecell{Frequency \\$\frac{\Omega_{\rm ac}}{2\pi}$ (GHz)} & Ref. \\[3mm]
         \hline
         QD \& SAW & (B) & 0.5 & 1 & \cite{gell2008modulation}\\
         QD \& mech. res.& (A) & 0.15 & $5.3\times 10^{-4}$ & \cite{yeo2014strain}\\
         QD \& SAW & (B) & 0.15 & 0.9 & \cite{schulein2015fourier}\\
         NV center  \& mech. res. & (A) & 0.04 & $8.7\times 10^{-4}$ & \cite{lee2016strain}\\
         hBN color center  \& SAW & (B) & 0.5 & 0.3 & \cite{lazic2019dynamically}\\
         hBN color center  \& SAW & (B) & 0.9 & 0.47 & \cite{iikawa2019acoustically}\\
         QD \&  SAW + Lamb mode & (B) & 1.3 & 0.37 &\cite{vogele2020quantum}\\
         GaAs emission center \&  SAW & (B) & 0.3 & 3.6 & \cite{yuan2021remotely}\\
         QW \&  BAW & (C) & 7 & 7 & \cite{kuznetsov2021electrically}\\
         QW \&  BAW & (C) & 1 & 6.5 & \cite{crespo2022ghz}\\
         QD \&  SAW & (B) & 0.4 & 0.4 &\cite{lienhart2023heterogeneous}\\
         NV center  \& mech. res. & (A) & $5\!\times\!10^{-3}$ & 0.028 & \cite{li2023mechanical}\\
         QW \&  BAW & (C) & 15 & 7 & \cite{sesin2023giant}\\
         QD \&  mech. res. & (A) & $5\!\times\!10^{-3}$ & 0.4 &\cite{spinnler2024quantum}\\
         TMDC \&  SAW & (B) & 0.5 & 0.3 &\cite{patel2024surface}\\
         TMDC \&  SAW & (B) & 0.29 & 0.49 &\cite{mohajerani2025acoustic}\\
         QD \& SAW & (B) & $1.45\!\times\!10^{-2}$ & 3.53& \cite{zhan2025dynamical}\\
         QD \& BAW & (C) & $3$ & $5.6$& \cite{kuznetsov2025deterministic}
    \end{tabular}
    \caption{Acousto-optical platforms with typical (approximate) PL energy modulation amplitudes $\hbar A_{\rm ac}$ and acoustic frequencies $\Omega_{\rm ac}$ in chronological publication order. The data are used for Fig.~\ref{fig:5}.}
    \label{tab:platforms}
\end{table}

{\it (c) Emitter-based acousto-optics:}
There already exist a few realizations of hybrid acousto-optical platforms as discussed in the introduction. Table~\ref{tab:platforms} collects typical examples of platforms studied so far. Except for Ref.~\cite{zhan2025dynamical}, where an explicit value for $A_{\rm ac}$ is given, we quantify the efficiency of the reported acoustics-emitter coupling by evaluating the modulation amplitude $A_{\rm ac}$ of the photoluminescence (PL) spectral broadening induced by a dynamic strain field~\cite{weiss2018interfacing,wigger2021remote}.

To assess and compare different acoustic platforms in combination with various single-photon emitter types, we collect key parameters of experimental reports in Fig.~\ref{fig:5} on a double-logarithmic scale. For each emitter type (different symbols in Fig.~\ref{fig:5}) we normalize the acoustic parameters $\Omega_{\text{ac}}$ and $A_{\text{ac}}$ from Tab.~\ref{tab:platforms} to typical mean excited state decay rates $\gamma_{\text{xd}}=1/T_1$ from Tab.~\ref{tab:lifetimes}. The parameter region for potential Floquet engineering is marked in green in Fig.~\ref{fig:5}. In this region, which satisfies the conditions $\Omega_{\rm ac}/\gamma_{\rm xd}\gtrsim 1$ and $A_{\rm ac}/\gamma_{\rm xd}\gtrsim 1$, anti-crossing (Fig.~\ref{fig:2}) and line-suppression (Fig.~\ref{fig:3}) from acousto-optical double-dressing in RF spectra at resonance $\Omega_{\rm ac}\approx\Omega_{\rm R}$ is possible. The parameters corresponding to our calculations in Fig.~\ref{fig:2} are also marked as crosses for reference.  In the following, we discuss the main advantages and shortcomings of the three acoustic platforms in the context of acousto-optical Floquet engineering:

\begin{figure}[t] 
	\centering
\includegraphics[width=0.9\linewidth]{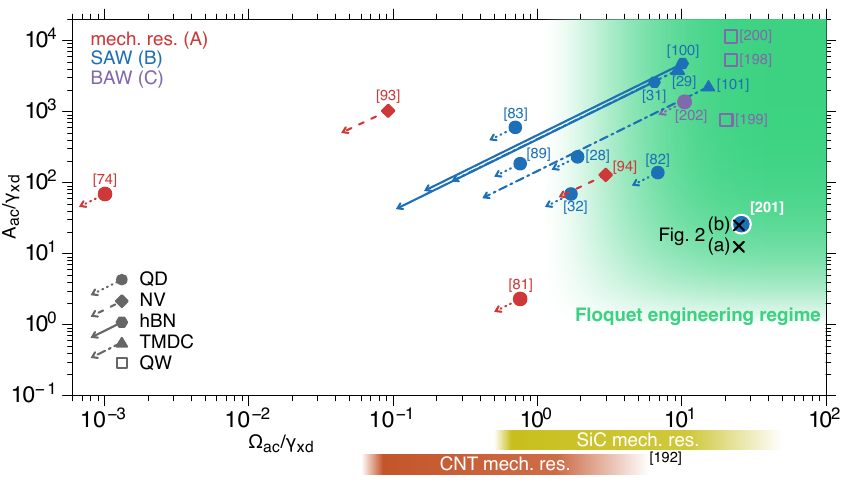}
\caption{Parameter study of different acoustic platforms coupled to single-photon emitters. The underlying parameters for the data points (dots) are given in Tabs.~\ref{tab:lifetimes} and \ref{tab:platforms}, except for the quantum well (QW) exciton lifetime, which is here taken to be $T_1=0.5$~ns~\cite{martinez1993temperature}. Also for Ref.~\cite{zhan2025dynamical} (blue circle with white border), which is the first successful experimental demonstration of acousto-optical Floquet engineering of a solid state single photon emitter, we used the lifetime $T_1=1.19$~ns, given therein. The open symbols correspond to emitter platforms that cannot be described by TLSs, i.e., by the theory presented in this work. The green shaded area marks the parameter range that has the potential for acousto-optical Floquet engineering, i.e., $\Omega_{\rm ac}/\gamma_{\rm xd}\gtrsim 1$, $A_{\rm ac}/\gamma_{\rm xd}\gtrsim 1$. The detrimental effect of additional pure dephasing discussed in the context of Tab.~\ref{tab:dephasing} is obtained by replacing the lifetimes with the dephasing times $T_1\rightarrow T_2$, i.e., by replacing $\gamma_{\rm xd}$ with $\gamma$ in the calculation of the parameter ratios, as denoted by the arrows attached to the data points.}
\label{fig:5}
\end{figure}

(A) (Nano)mechanical resonators have the advantage of high Q-factors allowing for large strain/acoustic amplitudes~\cite{engelsen2024ultrahigh}. However, the typical frequencies are in the range between hundreds of kHz and hundreds of MHz~\cite{xu2022nanomechanical,yuan2019frequency,tanos2024high} and are usually too low, compared to the GHz Rabi splittings in Mollow triplets, to overcome the (lifetime-induced) linewidths (see Tab.~\ref{tab:Mollow}). Thus, mechanical resonator systems (red symbols, Fig.~\ref{fig:5}), often have too small frequencies to satisfy the resonance condition of $\Omega_{\rm ac}\approx\Omega_{\rm R}$ for optically driven emitters with a sufficiently large Rabi splitting $\Omega_{\rm R}\gtrsim \gamma_{\rm xd}$.

(B) Operation frequencies of piezoelectrically generated  SAWs and related guided acoustic waves are mainly determined by the electrode design and periodicity of the employed interdigital transducer (IDT). Here, high operation frequencies translate to short acoustic wavelengths which directly connect to fundamental challenges. The spatial resolution of state-of-the-art nanofabrication limits the acoustic wavelengths to the low GHz range~\cite{hess2002surface}. Furthermore, a fundamental drawback of this platform is the SAW penetration depth, which is approximately given by the wavelength~\cite{maradudin1976attenuation,weiss2018interfacing}.  Thus, the optimum position for the emitters shifts closer to the surface for increasing operation frequency, which impairs the coherent properties of the optically active emitters~\cite{wang2004optical}. Similar difficulties arise for guided modes in heterogeneously integrated thin films~\cite{nysten2017multi} and suspended geometries~\cite{vogele2020quantum,weiss2016surface,spinnler2024single,rosinski2026coupling} interacting with QDs. However, there are also additional constraints for both of these guided mode platforms: For the former, acoustic waveguiding determines the strain profile at the position of the QD and the optimum coupling has to be carefully assessed. For the latter, the center of the membrane is a stress-neutral plane for the dominant flexural Lamb modes~\cite{vogele2020quantum}. Thus, an emitter placed there is modulated predominantly by a shear deformation potential which is typically weaker than that for normal (diagonal) strain elements which however vanish in such a system~\cite{rosinski2026coupling}. This reduced coupling can be compensated by larger mechanical displacements compared to substrate bound modes analogous to conventional (nano)mechanical resonators~(A) or by shifting the QD away from the center. 

Still we find that SAW platforms (blue symbols, Fig.~\ref{fig:5}) can reach frequencies $\Omega_{\rm ac}$ and acoustic modulation amplitudes $A_{\rm ac}$ in the required range for generating acousto-optical double dressing. Indeed, recently the features presented in Fig.~\ref{fig:2} have been measured~\cite{zhan2025dynamical}. To the best of our knowledge this was the first experimental demonstration of acousto-optical Floquet engineering of a solid state single photon emitter, i.e., a QD driven by a SAW, via anti-crossings and line suppressions in RF spectra. The corresponding parameter ratios are presented in Fig.~\ref{fig:5} (blue circle with white border). Note that the QD in Ref.~\cite{zhan2025dynamical} exhibits additional inhomogeneous broadening, not included in the simple parameter study in Fig.~\ref{fig:5}. Nevertheless, this was not strong enough to prevent the resolution of the anti-crossing.

(C) Finally, BAWs are modes which extend into the substrate perpendicular to the surface normal. Thus, BAWs overcome the limitations arising from the surface-confined nature of SAWs and other guided acoustic waves. In BAW devices, emitters can be buried deep below the sample surface and acoustic frequencies can reach tens of GHz with direct piezoelectric generation~\cite{machado2019generation} or even the THz range with ultrafast optical excitations~\cite{berstermann2009terahertz,van2015nonlinear} (Tab.~\ref{tab:acoustics}). The latter form of BAW-generation is however less relevant to potential Floquet engineering due to the inherent lack of time-periodicity. The accessible high frequencies, especially in the GHz range, are a major advantage of BAWs since they are fully compatible with typical Rabi splittings (Tab.~\ref{tab:Mollow}) and can easily exceed the linewidth of QDs and other emitters (Tab.~\ref{tab:lifetimes}). The strength of the acoustic modulation of embedded emitters can furthermore be large enough to achieve acousto-optical double dressing (violet symbols, Fig.~\ref{fig:5})

\begin{table}[t]
    \centering
    \begin{tabular}{cccc}
         Emitter type & \makecell{Dephasing\\ time $T_2$} & Detection method & Ref.\\
         \hline
         Epitaxial QDs & 200 ps & four-wave mixing &\cite{mermillod2016dynamics}\\
         NV centers in diamond & 8 ns & coherent control &\cite{robledo2010control}\\
         Color centers in hBN & 55 ps & coherent control &\cite{preuss2022resonant}\\
         Strain-induced emitters in TMDCs & 13.5 ps & Michelson interferometry &\cite{von2023temperature}
    \end{tabular}
    \caption{Typical dephasing times at cryogenic temperatures for the emitter types covered in Fig.~\ref{fig:5}. The dephasing rate is $\gamma=1/T_2$ such that the emitter coherence decays with $\exp(-t/T_2)$.}
    \label{tab:dephasing}
\end{table}

While we assumed lifetime-limited emitters, i.e., a vanishing pure dephasing $\gamma_{\rm pd}=0$, in the discussion so far, in the following we briefly comment on the impact of additional pure dephasing for the emitter types presented in Fig.~\ref{fig:5}. Typical dephasing times $T_2=1/\gamma$ at cryogenic temperatures are summarized in Tab.~\ref{tab:dephasing}. When comparing these to the lifetimes $T_1=1/\gamma_{\rm xd}$ in Tab.~\ref{tab:lifetimes}, we find $T_2<T_1$ throughout. This implies that the incoherent linewidth $\gamma_r$ of the RF spectrum, which lies between $\gamma_{\rm xd}$ and $\gamma=(\gamma_{\rm xd}+\gamma_{\rm pd})/2$ [see Eq.~\eqref{eq:floquet_rates}], has a maximum value of $\gamma=1/T_2$ for all considered emitter types, instead of $\gamma_{\rm xd}=1/T_1$ as assumed previously. Regarding the discussion on the feasibility of Floquet engineering in Fig.~\ref{fig:5}, this implies that a strong pure dephasing changes the relevant parameter ratios on the two axes of the figure by replacing $\gamma_{\rm xd}\rightarrow\gamma$ in the denominators. This leads to a reduction of the feasibility for all emitter types, determined by the ratio $\gamma_{\rm xd}/\gamma=T_2/T_1<1$. This reduction of the feasibility is marked in Fig.~\ref{fig:5} by arrows pointing from the previously discussed data points away from the Floquet engineering regime. The length of these arrows quantifies the typical feasibility reduction with longer arrows implying a smaller ratio of $T_2/T_1$. We find that the additional influence of pure dephasing has detrimental effects on the relatively novel color centers in hBN and the strain-induced emitters in TMDCs, while the impact on the technologically more mature QDs and NV centers is significantly smaller.

Finally we discuss the potential of optical cavity integration for the SAW and BAW platforms, since these provide suitable acoustic frequency ranges for Floquet engineering as discussed above. Cavity integration might be necessary for achieving efficient light matter coupling and reaching a sufficiently large Rabi frequency. For SAWs or other guided modes (B) integration with an optical cavity may limit the acoustic amplitudes $A_{\rm ac}$ accessible for these surface-confined modes. Using, for example, planar DBR cavities, the emitters would need to be positioned at larger distances from the sample surface, where the SAWs are confined, leading to a less efficient acoustic driving. In contrast, for BAWs (C) the integration with photonic and optomechanical cavities formed by DBRs is straightforward~\cite{chafatinos2020polariton,kuznetsov2021electrically} and a single DBR cavity can even be used to confine both the acoustic and the optical modes~\cite{fainstein2013strong}. Note that while optical cavity integration yields larger accessible Rabi frequencies, it also has a detrimental effect on the Floquet engineering potential due to the decrease of the $T_1$ lifetime via the Purcell effect~\cite{purcell1946spontaneous,vahala2003optical,noda2007spontaneous}. This would shift all datapoints in Fig.~\ref{fig:5} towards the lower left corner similar to the additional effect of pure dephasing discussed earlier.




Based on this parameter analysis we can draw the following conclusions: (i) BAWs offer modulation frequencies and amplitudes that can exceed typical Mollow-triplet splittings~\cite{vamivakas2009spin,flagg2009resonantly,ates2009post,fischer2016self} (see also violet symbols in Fig.~\ref{fig:5} lying inside the Floquet engineering regime) and the efficiency of optical driving can be further enhanced by DBR cavity integration~\cite{chafatinos2020polariton,kuznetsov2021electrically}. In fact, resonant acousto-optical hybridization has been reached with BAWs coupled to quantum well excitons with continuous energy dispersions, where phonon lasing was demonstrated successfully~\cite{chafatinos2020polariton}. Several studies employing BAWs in Fig.~\ref{fig:5} (violet squares) stem from such quantum well excitons, where the continuous energy spectrum prohibits the application of the presented theory using a two-level model. To make this key difference clear, we use open symbols in the figure. Replacing quantum wells by layers of QDs, being an established Mollow triplet system (Tab.~\ref{tab:Mollow}) with relatively small impact due to pure dephasing (Tabs.~\ref{tab:lifetimes} and \ref{tab:dephasing}), renders an obvious platform to test the proposed acousto-optical Floquet engineering~\cite{kuznetsov2025deterministic}. (ii) Guided acoustic wave platforms require more elaborate design optimization compared to BAWs due to their lower operation frequencies. In particular heterogeneously integrated and suspended device architectures can overcome the limitations of monolithic systems. While acousto-optical double dressing has been demonstrated successfully for a SAW-based QD system~\cite{zhan2025dynamical} the challenging integration with optical cavities might limit either the accessible Rabi frequencies or the accessible acoustic modulation amplitudes. (iii) Nanomechanical resonators are limited by low operation frequencies demonstrated in most platforms. Recent work showed that the operation frequencies can be pushed towards the GHz regime by shrinking the geometric dimensions~\cite{tanos2024high}. Thus, further reduction of the size of resonators may eventually take also this platform into the parameter range necessary for Floquet engineering. 

Finally, we assess alternative platforms for which only  $\Omega_{\rm ac}/\gamma_{\rm xd}$ can be reported so far. These complement the platforms collected in the main plot of Fig.~\ref{fig:5}, where acousto-optical coupling has already been explored. The two bars below the main plot mark the range of $\Omega_{\rm ac}/\gamma_{\rm xd}$ for carbon nanotube (CNT, orange) and SiC-based (yellow) mechanical resonators. Here we use independently reported values for $\Omega_{\rm ac}$ (cf. Tab.~\ref{tab:acoustics}) and $\gamma_{\rm xd}$  (cf. Tab.~\ref{tab:lifetimes}). Clearly, both systems can reach sufficiently high acoustic frequencies $\Omega_{\rm ac}$ and, thus, the experimental realization is worthwhile and may lead to two additional platforms for acousto-optical Floquet engineering with individual strengths and advantages over the more established platforms.

\section{Conclusions and outlook}\label{sec:conclusions}
In summary, we have studied a hybrid system consisting of an optically driven two-level system (TLS) and an acoustic modulation of the transition energy. While strong optical driving alone results in the development of the Mollow triplet in the scattering spectrum, strong acoustic modulation in the case of weak optical driving leads to phonon sidebands. The combination of strong optical driving and strong acoustic modulation results in non-trivial features in resonance fluorescence (RF) spectra, whose origin was investigated in great detail in the present work. We have applied Floquet theory to develop analytical expressions for the RF spectrum of the TLS. Using perturbative and non-perturbative techniques, we were able to explain the underlying acousto-optical double-dressing physics leading to complex spectral features combining line crossings, anti-crossings, and line suppressions. Using the presented simulations and analytical insights, we could predict crucial system parameters for the light-emitter and the acoustics-emitter couplings, that would allow for full control over the hybrid states of the acousto-optical system. Realizing such a platform would then make fully-fledged Floquet engineering of the light scattering process possible.

To set this theoretical prediction into the context of existing acousto-optical platforms, we have performed a feasibility study in Sec.~\ref{sec:platforms}, using state-of-the-art system parameters of different acousto-optical approaches. We conclude that typical mechanical resonator systems have too small acoustic frequencies to match the required Rabi frequencies for efficient acousto-optical double-dressing. Surface acoustic wave (SAW) platforms can reach the required frequencies, but integration with optical cavity infrastructures either sets boundaries to the light-emitter coupling efficiency and therefore Mollow triplet resolution, or to the strength of the acoustic modulation via such surface-confined modes. However, recently acousto-optical Floquet engineering has been performed successfully for the first time with a SAW-modulated quantum dot (QD)~\cite{zhan2025dynamical} using a confocal microscopy setup, i.e., without relying on optical cavity integration. This experimental advance underpins the acousto-optical Floquet engineering potential of SAW-based QD platforms. Finally, bulk acoustic waves (BAWs) allow to combine naturally high acoustic frequencies in the GHz range and compatibility with optical cavities. Consequently, we expect a combination of BAWs and  semiconductor quantum dots~\cite{kuznetsov2025deterministic}, being an established Mollow triplet system, to be a suitable platform for acousto-optical Floquet engineering.

Such hybrid control of single-photon emitters paves the way for several quantum technological applications. The possibility of optical frequency multiplexing~\cite{khodadad2025frequency} and temporal switching of single-photon emission~\cite{xiong2016active} are examples of such development. Another extension of the hybrid character could be the additional application of electric field pulses~\cite{michaelis2010coherent,pustiowski2015independent,widhalm2018ultrafast} introducing further control parameters. Finally, such a hybrid Floquet device represents a key step towards full quantum acousto-optics in the visible spectral range, where both the involved light and the acoustic wave show photon and phonon characteristics, respectively.


\ack
This project was funded by the Deutsche Forschungsgemeinschaft (DFG, German Research Foundation), grant no. 465136867, and by the Narodowe Centrum Nauki (NCN, Polish National Science Centre), grant no. 2023/50/A/ST3/00511. DG, DW, MW, HJK, TK, and PM are supported by the German Federal Ministry of Education and Research via the Research Group Linkage Program of the Alexander von Humboldt Foundation.\\[5mm]
\textbf{Funding:} German Federal Ministry of Education and Research: Research Group Linkage Program of the Alexander von Humboldt Foundation.\\
National Science Centre, Poland, grant no. 2023/50/A/ST3/00511.\\
Deutsche Forschungsgemeinschaft, grant no. 465136867.


\newpage
\begin{appendices}

\section{Impact of the phonon environment on acousto-optically modulated emitters}\label{app:phonons}
The model of the optically driven, acoustically modulated two-level emitter discussed in Sec.~\ref{sec:model} does not contain microscopic coupling to phonons of the bulk material, in which any solid state emitter is embedded in. In the following we discuss the impact of such a phonon environment. We find that for cryogenic temperatures and typical parameters, describing the coupling of solid state emitters to bulk acoustic phonons, the phonons provide a small constant background to RF spectra, which is negligible in the parameter range studied here, as well as renormalizations of parameters, while their dissipative impact on the driven two-level emitter is negligible. The reason for this is that from the point of view of the phonon environment, the two-level emitter is modulated adiabatically by the optical and acoustic driving for typical parameters given in Sec.~\ref{sec:platforms}.
\subsection{Independent boson model and polaron frame}
We consider coupling of the emitter to an arbitrary number of phonon modes with frequencies $\Omega_n$ via the well-known independent boson model by adding the terms~\cite{mahan, krummheuer2002theory,roy2011influence, nazir2016modelling, groll2020four, preuss2022resonant}
\begin{subequations}
\begin{align}
    H_{\mathrm{ph}}&=\sum_n \hbar\Omega_n b_n^{\dagger}b_n^{}\,,\\
    H_{\mathrm{IB}}&=\ket{x}\bra{x}\sum_n \hbar {g_n}\qty(b_n^{\dagger}+b_n^{})
\end{align}
\end{subequations}
to the Hamiltonian $H(t)$ appearing in Eqs.~\eqref{eq:Hamilton} and \eqref{eq:lindblad}. Here, $g_n$ is the coupling constant of the phonon mode with quantum numbers collectively denoted by $n$ and $b_n^{(\dagger)}$ are the corresponding bosonic operators describing annihilation (creation) of phonons. In the absence of optical driving the full Hamiltonian 
\begin{equation}
    H_{\mathrm{tot}}(t)=H(t)+H_{\mathrm{ph}}+H_{\mathrm{IB}}\label{eq:H_tot}
\end{equation}
can be diagonalized by the unitary polaron transformation~\cite{mahan, roy2011influence, nazir2016modelling, groll2020four, preuss2022resonant} 
\begin{equation}
    T_{\mathrm{P}}=\ket{g}\bra{g}+\ket{x}\bra{x}B_+\label{eq:T_P}
\end{equation}
with the multi-mode displacement operators
\begin{align}
    B_{\pm}=\exp\qty[\pm\sum_n \frac{g_n}{\Omega_n}\qty(b_n^{\dagger}-b_n^{})]\,,\qquad B_-=B_+^{\dagger}\,.
\end{align}
Without optical driving we obtain
\begin{equation}
    H_{0,\mathrm{P}}(t)=T_{\mathrm{P}}\eval{H_{\mathrm{tot}}(t)}_{\Omega_\mathrm{R}=0} T_{\mathrm{P}}^{\dagger}=\hbar\frac{\Delta_{\mathrm{P}}(t)}{2}\qty(\ket{g}\bra{g}-\ket{x}\bra{x})+H_{\mathrm{ph}}\,,\label{eq:H_0_P}
\end{equation}
where we dropped an additional term proportional to the identity operator $\ket{g}\bra{g}+\ket{x}\bra{x}$ of the two-level emitter, since it does not contribute to the system's density matrix dynamics. The result in Eq.~\eqref{eq:H_0_P} corresponds to a decoupling of the two-level emitter from the phonon modes, however leading to an energy renormalization via the polaron shift
\begin{equation}\label{eq:polaron_shift}
    \Delta_{\mathrm{P}}=\Delta + \Omega_{\mathrm{P}}\,,\qquad \Omega_{\mathrm{P}}=\sum_n \frac{g_n^2}{\Omega_n}\,.
\end{equation}
In this polaron frame, the two energy levels $\ket{g}$ and $\ket{x}$ describe the ground and excited state of the dressed polaron instead of the bare exciton due to the coupling to the phonon modes in the original frame. Furthermore, the optical driving term transforms into the polaron frame as
\begin{align}\label{eq:H_I_P}
    H_{I,\mathrm{P}}&=T_{\mathrm{P}}H_{\mathrm{tot}}(t)T_{\mathrm{P}}^{\dagger}-H_{0,\mathrm{P}}(t)=T_{\mathrm{P}}\hbar\frac{\Omega_{\rm  R}}{2}\qty(\ket{x}\bra{g}+\ket{g}\bra{x})T_{\mathrm{P}}^{\dagger}\notag\\
    &=\hbar\frac{\Omega_{\rm  R}}{2}B_+\ket{x}\bra{g}+h.c.
\end{align}
In the presence of optical driving the polaron and the phonon modes are never fully decoupled. Considering the phonons to be in an initial thermal state, we can separate the impact of the phonons on the optical driving of the polaron into two parts. The thermal average over the phonon degrees of freedom of the previous Hamiltonian yields the direct optical driving of the polaron
\begin{align}\label{eq:H_I_P_d}
    H_{I,\mathrm{P}}^{(\mathrm{d})}=\hbar\frac{\Omega_{\rm  R}}{2}B\ket{x}\bra{g}+h.c.
\end{align}
with the thermal average of the multi-mode displacement operator~\cite{roy2011influence, nazir2016modelling}
\begin{align}\label{eq:def_B}
    B=\expval{B_{\pm}}=\exp{-\frac{1}{2}\int\limits_0^{\infty}\dd\Omega\,\frac{J(\Omega)}{\Omega^2}[2n(\Omega)+1]}\,.
\end{align}
The thermal occupation of the phonon modes with frequency $\Omega$ at temperature $T$ is here given by
\begin{align}
    n(\Omega)=\frac{1}{e^{\beta\hbar\Omega}-1}\,,\qquad \beta=(k_{\mathrm{B}}T)^{-1}\,,
\end{align}
where $k_{\mathrm{B}}$ is Boltzmann's constant and we introduced the phonon spectral density~\cite{roy2011influence, vagov2011real, nazir2016modelling, groll2020four, preuss2022resonant}
\begin{align}
    J(\Omega)=\sum_n g_n^2\delta(\Omega-\Omega_n)\,,
\end{align}
which describes the effective coupling of all phonon modes with frequency $\Omega$ to the two-level emitter. This quantity determines all relevant properties of the model, e.g., the polaron shift [see Eq.~\eqref{eq:polaron_shift}]
\begin{equation}
    \Omega_{\mathrm{P}}=\int\limits_0^{\infty}\dd\Omega\,\frac{J(\Omega)}{\Omega}\,.
\end{equation}
For the coupling of solid state emitters to longitudinal acoustic phonons, e.g., via the deformation potential, a typical spectral density is of the form~\cite{ramsay2010phonon,roy2011influence, nazir2016modelling,groll2020four, preuss2022resonant}
\begin{equation}
    J(\Omega)=\alpha\Omega^3 f\qty(\frac{\Omega}{\Omega_{\mathrm{c}}})\label{eq:spec_dens_qd}
\end{equation}
with a cutoff function obeying
\begin{equation}
    f(0)=1\,,\qquad f(x\ll 1)\approx1\,,\qquad f(x\gg 1)\approx0\,.
\end{equation}
Here, $\alpha$ is a measure for the overall coupling strength and $\Omega_{\mathrm{c}}$ is the cutoff frequency, i.e., modes with higher frequencies couple inefficiently to the emitter, while modes with smaller frequencies couple more efficiently. However for sufficiently small frequencies the low energy behavior
\begin{equation}
    J(\Omega\ll\Omega_{\mathrm{c}})\approx\alpha\Omega^3
\end{equation}
has a super-ohmic character for the coupling to bulk acoustic modes in three dimensional systems. Typical cutoff frequencies are $\hbar\Omega_{\mathrm{c}}\sim1~$meV for quantum dots~\cite{ramsay2010phonon,roy2011influence, nazir2016modelling,groll2020four} or $\hbar\Omega_{\mathrm{c}}\sim 10~$meV for color centers in hBN and diamond~\cite{alkauskas2014first,goldman2015phonon,norambuena2020quantifying,preuss2022resonant}. In the latter cases the comparatively larger cutoff frequency stems from the small size of these color centers compared with quantum dots. We will here assume a cutoff frequency on the order of $\hbar\Omega_{\mathrm{c}}\sim 1$~meV, which sets the strictest requirement on adiabaticity of the acousto-optical modulation discussed in the following.

The thermal average of the phonon displacement operator from Eq.~\eqref{eq:def_B} fulfills $B<1$ leading to a renormalization of the Rabi frequency $\Omega_{\mathrm{R}}\rightarrow B\Omega_{\mathrm{R}}<\Omega_{\mathrm{R}}$ in Eq.~\eqref{eq:H_I_P_d} compared with Eq.~\eqref{eq:Hamilton}, i.e., the phonon-dressed polaron is driven less efficiently than the bare exciton~\cite{ramsay2010phonon, nazir2016modelling, groll2020four}. The difference between the full optical driving Hamiltonian in the polaron frame [Eq.~\eqref{eq:H_I_P}] and the direct polaron driving [Eq.~\eqref{eq:H_I_P_d}] yields the phonon-assisted optical driving
\begin{align}
    H_{I,\mathrm{P}}^{\mathrm{(p-a)}}=\hbar\frac{\Omega_{\rm  R}}{2}(B_+-B)\ket{x}\bra{g}+h.c.
\end{align}
By definition the thermal average of this Hamiltonian vanishes, since $\expval{B_+-B}=0$. It describes the residual coupling between optical driving, the polaron and the phonons, while the direct optical driving from Eq.~\eqref{eq:H_I_P_d} does not contain any coupling to the phonon modes anymore.

The dynamics of the density matrix in the polaron frame
\begin{align}
    \rho_\mathrm{P}^{}=T_\mathrm{P}^{}\rho T_\mathrm{P}^{\dagger}
\end{align}
can be determined by transforming the Lindblad equation~\eqref{eq:lindblad}, using the replacement $H(t)\rightarrow H_{\mathrm{tot}}(t)$ [from Eq.~\eqref{eq:H_tot}]
\begin{align}\label{eq:lindblad_P}
    \dv{t}\rho_{\mathrm{P}}(t)=-\frac{i}{\hbar}\comm{H_{0,\mathrm{P}}(t)+H_{I,\mathrm{P}}^{(\mathrm{d})}+H_{I,\mathrm{P}}^{(\mathrm{p-a})}}{\rho_{\mathrm{P}}(t)}+\mathcal{D}_{\rm  xd, P}\qty[\rho_{\mathrm{P}}(t)]+\mathcal{D}_{\rm  pd}\qty[\rho_{\mathrm{P}}(t)]\,.
\end{align}
The pure dephasing dissipator is not impacted by the polaron transformation, while the excited state decay dissipator changes to
\begin{equation}
    \mathcal{D}_{\rm  xd, P}(\rho_{\mathrm{P}})=\gamma_{\rm  xd}\qty(B_-\ket{g}\bra{x}\rho_{\mathrm{P}}\ket{x}\bra{g}B_+-\frac{1}{2}\acomm{\ket{x}\bra{x}}{\rho_{\mathrm{P}}})\,.\label{eq:D_xd_P}
\end{equation}
As in the case of optical driving, the (radiative) excited state decay couples the phonon modes and the polaron decay dynamics. In a similar fashion we can define a direct polaron decay dissipator
\begin{equation}
    \mathcal{D}_{\rm  xd, P}^{\mathrm{(d)}}(\rho_{\mathrm{P}})=B^2\gamma_{\rm  xd}\qty(\ket{g}\bra{x}\rho_{\mathrm{P}}\ket{x}\bra{g}-\frac{1}{2}\acomm{\ket{x}\bra{x}}{\rho_{\mathrm{P}}})\,,\label{eq:D_xd_P_d}
\end{equation}
which is equivalent to the original one in Eqs.~\eqref{eq:dissipators} with the rate being renormalized by $\gamma_{\mathrm{xd}}\rightarrow\gamma_{\mathrm{xd}}B^2$, and which describes excited state decay of the polaron. Furthermore we can define a phonon-assisted decay dissipator $\mathcal{D}_{\rm  xd, P}^{\mathrm{(p-a)}}(\rho_{\mathrm{P}})=\mathcal{D}_{\rm  xd, P}(\rho_{\mathrm{P}})-\mathcal{D}_{\rm  xd, P}^{\mathrm{(d)}}(\rho_{\mathrm{P}})$ which describes the residual coupling between polaron and phonons via the excited state decay.

\subsection{Effective Lindblad dissipator for adiabatic acoustic modulation}
To remove the explicit impact of the phonon coupling from Eq.~\eqref{eq:lindblad_P}, we can derive a Lindblad master equation in the polaron frame, interpreting the phonon modes as a bath, inducing dissipation in the optically driven polaron system~\cite{breuer2002theory,roy2011influence, nazir2016modelling,groll2020four}. To this aim, we define the system Hamiltonian as
\begin{equation}
    H_{\mathrm{S}}(t)=H_{0,\mathrm{P}}(t)+H_{I,\mathrm{P}}^{(\mathrm{d})}-H_{\mathrm{ph}}=\hbar\frac{\Delta_{\mathrm{P}}(t)}{2}\qty(\ket{g}\bra{g}-\ket{x}\bra{x})+\hbar\frac{\Omega_{\rm  R}}{2}B\qty(\ket{x}\bra{g}+\ket{g}\bra{x})\,,
\end{equation}
the bath Hamiltonian as
\begin{equation}
    H_{\mathrm{B}}=H_{\mathrm{ph}}
\end{equation}
and the system-bath interaction as
\begin{align}
    H_{\mathrm{SB}}&=H_{I,\mathrm{P}}^{\mathrm{(p-a)}}=\hbar\frac{\Omega_{\rm  R}}{2}(B_+-B)\ket{x}\bra{g}+h.c.\notag\\
    &=\sum_{i=1,2}\hbar\frac{\Omega_{\rm  R}}{2} B_i\sigma_i\,.\label{eq:H_SB}
\end{align}
The system Hamiltonian $H_{\mathrm{S}}$ is simply the Hamiltonian given in Eq.~\eqref{eq:Hamilton} but including the renormalizations $\Delta\rightarrow\Delta_{\mathrm{P}}$ and $\Omega_{\mathrm{R}}\rightarrow B\Omega_{\mathrm{R}}$ of the detuning and the Rabi frequency, respectively, while the bath Hamiltonian describes the free dynamics of the phonons. The system-bath interaction contains the phonon operators
\begin{subequations}
\begin{align}
    B_1&=\frac{1}{2}\qty(B_++B_--2B)\,,\\
    B_2&=\frac{1}{2i}\qty(B_+-B_-)\,.
\end{align}
\end{subequations}
and the Pauli matrices [see Eqs.~\eqref{eq:Pauli}].

In the standard approach to deriving an effective Lindblad equation one determines the bath correlation functions~\cite{breuer2002theory,roy2011influence, nazir2016modelling,groll2020four}
\begin{equation}
    C_{ij}(\tau)=\expval{B_i^{\dagger}(\tau)B_j}\,,
\end{equation}
where the dynamics are governed by the bath Hamiltonian $H_\mathrm{B}$ and the average is taken with respect to the thermal phonon state. This leads to
\begin{subequations}\label{eq:bath_corrs}
\begin{align}
    C_{11}(\tau)&=\frac{B^2}{2}\qty[e^{\phi(\tau)}+e^{-\phi(\tau)}-2]\,,\\
    C_{22}(\tau)&=\frac{B^2}{2}\qty[e^{\phi(\tau)}-e^{-\phi(\tau)}]\,,\\
    C_{12}(\tau)&=C_{21}(\tau)=0\,.
\end{align}
\end{subequations}
These bath correlation functions are determined by the dephasing function
\begin{equation}
    \phi(\tau)=\int\limits_0^{\infty}\dd\Omega\,\frac{J(\Omega)}{\Omega^2}\qty{e^{-i\Omega\tau}\qty[n(\Omega)+1]+e^{i\Omega\tau}n(\Omega)}\label{eq:phi_tau}
\end{equation}
which in turn is entirely defined via the phonon spectral density and the thermal phonon occupation. For the spectral density in Eq.~\eqref{eq:spec_dens_qd}, the dephasing function decays from its initial value $\phi(0)=\ln(B^{-2})$ [see Eq.~\eqref{eq:def_B}] towards zero
\begin{equation}
    \phi(\tau\gg \Omega_{\mathrm{c}}^{-1})\approx 0\,.
\end{equation}
This in turn implies that the bath correlation functions from Eqs.~\eqref{eq:bath_corrs} vanish on the time scale of the bath $\tau_{\mathrm{B}}\sim \Omega_{\mathrm{c}}^{-1}$.

For a sufficiently weak coupling between system and thermal bath the system's dynamics on a coarse time scale compared with $\tau_{\mathrm{B}}$ can be described by an additional dissipator and phonon-induced Lambshift Hamiltonian in Eq.~\eqref{eq:lindblad_P}, removing the explicit appearance of $H_{\mathrm{SB}}=H_{I,\mathrm{P}}^{\mathrm{(p-a)}}$~\cite{breuer2002theory, nazir2016modelling}. These additional terms describe phonon-induced transitions between the eigenstates of $H_\mathrm{S}(t)$, i.e., between the optically dressed states with energies [see also Eq.~\eqref{eq:H_dressed_states} for eigenstates in the resonant case $\Delta_{\mathrm{P}}=0$]
\begin{equation}
    E_{\pm}(t)=\pm\frac{\hbar}{2}\sqrt{B^2\Omega_{\mathrm{R}}^2+\Delta_{\mathrm{P}}(t)^2}\,,\label{eq:E_dressed}
\end{equation}
as well as phonon-induced energy renormalizations of these eigenstates. Focusing on the transition rates for a phonon-induced transition with a frequency difference $\Delta \omega$, e.g., $\hbar\Delta \omega=E_+-E_-$, these rates can be calculated from the bath correlation functions in Eq.~\eqref{eq:bath_corrs} via~\cite{breuer2002theory, roy2011influence, nazir2016modelling, groll2020four}
\begin{align}
    \gamma_{ij}(\Delta \omega)&=\frac{\Omega_{\mathrm{R}}^2}{4}\int\limits_{-\infty}^{\infty}\dd\tau\, C_{ij}(\tau)e^{i\Delta \omega\tau}\,.\label{eq:gamma_ph_1}
\end{align}
The energy renormalization parameters can be obtained from the bath correlation functions in a similar fashion. One important prerequisite for this effective description is that the system's explicit time-dependence, stemming from the acoustic modulation via $\Delta_{\mathrm{P}}(t)$, changes adiabatically compared with the bath correlation time $\tau_\mathrm{B}$. Otherwise we cannot simply apply the standard derivation of the Lindblad equation~\cite{breuer2002theory} to such a time-dependent problem. This implies
\begin{equation}
    \Omega_{\mathrm{ac}}\ll \Omega_{\mathrm{c}}
\end{equation}
for the acoustic detuning modulation considered explicitly in Eq.~\eqref{eq:detuning_sinus}. Since in the main text we focus heavily on the properties of the system in acousto-optical resonance with $\Omega_{\mathrm{R}}\sim \Omega_{\mathrm{ac}}$ we can furthermore state
\begin{equation}
    \Omega_{\mathrm{R}}\ll \Omega_{\mathrm{c}}
\end{equation}
for our purposes, which is in line with the assumption of a sufficiently weak system-bath coupling [see Eq.~\eqref{eq:H_SB}]. For not too strong acoustic driving, i.e., $\Delta(t)\sim   \Omega_{\mathrm{R}}$, the energy difference between the optically dressed states in Eq.~\eqref{eq:E_dressed} is on the order of the Rabi frequency $|\Delta \omega|\sim \Omega_{\mathrm{R}}$, implying
\begin{equation}
    |\Delta \omega|\ll \Omega_{\mathrm{c}}\,.
\end{equation}
In fact these considerations are justified given that typical Rabi frequencies and acoustic modulation frequencies discussed in Sec.~\ref{sec:platforms} are on the order of $\Omega_{\mathrm{R}}/(2\pi)\approx 5$~GHz, i.e., $\hbar\Omega_{\mathrm{R}}\approx 0.02$~meV, while typical cutoff frequencies for acoustic phonon baths described by Eq.~\eqref{eq:spec_dens_qd} are $\Omega_{\mathrm{c}}\sim 1$~meV or higher~\cite{ramsay2010phonon,roy2011influence, alkauskas2014first,goldman2015phonon,nazir2016modelling,groll2020four, norambuena2020quantifying,preuss2022resonant}.

We can make further progress in Eq.~\eqref{eq:gamma_ph_1} by assuming not too strong coupling between system and bath, i.e., $e^{\phi(\tau)}\approx 1+\phi(\tau)$ in Eqs.~\eqref{eq:bath_corrs}, and sufficiently high temperatures $k_{\mathrm{B}}T\gg \hbar\Omega_{\mathrm{R}}$, i.e., $n(|\Delta\omega|)\gg 1$, such that the only non-vanishing phonon-induced dissipation rate is given by
\begin{equation}
    \gamma_{22}(\Delta \omega)\approx \frac{\Omega_{\mathrm{R}}^2}{4}B^2\int\limits_{-\infty}^{\infty}\dd\tau\, \phi(\tau)e^{i\Delta \omega\tau}\approx\frac{\pi}{2} \Omega_{\mathrm{R}}^2B^2 \frac{J\qty(|\Delta \omega|)}{|\Delta \omega|^2}n(|\Delta \omega|)\,.\label{eq:gamma_ph_2}
\end{equation}
The high temperature assumption is justified even for cryogenic temperatures, since $k_{\mathrm{B}}T\approx 0.35~$meV at $T=4$~K, which is an order of magnitude larger than typical Rabi frequencies $\hbar\Omega_{\mathrm{R}}\approx 0.02$~meV considered here. To evaluate the numerical value of the phonon-induced dissipation rate, we consider approximately resonant driving $\Delta_{\mathrm{P}}\ll B\Omega_{\mathrm{R}}$ as an example, such that the renormalized transition frequency between the dressed states is given by $\Delta\omega\approx B\Omega_{\mathrm{R}}$, implying 
\begin{equation}
    \gamma_{22}(B\Omega_{\mathrm{R}})\approx \frac{\pi}{2} J(B\Omega_{\mathrm{R}})n(B\Omega_{\mathrm{R}})\,,\label{eq:gamma_ph_3}
\end{equation}
i.e., the dissipation rate is determined by the number of phonons that have a frequency resonant to the dressed state splitting $\Delta\omega=B\Omega_{\mathrm{R}}$, as well as the spectral density at the corresponding transition frequency~\cite{nazir2016modelling}. At $T=4$~K with a renormalized Rabi frequency of $B\Omega_{\mathrm{R}}/(2\pi)=5$~GHz we can furthermore approximate the thermal occupation due to the comparatively high temperature, such that
\begin{equation}
    \gamma_{22}(B\Omega_{\mathrm{R}})\approx \frac{\pi}{2} \alpha (B\Omega_{\mathrm{R}})^2 \frac{k_{\mathrm{B}}T}{\hbar} \,,\label{eq:gamma_ph_4}
\end{equation}
for the spectral density given in Eq.~\eqref{eq:spec_dens_qd} with $B\Omega_{\mathrm{R}}\ll\Omega_{\mathrm{c}}$. Using parameters typical for quantum dots or color centers in hBN and diamond, i.e., $\alpha\sim 10^{-2}$~ps$^2$ (or smaller)~\cite{ramsay2010phonon, alkauskas2014first,goldman2015phonon,nazir2016modelling,groll2020four, norambuena2020quantifying,preuss2022resonant}, and a renormalized Rabi frequency of $B\Omega_{\mathrm{R}}/(2\pi)=5$~GHz, we obtain a phonon-induced dissipation rate of $\gamma_{22}\sim 10^{-2}~\text{ns}^{-1}$. This is significantly smaller than typical excited state decay rates or pure dephasing rates considered in Sec.~\ref{sec:platforms}, allowing us to ignore the dissipative effect of the phonon bath via the system-bath coupling in Eq.~\eqref{eq:H_SB} for adiabatic acoustic driving and a correspondingly small Rabi frequency. This implies that we can approximate the effective dynamics of the polaron by Eq.~\eqref{eq:lindblad_P} with $H_{I,\mathrm{P}}^{(\mathrm{p-a})}\rightarrow 0$ and the relatively small impact of the phonons on the polaron allows us furthermore to replace the full excited state decay disspator $\mathcal{D}_{\rm  xd, P}$ in Eq.~\eqref{eq:D_xd_P} by the direct polaron decay dissipator $\mathcal{D}_{\rm  xd, P}^{(\mathrm{d})}$ in Eq.~\eqref{eq:D_xd_P_d}. This effectively decouples phonon and polaron dynamics due to the adiabatic nature of the acousto-optical modulation~\cite{machnikowski2007quantum,wigger2014energy} at cryogenic temperatures and renders the action of Eq.~\eqref{eq:lindblad_P} on the two-level emitter, i.e., the master equation of the polaron, to be approximately identical to the Lindblad equation~\eqref{eq:lindblad} in the main text with the parameter renormalizations
\begin{equation}\label{eq:parameter_renorm_phon}
    \Delta\rightarrow\Delta_{\mathrm{P}}\,,\qquad \gamma_{\mathrm{xd}}\rightarrow B^2\gamma_{\mathrm{xd}}\,,\qquad \Omega_{\mathrm{R}}\rightarrow B\Omega_{\mathrm{R}}\,.
\end{equation}
The dynamics of the phonons are still governed by $H_{\mathrm{ph}}$.
\subsection{Impact of the phonon bath on the RF spectrum}
The RF spectrum in Eq.~\eqref{eq:I_RF} is determined entirely by the two-time correlation function $C(t+\tau,t)$ from Eq.~\eqref{eq:def_C}. In the following we call the full correlation function, including the impact of coupling to a phonon bath, $\widetilde{C}(t+\tau,t)$. Going to the polaron frame via the unitary operator in Eq.~\eqref{eq:T_P} leads to
\begin{equation}
    \widetilde{C}(t+\tau,t)=\Tr\qty{\sigma_+B_+\mathcal{V}_{\mathrm{P}}(t+\tau,t)\qty[\sigma_-B_-\rho_{\mathrm{P}}(t)]}
\end{equation}
where $\rho_{\mathrm{P}}$ is the density matrix in the polaron frame, whose dynamics are governed by the Lindblad equation~\eqref{eq:lindblad_P}, which also defines the corresponding time evolution super-operator in the polaron frame $\mathcal{V}_{\mathrm{P}}$ analog to Eq.~\eqref{eq:action_V}. As discussed in the previous section, due to the adiabatic nature of the acousto-optical modulation, the Lindblad equation in the polaron frame can be approximated by a Lindblad equation for the two-level emitter (polaron) of the form in Eq.~\eqref{eq:lindblad} and additional free dynamics of the phonons. This allows us to separate the full correlation function $\widetilde{C}$ into a pure polaron correlation function, given by the correlation function for the two-level emitter $C$ from the main text [with additional parameter renormalizations from Eq.~\eqref{eq:parameter_renorm_phon}], as well as a pure phonon correlation function
\begin{equation}
\widetilde{C}(t+\tau,t)=C(t+\tau,t)\expval{B_+(\tau)B_-}\,.\label{eq:C_full}
\end{equation}
Here we used that the phonon correlation function is evaluated in a thermal state, such that it only depends on the time difference $\tau$. This phonon correlation function evaluates to~\cite{nazir2016modelling}
\begin{equation}
    \expval{B_+(\tau)B_-}=B^2e^{\phi(\tau)}\approx B^2+B^2\phi(\tau)
\end{equation}
with the thermal expectation value of the phonon displacement operators from Eq.~\eqref{eq:def_B} and the dephasing function from Eq.~\eqref{eq:phi_tau}. In the second step we assumed sufficiently weak phonon coupling, i.e., negligible two-phonon processes. Calculating the RF spectrum in Eq.~\eqref{eq:I_RF} using the full correlation function from Eq.~\eqref{eq:C_full} yields
\begin{subequations}
    \begin{align}
        I_{\rm RF}(\omega_l+\omega;\Gamma)&=I_{\rm RF}^{(0)}(\omega_l+\omega;\Gamma)+I_{\rm RF}^{(1)}(\omega_l+\omega;\Gamma)+\mathcal{O}(\phi^2)\,,\\
        I_{\rm RF}^{(0)}(\omega_l+\omega;\Gamma)&=B^2 2\text{Re}\left[\int\limits_0^{\infty}\dd\tau\int\limits_0^T\frac{\dd t}{T}\,C(t+\tau,t)e^{-i\omega\tau-\Gamma\tau}\right]\,,\label{eq:I_RF_0}\\
        I_{\rm RF}^{(1)}(\omega_l+\omega;\Gamma)&=B^2 2\text{Re}\left[\int\limits_0^{\infty}\dd\tau\int\limits_0^T\frac{\dd t}{T}\,C(t+\tau,t)\phi(\tau)e^{-i\omega\tau-\Gamma\tau}\right]\,,
    \end{align}
\end{subequations}
where higher order terms stem from higher powers of $\phi(\tau)$ in the expansion of the phonon correlation function. The zeroth order contribution to the RF spectrum $I_{\rm RF}^{(0)}$ is simply the renormalized version of Eq.~\eqref{eq:I_RF} with an additional rescaling by $B^2$ due to the phonon-induced screening of the polaron [see Eq.~\eqref{eq:parameter_renorm_phon}]. This part contains the so-called zero-phonon lines~\cite{duke1965phonon,krummheuer2002theory, wigger2019phonon, preuss2022resonant}. The first order contribution $I_{\rm RF}^{(1)}$ contains the phonon sidebands due to one-phonon processes, while the higher order contributions describe the impact of higher order phonon processes.  We can write the first order contribution as a convolution
\begin{align}
    I_{\rm RF}^{(1)}(\omega_l+\omega;\Gamma)=I_{\rm RF}^{(0)}(\omega_l+\omega;\Gamma)*\tilde{\phi}(\omega)\label{eq:I_RF_1}
\end{align}
such that to each peak in the zeroth order spectrum, an additional sideband of the form
\begin{align}
    \tilde{\phi}(\omega)=\frac{1}{2\pi}\int\limits_{-\infty}^{\infty}\dd\tau e^{-i\omega\tau}\phi(\tau)
\end{align}
is attached. We are interested in the form of this sideband at $\omega\sim\Omega_{\mathrm{ac}}$, i.e., where it overlaps with the sidebands due to the acoustic modulation, e.g., in Fig.~\ref{fig:2}. For $\omega\sim\Omega_{\mathrm{R}}$ and cryogenic temperatures on the order of $T=4$~K, we can apply Eqs.~\eqref{eq:gamma_ph_2} and \eqref{eq:gamma_ph_4} to find that 
\begin{align}
    \tilde{\phi}(\omega)\approx \frac{J\qty(|\omega|)}{|\omega|^2}n(|\omega|)\approx \frac{\alpha}{\hbar} k_{\mathrm{B}}T\,,\label{eq:phi_omega}
\end{align}
i.e., the sideband produces a constant background. This is of course only true close to each zero-phonon line contained in $I_{\rm RF}^{(0)}$, i.e., at a frequency distance on the order of $\sim\Omega_{\mathrm{ac}}$, and stems from the adiabatic acoustic modulation $\Omega_{\mathrm{ac}}\ll\Omega_{\mathrm{c}}$, as well as the high temperature limit $k_{\mathrm{B}}T\gg\hbar\Omega_{\mathrm{ac}}$. To estimate the impact of this background, we calculate its intensity via Eqs.~\eqref{eq:I_RF_0}, \eqref{eq:I_RF_1} and \eqref{eq:phi_omega}
\begin{align}
    I_{\rm RF}^{(1)}(\omega_l+\omega;\Gamma)&\approx\frac{\alpha}{\hbar} k_{\mathrm{B}}T \int\dd\omega'\,I_{\rm RF}^{(0)}(\omega_l+\omega-\omega';\Gamma)\notag\\
    &=\frac{\alpha}{\hbar} k_{\mathrm{B}}T B^2 \int\dd\omega\,2\text{Re}\left[\int\limits_0^{\infty}\dd\tau\int\limits_0^T\frac{\dd t}{T}\,C(t+\tau,t)e^{-i\omega\tau-\Gamma\tau}\right]\notag\\
    &=\frac{\alpha}{\hbar} k_{\mathrm{B}}T B^2 4\pi \text{Re}\left[\int\limits_0^{\infty}\dd\tau\int\limits_0^T\frac{\dd t}{T}\,C(t,t)\delta(\tau)\right]\notag\\
    &=\frac{\alpha}{\hbar} k_{\mathrm{B}}T B^2 2\pi\int\limits_0^T\frac{\dd t}{T}\,C(t,t)\,.
\end{align}
The intensity of the sideband is determined by the average occupation, i.e., the average over $C(t,t)=\expval{\ket{x}\bra{x}}(t)$, which can be unity at most. Thus the first order phonon sidebands produce a constant background with the upper bound
\begin{equation}
    I_{\rm RF}^{(1)}(\omega_l+\omega;\Gamma)<2\pi\frac{\alpha}{\hbar}k_{\mathrm{B}}TB^2\,.\label{eq:background}
\end{equation}
For reference the intensity of the peaks in the zeroth order spectrum can be estimated from Eqs.~\eqref{eq:I_RF_0} and \eqref{eq:I_RF_full} as $\sim B^2(\Gamma+\gamma_r)^{-1}$. Note that both the zeroth order spectrum, as well as the phonon background in Eq.~\eqref{eq:background} are renormalized by the same factor $B^2$, such that we can ignore it for the direct comparison. We thus simply set it to $B=1$ for the remaining discussion. The height of the peaks is then determined by $\sim (\Gamma+\gamma_r)^{-1}$, which for ideal detection conditions with $\Gamma\rightarrow 0$ becomes $\sim\gamma_r^{-1}<\min(T_1,T_2)$ with $T_1=1/\gamma_{\mathrm{xd}}$ and $T_2=1/\gamma$ [see Eq.~\eqref{eq:floquet_rates}]. For typical emitters we find a dephasing time $T_2$ that is smaller than the (radiative) lifetime $T_1$ (see Tabs.~\ref{tab:lifetimes} and \ref{tab:dephasing}). Typical values are $T_2\sim 10-1000$~ps, while the upper bound of the phonon background in Eq.~\eqref{eq:background} is on the order of $I_{\rm RF}^{(1)}\sim 10^{-2}$~ps for cryogenic temperatures on the order of $T=4~$K and $\alpha\sim 10^{-2}$~ps$^2$~\cite{ramsay2010phonon, alkauskas2014first,goldman2015phonon,nazir2016modelling,groll2020four, norambuena2020quantifying,preuss2022resonant}. From this estimate we find that the phonon background is negligible compared to the zeroth order signal, as it is expected to be at least three orders of magnitude smaller for typical parameters considered in Sec.~\ref{sec:platforms}. We therefore do not expect a significant impact on the RF signals, e.g., shown in Fig.~\ref{fig:2}. Indeed recently the anticrossing structures predicted in Fig.~\ref{fig:2} have been measured in Ref.~\cite{zhan2025dynamical} with no significant visible impact from a phonon background and the results have been explained using a simple two-level model analog to the one presented in Sec.~\ref{sec:model}.
\section{General properties of Floquet eigenvalues of the periodically driven dissipative two-level system}\label{app:floquet_eigen_diss}

\subsection{System and Liouville space representation}\label{app:liouville_rep}
	We consider the most general form of a periodically driven TLS with the Hamiltonian
	\begin{equation}\label{eq:Hamilton_general}
		H(t)=\sum_{i=1}^3 h_i(t)\sigma_i
	\end{equation}
	written in terms of the Pauli matrices in the basis $\qty{\ket{g},\ket{x}}$~\cite{allen1987optical,breuer2002theory}
	\begin{align}\label{eq:Pauli}
		\sigma_1&=\ket{x}\bra{g}+\ket{g}\bra{x}\,,\notag\\ \sigma_2&=i\ket{x}\bra{g}-i\ket{g}\bra{x}\,,\notag\\ \sigma_3&=\ket{g}\bra{g}-\ket{x}\bra{x}\,.
	\end{align}
 For $H(t)$ to be hermitian, the functions $h_i(t)$ have to be real. The acoustically modulated optically driven TLS discussed in the main text [Eq.~\eqref{eq:Hamilton}] is recovered for $h_1(t)=\hbar\Omega_{\rm R}/2$, $h_2(t)=0$ and $h_3(t)=\hbar\Delta(t)/2$, however the following discussion is of a more general nature and the only restriction we place on the Hamiltonian here is time-periodicity $H(t+T)=H(t)$, i.e., $h_i(t+T)=h_i(t)$.  Note that with this definition of the Pauli matrices we have 
    \begin{align}
        \sigma_-=\ket{g}\bra{x}=\frac{1}{2}(\sigma_1+i\sigma_2)=\sigma_+^{\dagger}\,
    \end{align}
    for the ladder operator appearing in the calculation of the RF spectrum, e.g., in Eq.~\eqref{eq:I_RF_full}.
    
	In addition to the unitary dynamics induced by the Hamiltonian in Eq.~\eqref{eq:Hamilton_general}, we consider excited state decay and pure dephasing of the TLS with the rates $\gamma_{\rm xd}$ and $\gamma_{\rm pd}$, respectively. The dynamics of the TLS are then described by the Lindblad equation~\eqref{eq:lindblad}.

    Together with the identity operator $\sigma_0=\ket{g}\bra{g}+\ket{x}\bra{x}$, the Pauli matrices form a complete orthogonal basis for operators $A$ acting on the TLS Hilbert space, i.e., for states $\ketL{A}$ in four-dimensional Liouville space (see also Sec.~\ref{sec:liouville})~\cite{allen1987optical,breuer2002theory}, allowing one to decompose
	\begin{equation}
		\left|A\right)=\sum_{i=0}^3 \alpha_i \left|\sigma_i\right)\,.
	\end{equation}
    Note that Eq.~\eqref{eq:Hamilton_general} does not contain a contribution proportional to the identity $\sigma_0$, since this can always be removed by a simple unitary transformation.	The components $\alpha_i$ can be obtained by making use of the scalar product defined in Eq.~\eqref{eq:scalar_prod},
	\begin{equation}
		\left(\sigma_i|\sigma_j\right)=2\delta_{ij}\,,
	\end{equation}
	such that $\alpha_i=\left(\sigma_i|A\right)/2$. Using the orthonormal basis $\qty{\left|\sigma_i\right)/\sqrt{2}|i=0,1,2,3}$ we can represent the density matrix as
	\begin{equation}
		\left|\rho\right)=\sum_{i=0}^3 \varrho_i\frac{\left|\sigma_i\right)}{\sqrt{2}}\hat{=}\left(\begin{matrix}\varrho_0\\\varrho_1\\\varrho_2\\\varrho_3\end{matrix}\right)
	\end{equation}
	and the Lindblad super-operator in Eq.~\eqref{eq:lindblad} with the Hamiltonian in Eq.~\eqref{eq:Hamilton_general} as 
	\begin{equation}\label{eq:lindbladian_matrix}
		{\mathcal{L}}(t)\hat{=}\left(\begin{matrix}0&0&0&0\\0&-\gamma&-2h_3(t)/\hbar&2h_2(t)/\hbar\\0&2h_3(t)/\hbar&-\gamma&-2h_1(t)/\hbar\\\gamma_{\rm xd}&-2h_2(t)/\hbar&2h_1(t)/\hbar&-\gamma_{\rm xd}\end{matrix}\right),
	\end{equation}
	with the matrix elements given by $\frac{1}{2}\left(\sigma_i\right|{\mathcal{L}}\left|\sigma_j\right)=\frac{1}{2}\Tr\qty(\sigma_i^\dagger\mathcal{L}\qty[\sigma_j])$. We denote here the total dephasing rate by $\gamma=\qty(\gamma_{\rm xd}+\gamma_{\rm pd})/2$~\cite{allen1987optical,mukamel1995principles,groll2025fundamentals}. The vanishing of the first row in Eq.~\eqref{eq:lindbladian_matrix} is connected to the fact that the Lindblad equation~\eqref{eq:lindblad} preserves the trace of the density matrix~\cite{lindblad1976generators,breuer2002theory},
	\begin{equation}\label{eq:trace_preserve}
		\dv{t}\Tr[\rho(t)]=\left(\sigma_0|{\mathcal{L}}(t)|\rho(t)\right)=0\,.
	\end{equation}
	Therefore, the subspace spanned by the traceless Pauli matrices $\left|\sigma_i\right)$, $i=1,2,3$ is closed under time evolution via the Lindblad equation~\eqref{eq:lindblad}, as can be seen in Eq.~\eqref{eq:lindbladian_matrix}. On traceless states of the form 
	\begin{equation}
		\left|\rho\right)_{\rm Tr=0}=\sum_{i=1}^3\varrho_{i}\frac{\left|\sigma_i\right)}{\sqrt{2}}
	\end{equation}
	the Lindblad super-operator therefore acts as the $3\times3$-matrix
	\begin{equation}\label{eq:lindbladian_matrix_3d}
		{\mathcal{L}_{\rm Tr=0}}(t)\hat{=}\left(\begin{matrix}-\gamma&-2h_3(t)/\hbar&2h_2(t)/\hbar\\2h_3(t)/\hbar&-\gamma&-2h_1(t)/\hbar\\-2h_2(t)/\hbar&2h_1(t)/\hbar&-\gamma_{\rm xd}\end{matrix}\right)\,.
	\end{equation}

	\subsection{Constraints on Floquet eigenvalues}\label{app:floquet_eigen_constraints}
    As discussed in the context of Eq.~\eqref{eq:trace_eigen_floquet}, due to the preservation of the trace in the Lindblad equation [see Eq.~\eqref{eq:trace_preserve}] we have $\Tr(M_r)=\mu_r\Tr(M_r)$ for the Floquet eigenvalues $\mu_r$ and eigenstates $M_r$ obeying Eq.~\eqref{eq:eigen_floquet}, i.e., $\mathcal{T}\ketL{M_r}=\mathcal{V}(T,0)\ketL{M_r}=\mu_r\ketL{M_r}$. The Floquet eigenstates are therefore traceless or otherwise have to have eigenvalue $\mu_r=1$. The subspace of traceless states is spanned by the three Pauli matrices $\ketL{\sigma_i}$ with $i=1,2,3$ and closed under the Lindblad equation [see Eq.~\eqref{eq:lindbladian_matrix}], such that there are three traceless solutions $\ketL{M_r}$ with $r=1,2,3$. The fourth remaining solution $\ketL{M_0}$ then has the properties $\Tr(M_0)=\expL{\sigma_0|M_0}\neq 0$, $\mu_0=1$. 
    
    Using the infinitesimal time evolution super-operator, i.e., solving Eq.~\eqref{eq:V_eom},
    \begin{equation}
        \mathcal{V}(t+\epsilon,t)=1+\epsilon \mathcal{L}(t)+\mathcal{O}\qty(\epsilon^2)=e^{\epsilon\mathcal{L}(t)}+\mathcal{O}\qty(\epsilon^2)\,,
    \end{equation}
    we can find further restrictions on the possible Floquet eigenvalues, applying the semigroup property in Eq.~\eqref{eq:semigroup}
	\begin{align}
		\prod_r \mu_r&=\det\qty[{\mathcal{V}}(T,0)]=\lim\limits_{N\rightarrow\infty}\prod_{j=1}^N\det\qty[e^{\frac{T}{N}{\mathcal{L}}\qty(T-j\frac{T}{N})}+\mathcal{O}\qty(\frac{1}{N^2})]\notag\\
        &=\lim\limits_{N\rightarrow\infty}\prod_{j=1}^N\qty{e^{\frac{T}{N}\Tr\qty[{\mathcal{L}}\qty(T-j\frac{T}{N})]}+\mathcal{O}\qty(\frac{1}{N^2})}\notag\\
        &=\lim\limits_{N\rightarrow\infty}\qty[e^{(-2\gamma-\gamma_{\rm xd})\frac{T}{N}}]^N+\lim\limits_{N\rightarrow\infty}\mathcal{O}\qty(\frac{1}{N})\notag\\
        &=e^{-(2\gamma+\gamma_{\rm xd})T}\label{eq:prod_mu_r}\,.
	\end{align}
	Here the determinant and trace are acting on the $4\times4$ matrix representations in Liouville space [see Eq.~\eqref{eq:lindbladian_matrix}] and we made use of Jacobi's formula for the determinant of matrix exponentials~\cite{hall2013lie}.
	
	The Floquet eigenstates $\ketL{M_r}$ constitute a basis of the Liouville space (except at exceptional points~\cite{muller2008exceptional,minganti2019quantum}), however they are not necessarily orthonormal, only linearly independent. The time-periodic states $\left|V_r(t)\right)$ in Eq.~\eqref{eq:def_V_r} are then also linearly independent since the matrix $e^{i\delta_r t+\gamma_r t}{\mathcal{V}}(t,0)$ has determinant [analog to Eq.~\eqref{eq:prod_mu_r}]
	\begin{equation}
		\det[e^{i\delta_r t+\gamma_r t}{\mathcal{V}}(t,0)]=e^{i\delta_r t+\gamma_r t} e^{-(2\gamma+\gamma_{\rm xd})t}\neq 0
	\end{equation}
	and is therefore invertible.

	\subsubsection{Constraints on the damping rates $\gamma_r$}\label{app:damping_rates}
	We can determine an upper bound on the largest absolute eigenvalue in Eq.~\eqref{eq:eigen_floquet} on the closed subspace of the traceless solutions, i.e., the spectral radius $R_S$ of ${\mathcal{V}_{\Tr=0}}(T,0)$~\cite{horn2012matrix}, where the index $\Tr=0$ indicates that we consider the action of the single-period translation $\mathcal{V}(T,0)$ on the subspace spanned by $|\sigma_i)$, $i=1,2,3$, analog to Eq.~\eqref{eq:lindbladian_matrix_3d}. The spectral radius of a matrix generally obeys~\cite{horn2012matrix}
	\begin{equation}
		R_S(A)\leq ||A||_2\,,
	\end{equation}
where $||A||_2$ is the spectral norm of the matrix, which is given by the square root of the largest eigenvalue of $A^{\dagger}A$. Furthermore, the spectral norm is sub-multiplicative such that~\cite{horn2012matrix}
\begin{equation}
	||AB||_2\leq ||A||_2||B||_2\,,
\end{equation}
leading to
\begin{equation}
	R_S(AB)\leq ||A||_2||B||_2\,.
\end{equation}
Using the semigroup property from Eq.~\eqref{eq:semigroup} for the dynamics on the closed subspace of traceless states, we can write analog to Eq.~\eqref{eq:prod_mu_r}
\begin{equation}
	R_S\qty[{\mathcal{V}_{\Tr=0}}(T,0)]\leq \lim\limits_{N\rightarrow\infty}\prod_{j=1}^N||e^{\frac{T}{N}{\mathcal{L}_{\rm Tr=0}}(T-j\frac{T}{N})}||_2\,.
\end{equation}
Next we need to determine $||e^{\epsilon {\mathcal{L}_{\rm Tr=0}}(t)}||_2$ for $\epsilon=\mathcal{O}(1/N)$, i.e., the square root of the largest eigenvalue of 
\begin{align}
	\qty[e^{\epsilon {\mathcal{L}_{\rm Tr=0}}(t)}]^\dagger&\qty[e^{\epsilon {\mathcal{L}_{\rm Tr=0}}(t)}]\notag\\
    &=1+\epsilon\qty[{\mathcal{L}_{\rm Tr=0}}(t)^{\dagger}+{\mathcal{L}_{\rm Tr=0}}(t)]+\mathcal{O}\qty(\epsilon^2)\,.
\end{align}
Note that $^\dagger$ denotes here the complex conjugate and transpose of the $3\times 3$ matrix representations in the Floquet subspace of traceless states. Using Eq.~\eqref{eq:lindbladian_matrix_3d} it is easy to determine the possible eigenvalues, as the matrix in the previous equation is diagonal, yielding
\begin{align}
	||e^{\epsilon {\mathcal{L}_{\rm Tr=0}}(t)}||_2&=\sqrt{\max(e^{-2\gamma_{\rm xd}\epsilon}, e^{-2\gamma\epsilon})+\mathcal{O}\qty(\epsilon^2)}\notag\\
    &=e^{-\min(\gamma,\gamma_{\rm xd})\epsilon}+\mathcal{O}(\epsilon^2)\,.
\end{align}
This finally yields an upper bound for the eigenvalues of the traceless solutions $r=1,2,3$
\begin{equation}
	|\mu_r|\leq 	R_S\qty[{\mathcal{V}_{\Tr=0}}(T,0)]\leq e^{-\min(\gamma,\gamma_{\rm xd})T}\,.
\end{equation}
In an analog fashion we can also determine a lower bound by using that the inverse eigenvalue of a matrix can be determined from the eigenvalues of its inverse matrix, such that
\begin{align}
	|\mu_r|^{-1}&=|\mu_r^{-1}|\leq R_S\qty[{\mathcal{V}_{\Tr=0}}(T,0)^{-1}]\notag\\&\leq\lim\limits_{N\rightarrow\infty}\prod_{j=1}^N||e^{-\frac{T}{N}{\mathcal{L}_{\rm Tr=0}}(T-j\frac{T}{N})}||_2\,.
\end{align}
Here we used that the inverse of a matrix product is given by $(AB)^{-1}=B^{-1}A^{-1}$ together with the fact that the inverse of $\exp(\epsilon{\mathcal{L}_{\rm Tr=0}})$ is given by $\exp(-\epsilon{\mathcal{L}_{\rm Tr=0}})$. Analog to before we now have
\begin{align}
	||e^{-\epsilon {\mathcal{L}_{\rm Tr=0}}(t)}||_2&=\sqrt{\max(e^{2\gamma_{\rm xd}\epsilon}, e^{2\gamma\epsilon})+\mathcal{O}\qty(\epsilon^2)}\notag\\
    &= e^{\max(\gamma,\gamma_{\rm xd})\epsilon}+\mathcal{O}(\epsilon^2)\,,
\end{align}
leading to
\begin{equation}
	|\mu_r|^{-1}\leq e^{\max(\gamma,\gamma_{\rm xd})T}\,,\qquad r=1,2,3\,.
\end{equation}
In total we see that the eigenvalues of the traceless solutions $r=1,2,3$ of Eq.~\eqref{eq:eigen_floquet} obey 
\begin{equation}\label{eq:mu_r_bound}
	e^{-\max(\gamma,\gamma_{\rm xd})T}\leq |\mu_r|\leq e^{-\min(\gamma,\gamma_{\rm xd})T}\,.
\end{equation}
With the parametrization in Eq.~\eqref{eq:parametrization} this reads
\begin{equation}
	\min(\gamma,\gamma_{\rm xd})\leq \gamma_r \leq \max(\gamma,\gamma_{\rm xd})\,,\qquad r=1,2,3
\end{equation}
and $\mu_0=1$ implies $\gamma_0=0$. We can interpret $\gamma_r$ as the period-averaged damping rate, which has to lie in between $\gamma$ and $\gamma_{\rm xd}$ for the traceless states.

The traceless solutions are therefore damped, if $\min(\gamma,\gamma_{\rm xd})\geq\gamma_{\rm xd}/2>0$. We can see this using an equation analog to Eq.~\eqref{eq:propagation_rewritten_floquet}
\begin{equation}
	{\mathcal{V}}(0,-\infty)=\lim\limits_{N\rightarrow\infty}{\mathcal{V}}(0,-NT)=\lim\limits_{N\rightarrow\infty}\qty[{\mathcal{V}}(T,0)]^N\,.
\end{equation}
Propagating any initial density matrix written in the Floquet eigenstate basis
\begin{equation}
	\left|\rho\right(-\infty))=\sum_{r=0}^3 m_r\ketL{M_r}
\end{equation}
yields
\begin{align}
	\left|\rho(0)\right)&=\lim\limits_{N\rightarrow\infty}\qty[{\mathcal{V}}(T,0)]^N\sum_{r=0}^3 m_r\ketL{M_r}\notag\\
    &=\lim\limits_{N\rightarrow\infty}\sum_{r=0}^3 m_r\mu_r^N\ketL{M_r}\,.
\end{align}
For $\gamma_{\rm xd}>0$ we have $|\mu_r|<1$ for $r=1,2,3$ and therefore 
\begin{equation}
	\left|\rho(0)\right)=\lim\limits_{N\rightarrow\infty}m_0\mu_0^N\left|M_0\right)=m_0\left|M_0\right)\,,\label{eq:stationary_rho}
\end{equation}
i.e., the stationary state of the system is proportional to the only Floquet eigenstate with non-vanishing trace, since the states with vanishing trace are damped with at least the rate $\min(\gamma,\gamma_{\rm xd})\geq \gamma_{\rm xd}/2$.

\subsubsection{Symmetry properties of the Floquet frequencies $\delta_r$}\label{app:floquet_eigen_symm}
The Lindblad super-operator can be represented as a real matrix in Eq.~\eqref{eq:lindbladian_matrix}, which is then also true for the single-period translation $\mathcal{T}={\mathcal{V}}(T,0)$. This implies that, if a solution to Eq.~\eqref{eq:eigen_floquet} with eigenvalue $\mu_r$ exists, there also has to be a solution with eigenvalue $\mu_r^*$. The solution with non-vanishing trace obviously obeys this since $\mu_0=1=\mu_0^*$. However, this requirement leads to a restriction on the phases $\delta_rT$ of the traceless solutions [see Eq.~\eqref{eq:parametrization}]. Since all traceless solutions have $|\mu_r|<1$ for non-vanishing decay, the requirement that $\mu_r^*$ is an eigenvalue if $\mu_r$ is, has to be fulfilled within the subspace of traceless solutions, implying that for any solution with $\delta_r\neq 0$ we need another solution with $\delta_{r'}=-\delta_r\neq 0$. From Eq.~\eqref{eq:prod_mu_r} we also get an additional requirement
\begin{equation}
	e^{-(2\gamma+\gamma_{\rm xd})T}=\mu_0\prod_{r=1}^3\mu_r =e^{-i\sum_{r=1}^3\delta_r T}e^{-\sum_{r=1}^3\gamma_rT}\,,
\end{equation}
which translates to
\begin{align}
	\sum_{r=1}^3\delta_r&=0\,,\\
	\sum_{r=1}^3\gamma_r&=2\gamma+\gamma_{\rm xd}\,.
\end{align}
If we now have one solution with $\delta_r\neq 0$, there has to be another solution with $\delta_{r'}=-\delta_r\neq 0$, such that the remaining third solution has to obey $\delta_{r''}=0$ in order to fulfill these constraints. With this we have shown that at least one of the traceless solutions obeys $\delta_r=0$ while the other two have opposite frequencies $\delta_r=\pm\delta$. If these have a non-vanishing phase $\delta_r T\neq 0$ they have to have the same damping rate $\gamma_r$, such that $\mu_r$ and $\mu_r^*$ are both possible eigenvalues of Eq.~\eqref{eq:eigen_floquet}. 

\section{Unitary limiting case}\label{app:unitary_limit}
\begin{figure}[t] 
\centering
\includegraphics[width=0.6\linewidth]{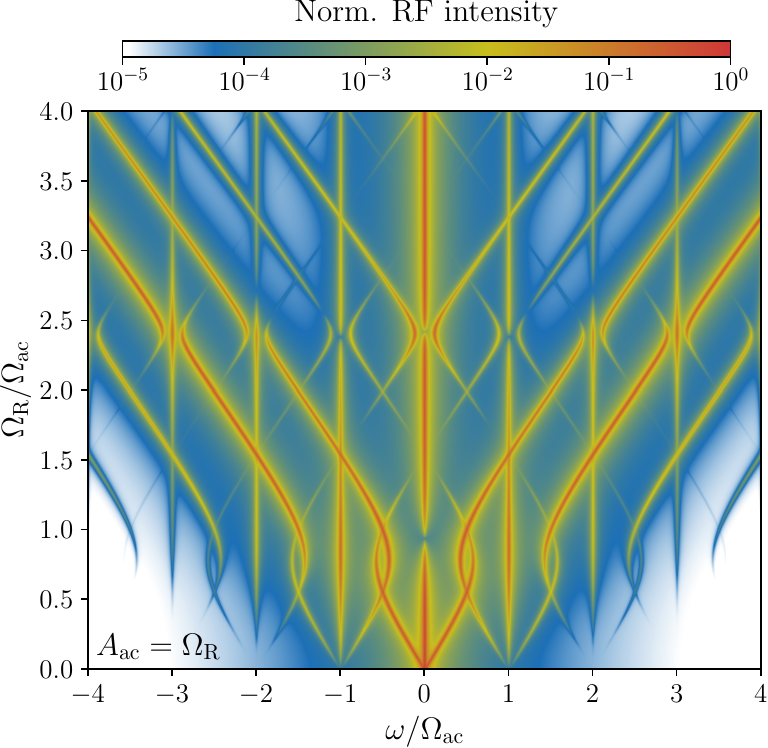}
\caption{Numerical simulation of RF spectrum in the unitary limit using Eq.~\eqref{eq:I_RF_unitary}. Parameters as in Fig.~\ref{fig:2}(c), except for $\gamma_{\rm xd}$ and $\gamma_{\rm pd}$ which are set to zero.}
\label{fig:appB}
\end{figure}

As stated in Eq.~\eqref{eq:stationary_rho}, for any arbitrarily small damping rate $\gamma_{\rm xd}>0$, the normalized, stationary density matrix is given by $\ketL{\rho(0)}=\ketL{M_0}/\expL{\sigma_0|M_0}$, i.e., by the Floquet eigenstate with non-vanishing trace. For $\gamma_{\rm xd}\ll \frac{1}{T}$ we see from Eq.~\eqref{eq:lindbladian_matrix} that the solution to Eq.~\eqref{eq:eigen_floquet} with non-vanishing trace is simply given by $\left|M_0\right)\approx\left|\sigma_0\right)$. If we take $\gamma_{\rm xd}\rightarrow 0$ this becomes exact, since then $\mathcal{L}(t)\ketL{\sigma_0}=0$, which automatically leads to $\mathcal{T}\ketL{\sigma_0}=1\ketL{\sigma_0}=\mu_0\ketL{\sigma_0}$. However, in this strict limit it is not guaranteed that the stationary state $\ketL{\rho(0)}\sim\ketL{\sigma_0}$ is reached, as $\gamma_r$ might be zero for some of the traceless states. To have a well-defined unitary limit, we therefore assume that there is an infinitesimal damping rate $\gamma_{\rm xd}>0$, such that 
\begin{equation}\label{eq:stationary_rho_unitary}
    \ketL{\rho(0)}=\frac{1}{2}\ketL{\sigma_0}+\mathcal{O}\qty(\frac{\gamma_{\rm xd}}{\Omega_{\rm ac}})\,,
\end{equation}
taking the limit $\gamma_{\rm xd,pd}\rightarrow 0$ at the end of the calculations. Looking at Eq.~\eqref{eq:I_RF}, the unitary limit describes the RF spectrum very well, if we have a sufficiently bad resolution of the spectrometer $\Gamma\gg\gamma_{\rm xd,pd}$, as well as weak dissipation over a single period $\frac{1}{T}\gg\gamma_{\rm xd,pd}$. The correlation function in Eq.~\eqref{eq:def_C} can then be approximated as
\begin{align}
    &C(t+\tau,t)\notag\\
    &=\Tr\qty[\sigma_+U(t+\tau,t)\sigma_- \rho(t)U^{\dagger}(t+\tau,t)]+\mathcal{O}(\gamma_{\rm xd,pd})\notag\\
    &=\frac{1}{2}\Tr\qty[\sigma_+U(t+\tau,t)\sigma_-U^{\dagger}(t+\tau,t)]+\mathcal{O}(\gamma_{\rm xd,pd})
\end{align}
where we used Eq.~\eqref{eq:stationary_rho_unitary} for $\rho(0)$, as well as the unitary time-evolution operator defined in Eq.~\eqref{eq:U_eom}. Using the definition of the time-periodic states $\ket{u_\alpha(t)}$ in Eq.~\eqref{eq:def_u_alpha} and the semigroup property in Eq.~\eqref{eq:semigroup_unitary}, one can easily show that
\begin{align}
    U(t+\tau,t)\ket{u_\alpha(t)}=e^{-i\epsilon_\alpha\tau}\ket{u_\alpha(t+\tau)}
\end{align}
analog to the non-unitary case in Eq.~\eqref{eq:action_time_evol_V_r}. The $\ket{u_\alpha(t)}$ form a complete orthonormal basis of the Hilbert space, as they are related to the unitary Floquet eigenstates $\ket{\psi_\alpha}$ from Eq.~\eqref{eq:eigen_floquet_unitary} via unitary transformation. We can therefore represent the unitary time-evolution operator via~\cite{breuer2002theory}
\begin{equation}
    U(t+\tau,t)=\sum_{\alpha} e^{-i\epsilon_\alpha\tau}\ket{u_{\alpha}(t+\tau}\bra{u_\alpha(t)}\,.
\end{equation}
The correlation function can finally be written as
\begin{align}
    &C(t+\tau,t)\notag\\
    &=\frac{1}{2}\sum_{\alpha\beta}e^{-i(\epsilon_\alpha-\epsilon_\beta)\tau}P_{\alpha\beta}(t)P_{\alpha\beta}(t+\tau)^*+\mathcal{O}(\gamma_{\rm xd,pd})
\end{align}
with the matrix elements $P_{\alpha\beta}(t)$ from Eq.~\eqref{eq:def_P_alpha_beta}. Inserted into Eq.~\eqref{eq:I_RF} and taking the limit $\gamma_{\rm xd,pd}\rightarrow 0$ this finally yields the unitary RF spectrum in Eq.~\eqref{eq:I_RF_unitary}. An example of such a spectrum, using the same parameters as in Fig.~\ref{fig:2}(c), except now we have $\gamma_{\rm xd}=\gamma_{\rm pd}=0$, is shown in Fig.~\ref{fig:appB}.

\section{Phonon-induced Bloch-Siegert shift}\label{app:bloch_siegert}
While App.~\ref{app:floquet_eigen_diss} and \ref{app:unitary_limit} considered a general periodic TLS Hamiltonian in Eq.~\eqref{eq:Hamilton_general}, we turn back to the periodically modulated optically driven TLS from Eq.~\eqref{eq:Hamilton}, i.e.,
\begin{equation}\label{eq:Hamilton_Pauli}
    H(t)=\hbar\frac{\Omega_{\rm R}}{2}\sigma_1+\hbar\frac{\Delta(t)}{2}\sigma_3\,,\qquad \Delta(t)=A_{\rm ac}\sin(\Omega_{\rm ac}t)
\end{equation}
with the Pauli matrices $\sigma_i$ given in Eq.~\eqref{eq:Pauli} and $\Delta(t)$ as in Eq.~\eqref{eq:detuning_sinus}. As discussed in the context of Fig.~\ref{fig:2}, the sinusoidal acoustic modulation leads to sideband replicas in the spectra. Considering that under weak optical driving, i.e., for $\Omega_{\rm R}\to 0$, the RF spectrum can be determined completely analytically~\cite{weiss2021optomechanical,wigger2021resonance}, we will perform a unitary transformation, aiming to partially remove the effect of the acoustic modulation from the Hamiltonian~\cite{kibis2009matter,lu2012effects,gladysz2025light}. The transformation brings the time evolution to the interaction picture with respect to $H_0(t)=\hbar\frac{\Delta(t)}{2}\sigma_3$ and is described by 
\begin{equation}
    U(t) = \exp\qty[-\frac{i}{\hbar}\int\limits_{-T/4}^t\dd\tau\,H_0(\tau)]
\end{equation}
(note that $H_0(t)$ commutes with itself at different times). This leads to the following interaction Hamiltonian in the interaction picture
\begin{align}\label{eq:trafo_BS}
    H_I^I(t)&=U^{\dagger}(t) [H(t)-H_0(t)]U(t)\notag\\
    &=\hbar\frac{\Omega_{\rm R}}{2} e^{-i\frac{A_{\rm ac}}{2\Omega_{\rm ac}}\cos(\Omega_{\rm ac}t)\sigma_3}\sigma_1 e^{i\frac{A_{\rm ac}}{2\Omega_{\rm ac}}\cos(\Omega_{\rm ac}t)\sigma_3}\,.
\end{align}
For convenience we chose $t=-T/4$ as the time when the original Schrödinger picture description and the interaction picture description coincide. Making use of the commutator relations $\comm{\sigma_3}{\sigma_1}=2i\sigma_2$ and $\comm{\sigma_3}{\sigma_2}=-2i\sigma_1$, we obtain~\cite{allen1987optical,breuer2002theory}
\begin{align}
    H_I^I(t)&=\hbar\frac{\Omega_{\rm R}}{2} \sum_{n=0}^{\infty} \frac{1}{n!}\qty[-i\frac{A_{\rm ac}}{2\Omega_{\rm ac}}\cos(\Omega_{\rm ac}t)]^n\comm{\sigma_3}{\sigma_1}^{(n)}\notag\\
    &=\hbar\frac{\Omega_{\rm R}}{2}\cos\qty[\frac{A_{\rm ac}}{\Omega_{\rm ac}}\cos(\Omega_{\rm ac}t)]\sigma_1\notag\\
    &+\hbar\frac{\Omega_{\rm R}}{2}\sin\qty[\frac{A_{\rm ac}}{\Omega_{\rm ac}}\cos(\Omega_{\rm ac}t)]\sigma_2\label{eq:BS_H_I_total}
\end{align}
where $\comm{A}{B}^{(n)}=\comm{A}{\comm{A}{B}^{(n-1)}}$ and $\comm{A}{B}^{(0)}=B$. The $\cos$-function contains all the even harmonics of $\Omega_{\rm ac}$, i.e., terms $\sim \exp(i2p\Omega_{\rm ac}t)$, with $p=1,2,..$, while the $\sin$-function contains the odd harmonics $\sim \exp[i(2p-1)\Omega_{\rm ac}t]$. As discussed in Sec.~\ref{sec:floquet_bands}, the odd harmonics lead to transitions between the optically dressed states, as can be seen here from $\sigma_2=i(\ket{+}\bra{-}-\ket{-}\bra{+})$ with the dressed states defined in Eq.~\eqref{eq:def_dressed_states}, while the even harmonics do not induce such transitions since $\sigma_1=\ket{+}\bra{+}-\ket{-}\bra{-}$. In addition to these time-dependent terms, the Hamiltonian in Eq.~\eqref{eq:BS_H_I_total} contains also a time-independent contribution (0th-order harmonic of the $\cos$ term)
\begin{align}\label{eq:BS_H_I_const}
    \frac{1}{T}\int\limits_0^T\dd t\, H_I^I(t)=\hbar\frac{\Omega_{\rm R}}{2}J_0\qty(\frac{A_{\rm ac}}{\Omega_{\rm ac}})(\ket{+}\bra{+}-\ket{-}\bra{-})
\end{align}
with $J_0(z)$ denoting the $0$-th Bessel function of the first kind~\cite{abramowitz1964handbook}. This time-independent part of the Hamiltonian describes how the transition frequency between the dressed states, i.e., the Rabi frequency, is renormalized. For weak acoustic driving this yields the renormalization
\begin{align}\label{eq:bloch-siegert}
    \Omega_{\rm R}\rightarrow \widetilde{\Omega}_{\rm R}&=\Omega_{\rm R}J_0\qty(\frac{A_{\rm ac}}{\Omega_{\rm ac}})\notag\\&=\Omega_{\rm R}\qty[1-\qty(\frac{A_{\rm ac}}{2\Omega_{\rm ac}})^2]+\mathcal{O}\qty[\qty(\frac{A_{\rm ac}}{\Omega_{\rm ac}})^4]\,,
\end{align}
which can be identified with a phonon-induced Bloch-Siegert shift~\cite{bloch1940magnetic,shirley1965solution,lu2012effects}. This changes all the resonances as a function of the acoustic modulation, as discussed in the context of Fig.~\ref{fig:2}, leading to the downward shift of the horizontal cuts in Fig.~\ref{fig:2}(c) compared to Fig.~\ref{fig:2}(a). 

\section{Non-perturbative calculations for transition dipole matrix elements $P_{\alpha\beta}$}\label{app:unitary_nonpert}
In the following we will show some properties of the transition dipole matrix elements $P_{\alpha\beta}$ from Eq.~\eqref{eq:def_P_alpha_beta} which explain, at least mathematically, the presence of certain line suppressions in the spectra presented in Fig.~\ref{fig:2}. The following relations between Pauli matrices and the Hamiltonian in Eq.~\eqref{eq:Hamilton_Pauli}
\begin{equation}\label{eq:sigma_1_H}
    \sigma_1=\frac{2}{\hbar}\pdv{\Omega_{\rm R}}H(t)\,
\end{equation}
and 
\begin{equation}\label{eq:sigma_2_H}
    \sigma_2=-\frac{i}{\hbar\Omega_{\rm R}}\comm{\sigma_3}{H(t)}\,
\end{equation}
are very helpful for this. In the last equation we made use of the SU(2) algebra of the Pauli matrices~\cite{allen1987optical}.  To make further progress, in the following we state some general symmetry properties of the system described by Eq.~\eqref{eq:Hamilton_Pauli}.
\subsection{Half cycle translation}
With the detuning from Eq.~\eqref{eq:detuning_sinus} we have
\begin{equation}\label{eq:transform_H_sigma_1}
    H(t\pm T/2)=\hbar\frac{\Omega_{\rm R}}{2}\sigma_1-\hbar\frac{\Delta(t)}{2}\sigma_3=\sigma_1 H(t)\sigma_1\,.
\end{equation}
Such a shift by a half period therefore corresponds to a unitary basis transformation with the operator $\sigma_1=\sigma_1^{-1}$. Regarding the time-evolution operator defined in Eq.~\eqref{eq:U_eom} this implies
\begin{equation}
    \sigma_1 U(t,t_0)\sigma_1=U(t\pm T/2,t_0\pm T/2)
\end{equation}
as both sides fulfill the same equation of motion and have the same initial condition at $t=t_0$. 

Using the transformation of the time evolution operator via $\sigma_1$, we can show using $\sigma_1^2=1$ and the semigroup property in Eq.~\eqref{eq:semigroup_unitary}, that
	\begin{align}\label{eq:sigma_1U_T_2}
		U(T,0)&=\sigma_1^2 U(T,T/2)\sigma_1^2U(T/2,0)\notag\\
        &=\sigma_1 U(T/2,0)\sigma_1U(T/2,0)\notag\\
        &=\qty[\sigma_1 U(T/2,0)]^2\,.
	\end{align}
	The information on the Floquet states is already contained in a half-period~\cite{chu1985recent}. It follows that the eigenstates $\ket{\psi_\alpha}$ to $U(T,0)$ defined in Eq.~\eqref{eq:eigen_floquet_unitary} are also eigenstates to $\sigma_1 U(T/2,0)=U(0,-T/2)\sigma_1$. From 
	\begin{align}
		\det[\sigma_1 U(T/2,0)]&=\det(\sigma_1)\det[U(T/2,0)]\notag\\
        &=-\det[U(T/2,0)]\notag\\
        &=-\exp{-i\int\limits_{0}^{T/2}\dd\tau \Tr[H(\tau)]}=-1
	\end{align}
	we can see that the product of the two eigenvalues has to be $-1$, i.e., $\sigma_1 U(T/2,0)$ is a generalized parity operator~\cite{ben1993effect,yan2019role}. Let $\lambda_\alpha$ be the eigenvalue of $\ket{\psi_\alpha}$ with respect to $\sigma_1 U(T/2,0)$. Then we have [using Eqs.~\eqref{eq:eigen_floquet_unitary} and \eqref{eq:sigma_1U_T_2}]
	\begin{equation}\label{eq:lambda_alpha}
		\lambda_\alpha^2=e^{-i\epsilon_\alpha T}\,,\qquad \lambda_1\lambda_2=-1\,,\qquad \alpha=1,2\,.
	\end{equation}
    Note that analog to the symmetry discussed in App.~\ref{app:floquet_eigen_symm} the Floquet frequencies in the unitary limit obey $\epsilon_1=-\epsilon_2$ due to
    \begin{equation}
        \prod_{\alpha}e^{-i\epsilon_\alpha T}=\det\qty[U(T,0)]=\exp{-i\int\limits_{0}^{T}\dd\tau \Tr[H(\tau)]}=1\,.
    \end{equation}
    In the case of degeneracy $\epsilon_1=\epsilon_2=-\epsilon_1=0$ we cannot distinguish between the two Floquet eigenstates using the operator $U(T,0)$, however we can still classify the states using the operator $\sigma_1 U(T/2,0)$, since in the case of such degeneracy we have $\lambda_\alpha=\pm 1$ from Eq.~\eqref{eq:lambda_alpha}. Without any constraint on generality we can write
\begin{equation}
		\lambda_\alpha=(-1)^\alpha e^{-i\epsilon_\alpha T/2}\,,\qquad\alpha=1,2\,.
	\end{equation}
 Finally, the action of $\sigma_1$ on the time-periodic Floquet states defined in Eq.~\eqref{eq:def_u_alpha} is given by
	\begin{align}
		\sigma_1\ket{u_\alpha(t)}&=e^{i\epsilon_\alpha t}\sigma_1 U(t,0)\ket{\psi_\alpha}\notag\\
        &=e^{i\epsilon_\alpha t}U(t-T/2,-T/2)\sigma_1\ket{\psi_\alpha}\notag\\
		&=e^{i\epsilon_\alpha t}U(t-T/2,0)U(0,-T/2)\sigma_1\ket{\psi_\alpha}\notag\\&=\lambda_\alpha e^{i\epsilon_\alpha T/2}\ket{u_\alpha(t-T/2)}\notag\\&=(-1)^\alpha\ket{u_\alpha(t-T/2)}\,.\label{eq:symmetry_u_1}
	\end{align}

\subsection{Time reversal}
In an analog fashion to Eq.~\eqref{eq:transform_H_sigma_1} we have 
\begin{equation}\label{eq:transform_H_sigma_2}
    \sigma_2H(t)\sigma_2=-H(t)\,,
\end{equation}
i.e., a unitary transformation with the Pauli matrix $\sigma_2$ corresponds to changing the sign of all energies. Since the two unitary floquet frequencies $\epsilon_\alpha$ have opposite sign and equal amplitude, this leads to $\sigma_2\ket{\psi_1}\sim\ket{\psi_2}$, as shown in the following.

Regarding the time-evolution operator defined in Eq.~\eqref{eq:U_eom} we have the transformation
\begin{equation}
    \sigma_2 U(t,t_0)\sigma_2=U(-t,-t_0)
\end{equation}
as both sides fulfill the same equation of motion and have the same initial condition for $t=t_0$. 

We can use this property together with the eigenvalue equation~\eqref{eq:eigen_floquet_unitary} of the Floquet states and the periodicity of the time-evolution operator [analog to  Eq.~\eqref{eq:V_symm}] to write
\begin{align}
    e^{-i\epsilon_\alpha T}\bra{\psi_\alpha}\sigma_2\ket{\psi_\alpha}&=\bra{\psi_\alpha}\sigma_2U(T,0)\ket{\psi_\alpha}\notag\\
    &=\bra{\psi_\alpha}U(-T,0)\sigma_2\ket{\psi_\alpha}\notag\\
    &=\bra{\psi_\alpha}\sigma_2U(0,-T)\ket{\psi_\alpha}^*\notag\\
    &=\bra{\psi_\alpha}\sigma_2U(T,0)\ket{\psi_\alpha}^*\notag\\
    &=e^{i\epsilon_\alpha T}\bra{\psi_\alpha}\sigma_2\ket{\psi_\alpha}\,.
\end{align}
For this relation to be fulfilled we either need $\bra{\psi_\alpha}\sigma_2\ket{\psi_\alpha}=0$, i.e.,
\begin{equation}\label{eq:action_sigma_2_psi}
    \sigma_2\ket{\psi_\alpha}=\kappa_\alpha\ket{\psi_\beta}\,,\qquad \beta\neq \alpha
\end{equation}
for a certain complex number $\kappa_\alpha$,
or we need $\epsilon_\alpha T=n_\alpha\pi$ with $n_\alpha\in\mathbb{Z}$. The latter case corresponds to degeneracy [see Eq.~\eqref{eq:eigen_floquet_unitary}], since the two Floquet frequencies $\epsilon_\alpha$ have to have the same amplitude but opposite sign. We can then always construct a linear combination to fulfill Eq.~\eqref{eq:action_sigma_2_psi}. The Pauli matrix $\sigma_2$ thus connects the two Floquet eigenstates $\ket{\psi_\alpha}$. For the periodic Floquet states we find
\begin{align}
    \sigma_2\ket{u_\alpha (t)}&=e^{i\epsilon_\alpha t}\sigma_2 U(t,0)\ket{\psi_\alpha}=e^{-i\epsilon_\beta t} U(-t,0)\kappa_\alpha \ket{\psi_\beta}\notag\\
    &=\kappa_\alpha \ket{u_\beta (-t)}\,,\qquad \beta\neq\alpha\,.\label{eq:symmetry_u_2}
\end{align}

\subsection{Symmetries of the Pauli matrix elements}
Defining, analog to Eq.~\eqref{eq:def_P_alpha_beta}, the matrix elements
\begin{equation}
    P_{\alpha\beta (j)}(t)=\frac{1}{2}\bra{u_\alpha(t)}\sigma_j\ket{u_\beta(t)}
\end{equation}
for $j=1,2$, we find using Eq.~\eqref{eq:symmetry_u_1}
\begin{align}
	P_{\alpha\beta(j)}(t)&=\frac{1}{2}\bra{u_\alpha(t)}\sigma_1^2\sigma_j\ket{u_{\beta}(t)}\notag\\
    &=(-1)^{j-1}\frac{1}{2}\bra{u_\alpha(t)}\sigma_1\sigma_j\sigma_1\ket{u_{\beta}(t)}\notag\\
    &=(-1)^{j-1+\alpha+\beta}P_{\alpha\beta(j)}(t-T/2)\,.\label{eq:p_trafo_sigma_1}
\end{align}
In the following we will use this property to show why certain lines get suppressed in the spectrum.

\subsection{Suppression of the center line for anti-crossing}\label{app:anticrossing_suppression}
According to the RF spectrum in the unitary limit in Eq.~\eqref{eq:I_RF_unitary} the center line at $\omega=0$ has an intensity proportional to 
\begin{equation}
    \sum_{\alpha=1,2} \qty|P_{\alpha\alpha}^{(0)}|^2\,.
\end{equation}
We can break down these transition dipole matrix elements into
\begin{equation}\label{eq:P_separate_sigmas}
    P_{\alpha\alpha}^{(0)}=\frac{1}{2T}\int\limits_0^T\dd t\, \bra{u_\alpha(t)}(\sigma_1+i\sigma_2)\ket{u_\alpha(t)}\,.
\end{equation}
The matrix element with $\sigma_2$ has to vanish since according to Eq.~\eqref{eq:p_trafo_sigma_1} $P_{\alpha\alpha (2)}^{(0)}=-P_{\alpha\alpha (2)}^{(0)}=0$ for the average over the matrix element with respect to $\sigma_2$. This implies
\begin{equation}
    P_{\alpha\alpha}^{(0)}=\frac{1}{2T}\int\limits_0^T\dd t\, \bra{u_\alpha(t)}\sigma_1\ket{u_\alpha(t)}\,.
\end{equation}
Using now Eq.~\eqref{eq:sigma_1_H} together with the eigenvalue equation~\eqref{eq:floquet_states_eom} of the periodic Floquet states, we obtain
\begin{align}
    P_{\alpha\alpha}^{(0)}&=\frac{1}{2T}\int\limits_0^T\dd t\,\bra{u_\alpha(t)}\sigma_1\ket{u_\alpha(t)}\notag\\
    &=\frac{1}{\hbar T}\int\limits_0^T\dd t\,\,\bra{u_\alpha(t)}\qty[\pdv{H(t)}{\Omega_{\rm R}}]\ket{u_\alpha(t)}=\pdv{\epsilon_\alpha}{\Omega_{\rm R}}
\end{align}
courtesy of the Hellmann-Feynman theorem~\cite{chu1985recent}. To show this relation more explicitly, one can rewrite the left-hand side to
\begin{align}
    P_{\alpha\alpha}^{(0)}&=\pdv{\Omega_{\rm R}}\frac{1}{\hbar T}\int\limits_0^T\dd t\,\,\bra{u_\alpha(t)}H(t)\ket{u_\alpha(t)}\notag\\
    &-\frac{2}{\hbar T}\int\limits_0^T\dd t\,\,\Re\qty[\bra{u_\alpha(t)}H(t)\pdv{\Omega_{\rm R}}\ket{u_\alpha(t)}]
\end{align}
and use Eq.~\eqref{eq:floquet_states_eom} as well as partial integration in $t$ to obtain the right-hand side. This implies that the center line gets dim whenever the Floquet frequencies, as a function of the Rabi frequency $\Omega_{\rm R}$ at constant acoustic modulation amplitude $A_{\rm ac}$, have an extremum. When considering Fig.~\ref{fig:4}, it becomes clear that this is the case whenever there is a level repulsion at the edge of the first BZ, leading to an anti-crossing in the spectra in Fig.~\ref{fig:2}. Note that in Figs.~\ref{fig:2} and \ref{fig:4}, both the acoustic modulation amplitude $A_{\rm ac}$ as well as the Rabi frequency $\Omega_{\rm R}$ are varied at the same time, such that the position of the line suppressions is slightly different from the position of the level repulsions.

\subsection{Dim lines crossing the center line}\label{app:crossing_suppression}
Next we will show that the lines crossing the center line in Fig.~\ref{fig:2} have to vanish at crossing. According to Eq.~\eqref{eq:I_RF_unitary}, their intensity is determined by $P_{12}^{(0)}$ and $P_{21}^{(0)}$. Analog to Eq.~\eqref{eq:P_separate_sigmas}, we can break these amplitudes down into a contribution with the Pauli matrix $\sigma_1$ and one with the Pauli matrix $\sigma_2$. Making use of Eq.~\eqref{eq:p_trafo_sigma_1} we see that only the contribution due to $\sigma_2$ is non-vanishing. Together with Eq.~\eqref{eq:sigma_2_H} this leads to
\begin{align}
    P_{12}^{(0)}&=\frac{i}{2T}\int\limits_0^T\dd t\, \bra{u_1(t)}\sigma_2\ket{u_2(t)}=-P_{21}^{(0)*}\notag\\
    &=\frac{1}{2T\hbar\Omega_{\rm R}}\int\limits_0^T\dd t\, \bra{u_1(t)}\comm{\sigma_3}{H(t)}\ket{u_2(t)}
\end{align}
Using $i\sigma_3=\sigma_1\sigma_2=-\sigma_2\sigma_1$ and Eq.~\eqref{eq:symmetry_u_2}, we can rewrite this expression into
\begin{align}
    P_{12}^{(0)}&=\frac{i}{2T\hbar\Omega_{\rm R}}\int\limits_0^T\dd t\, \bra{u_1(t)}\sigma_2\sigma_1 H(t)+H(t)\sigma_1\sigma_2\ket{u_2(t)}\notag\\
    &=\frac{i\kappa_2}{2T\hbar\Omega_{\rm R}}\int\limits_0^T\dd t\, [\bra{u_2(-t)}\sigma_1 H(t)\ket{u_2(t)}\notag\\
    &\qquad\qquad\qquad+\bra{u_1(t)}H(t)\sigma_1\ket{u_1(-t)}]\,,
\end{align}
making use of the fact that $\kappa_2=\bra{\psi_1}\sigma_2\ket{\psi_2}=\kappa_1^*$.  Next we can make a substitution $t\rightarrow -t$ in the first term and using $H(-t)=\sigma_1 H(t)\sigma_1$ for $H(t)$ from Eq.~\eqref{eq:Hamilton_Pauli}, as well as the periodicity of all terms we obtain
\begin{align}
    P_{12}^{(0)}&=\frac{i\kappa_2}{2T\hbar\Omega_{\rm R}}\int\limits_0^T\dd t\, \sum_{\alpha=1,2}\bra{u_\alpha(t)}H(t)\sigma_1\ket{u_\alpha(-t)}\,.
\end{align}
Using Eq.~\eqref{eq:floquet_states_eom} and its adjoint, we can write 
\begin{equation}
	\dv{t} \bra{u_\alpha(t)}\sigma_1\ket{u_\alpha(-t)}=2i\bra{u_\alpha(t)}\qty[\frac{1}{\hbar}H(t)-\epsilon_\alpha]\sigma_1\ket{u_\alpha(-t)}\,.
\end{equation}
With this we obtain
\begin{align}
	 P_{12}^{(0)}&=\frac{\kappa_2}{4T\Omega_{\rm R}}\int\limits_0^T\dd t\,\sum_{\alpha=1,2} \qty[\dv{t}+2i\epsilon_\alpha]\bra{u_\alpha(t)}\sigma_1\ket{u_\alpha(-t)}\,.
\end{align}
The term with $\dv{t}$ vanishes due to the periodicity of the Floquet states. Using $\epsilon_2=-\epsilon_1$ [see discussion after Eq.~\eqref{eq:eigen_floquet_unitary}], this leads to
\begin{align}
    |P_{12}^{(0)}|\sim|\epsilon_\alpha|\,.
\end{align}
This means that all lines that cross the center line, i.e., with frequencies given by $\omega=\pm (\epsilon_2-\epsilon_1)$ according to Eq.~\eqref{eq:I_RF_unitary}, are completely suppressed at the crossing, since then $0=\omega=(\epsilon_2-\epsilon_1)$ implies $\epsilon_1=\epsilon_2=-\epsilon_1=0$, leading to a vanishing of the intensity.

\section{Unitary quasi-degenerate perturbation theory: Floquet eigenvalues at harmonic resonances}\label{app:pt}
In the following we will discuss the perturbative solution of Eqs.~\eqref{eq:eigen_floquet_hamiltonian} and \eqref{eq:Hamiltonian_Floquet}. Before considering the specific structure of the Hamiltonian, we will discuss Brillouin-Wigner perturbation theory in general~\cite{taylor2002quantum,englert2024lectures}. We consider an eigenvalue equation of the form in Eq.~\eqref{eq:eigen_floquet_hamiltonian}
\begin{equation}\label{eq:pt_eveq}
    (\mathbb{H}_0+\mathbb{V})\left.\ket{\Psi}\right>=\hbar\epsilon\left.\ket{\Psi}\right>
\end{equation}
with the unperturbed solutions
\begin{equation}\label{eq:pt_eveq_unpert}
    \mathbb{H}_0\left.\ket{\Phi_\alpha}\right>=\hbar\epsilon^{(0)}_{\alpha}\left.\ket{\Phi_\alpha}\right>\,.
\end{equation}
Next we introduce the notion of quasi-degenerate subspaces. Assume that we are interested in the effect of the perturbation $\mathbb{V}$ on a certain unperturbed eigenstate $\left.\ket{\Phi_{\alpha_0}}\right>$ with quantum number $\alpha_0$. We define the quasi-degenerate subspace of this state, called q.d.$(\alpha_0)$ as 
\begin{equation}
    \alpha\in {\rm q.d.}(\alpha_0)\Leftrightarrow |\epsilon^{(0)}_\alpha-\epsilon^{(0)}_{\alpha_0}|\leq \delta\,,
\end{equation}
where $\delta>0$ is the desired accuracy, i.e., it defines what is called \textit{quasi}-degenerate. At $\delta\rightarrow 0$ we would include only truly degenerate states. This separation is especially helpful, if  all states not in this quasi-degenerate subspace are far away in energy, i.e.,
\begin{equation}
    \alpha\neq {\rm q.d.}(\alpha_0)\Rightarrow |\epsilon_{\alpha}^{(0)}-\epsilon_{\alpha_0}^{(0)}|\gg \delta\,.\label{eq:not_quasi_degenerate}
\end{equation}
In such a situation, Brillouin-Wigner perturbation theory, as applied in the following, is helpful in approximating the correct eigenvalues $\hbar\epsilon$ in Eq.~\eqref{eq:pt_eveq}.

We define the projection onto the considered quasi-degenerate state subspace as 
\begin{equation}
    \mathbb{P}_{\alpha_0}=\sum_{\alpha\in {\rm q.d.}(\alpha_0)} \left.\ket{\Phi_\alpha}\right>\left<\bra{\Phi_\alpha}\right.\,,
\end{equation}
such that
\begin{equation}
    \comm{\mathbb{P}_{\alpha_0}}{\mathbb{H}_0}=0\,.
\end{equation}
Any solution $\left.\ket{\Psi}\right>$ of Eq.~\eqref{eq:pt_eveq} can generally be written as
\begin{equation}\label{eq:pt_psi}
    \left.\ket{\Psi}\right>=\mathbb{P}_{\alpha_0}\left.\ket{\Psi}\right>+(1-\mathbb{P}_{\alpha_0})\left.\ket{\Psi}\right>\,\,.
\end{equation}
In the following we denote the part of the full eigenstate $\left.\ket{\Psi}\right>$ that lies in the the considered quasi-degenerate subspace q.d.$(\alpha_0)$ as
\begin{equation}
    \left.\ket{\Phi}\right>=\mathbb{P}_{\alpha_0}\left.\ket{\Psi}\right>\,.
\end{equation}
We will be interested in how the perturbation changes the unperturbed states from q.d.$(\alpha_0)$. Applying perturbation theory works especially well if we can assume that 'most' of $\left.\ket{\Psi}\right>$ still lies in this subspace. By this we mean that we are interested in situations, where the norm of $\left.\ket{\Phi}\right>$, i.e., the part lying in q.d.$(\alpha_0)$, is much larger than the norm of the remaining part $\left.\ket{\Psi}\right>-\left.\ket{\Phi}\right>=(1-\mathbb{P}_{\alpha_0})\left.\ket{\Psi}\right>$ lying outside of the considered quasi-degenerate subspace.  This remaining part of $\left.\ket{\Psi}\right>$ can be rewritten using Eq.~\eqref{eq:pt_eveq}
\begin{align}
    (1-\mathbb{P}_{\alpha_0})(\mathbb{H}_0+\mathbb{V})\left.\ket{\Psi}\right>&=\hbar\epsilon (1-\mathbb{P}_{\alpha_0}) \left.\ket{\Psi}\right>\notag\\
    \Rightarrow (1-\mathbb{P}_{\alpha_0})\mathbb{V}\left.\ket{\Psi}\right>&=(\hbar\epsilon-\mathbb{H}_0)(1-\mathbb{P}_{\alpha_0}) \left.\ket{\Psi}\right>\notag\\
    \Rightarrow \mathbb{G}_{\alpha_0}\mathbb{V}\left.\ket{\Psi}\right>&=(1-\mathbb{P}_{\alpha_0}) \left.\ket{\Psi}\right>\,,\label{eq:psi_remaining}
\end{align}
where
\begin{equation}\label{eq:pt_G_0}
    \mathbb{G}_{\alpha_0}=(\hbar\epsilon-\mathbb{H}_0)^{-1}(1-\mathbb{P}_{\alpha_0})
\end{equation}
defines the Green's function~\cite{taylor2002quantum,grosso2013solid}. The remaining part is thus 'small' if the perturbation $\mathbb{V}$ is weak, consistent with the perturbative approach, and if the eigenvalues of $\mathbb{G}_{\alpha_0}$
\begin{equation}
    \mathbb{G}_{\alpha_0}\left.\ket{\Phi_{\alpha}}\right>=\left\lbrace\begin{matrix}0&\alpha\in {\rm q.d.}(\alpha_0)\\
    \frac{1}{\hbar\epsilon-\hbar\epsilon_{\alpha}^{(0)}}\left.\ket{\Phi_{\alpha}}\right> & \alpha\notin {\rm q.d.}(\alpha_0)\end{matrix}\right.\label{eq:eigenvalue_green}
\end{equation}
are small. Since we want to investigate especially those solutions $\left.\ket{\Psi}\right>$ that lie in q.d.$(\alpha_0)$ at vanishing perturbation, i.e., $|\epsilon-\epsilon_{\alpha_0}^{(0)}|\leq \delta$ for $\mathbb{V}\rightarrow 0$, we can assume $(\epsilon-\epsilon_{\alpha}^{(0)})^{-1}\ll\delta$ in Eq.~\eqref{eq:eigenvalue_green} for sufficiently weak perturbation if Eq.~\eqref{eq:not_quasi_degenerate} holds, i.e., if the states not in the degenerate subspace $\alpha\notin{\rm q.d.}(\alpha_0)$ are far away in energy. In such a situation 'most' of $\left.\ket{\Psi}\right>$ will then automatically lie in q.d.$(\alpha_0)$.

Note that by definition we have $\mathbb{P}_{\alpha_0}\mathbb{G}_{\alpha_0}=\mathbb{G}_{\alpha_0}\mathbb{P}_{\alpha_0}=0$. Now we can finally write Eq.~\eqref{eq:pt_psi} using Eq.~\eqref{eq:psi_remaining} as
\begin{equation}\label{eq:Psi-Phi-G0V}\left.\ket{\Psi}\right>=\left.\ket{\Phi}\right>+\mathbb{G}_{\alpha_0}\mathbb{V}\left.\ket{\Psi}\right>\,,
\end{equation}
which is formally solved via
\begin{equation}\label{eq:Psi-formal}
    \left.\ket{\Psi}\right>=\sum_{n=0}^{\infty}(\mathbb{G}_{\alpha_0}\mathbb{V})^n\left.\ket{\Phi}\right>\,.
\end{equation}
Projecting $\left<\bra{\Phi}\right.$ onto Eq.~\eqref{eq:pt_eveq} leads to
\begin{equation}
    \hbar\epsilon\expval{\expval{\Phi|\Psi}}=\expval{\bra{\Phi}\mathbb{H}_0+\mathbb{V}\ket{\Psi}}\,.
\end{equation}
We can of course normalize $\left.\ket{\Phi}\right>$, ignoring proper normalization of $\left.\ket{\Psi}\right>$ for now, as is usually done in perturbation theory~\cite{taylor2002quantum,englert2024lectures}. Then by definition in Eq.~\eqref{eq:pt_psi}, we have $\expval{\expval{\Phi|\Psi}}=\expval{\expval{\Phi|\Phi}}=1$, i.e., using Eq.~\eqref{eq:Psi-formal} we obtain
\begin{equation}\label{eq:pt_E-E_0}
    \hbar\epsilon=\left<\bra{\Phi}\right.\qty(\mathbb{H}_0+\mathbb{V})\sum_{n=0}^{\infty}(\mathbb{G}_{\alpha_0}\mathbb{V})^n\left.\ket{\Phi}\right>\,.
\end{equation}
Next, take any state $\left.\ket{\Phi'}\right>$ from the quasi-degenerate subspace q.d.$(\alpha_0)$, but orthogonal to $\left.\ket{\Phi}\right>$. Since $\left<\bra{\Phi'}\right. = \left<\bra{\Phi'}\right.\mathbb{P}_{\alpha_0}$, we have 
$\left<\bra{\Phi'}\right.\mathbb{G}_{\alpha_0}=0$ due to $\mathbb{P}_{\alpha_0}\mathbb{G}_{\alpha_0}=0$. From Eq.~\eqref{eq:Psi-Phi-G0V} it then follows that $\expval{\expval{\Phi'|\Psi}}=\expval{\expval{\Phi'|\Phi}}=0$. Therefore, projecting Eq.~\eqref{eq:pt_eveq} on the state $\left<\bra{\Phi'}\right.$ we find 
\begin{align}
0&=\left<\bra{\Phi'}\right.\mathbb{H}_0+\mathbb{V}\left.\ket{\Psi}\right>\notag\\
&=\left<\bra{\Phi'}\right.\qty(\mathbb{H}_0+\mathbb{V})\sum_{n=0}^{\infty}(\mathbb{G}_{\alpha_0}\mathbb{V})^n\left.\ket{\Phi}\right>\,,
\end{align}
where we have used Eq.~\eqref{eq:Psi-formal}.
This implies that we can find the contributions $\left.\ket{\Phi}\right>$ from the degenerate subspace q.d.$(\alpha_0)$ to the correct eigenstates $\left.\ket{\Psi}\right>$ lying approximately within this subspace via diagonalization of the hermitian matrix
\begin{align}\label{eq:pt_matrix}
    M_{\alpha\beta}(\alpha_0)&=\left<\bra{\Phi_\alpha}\right.\qty(\mathbb{H}_0+\mathbb{V})\sum_{n=0}^{\infty}(\mathbb{G}_{\alpha_0}\mathbb{V})^n\left.\ket{\Phi_\beta}\right>\notag\\
    &=\hbar\epsilon_{\alpha}^{(0)}\delta_{\alpha\beta}+\left<\bra{\Phi_\alpha}\right.\mathbb{V}\sum_{n=0}^{\infty}(\mathbb{G}_{\alpha_0}\mathbb{V})^n\left.\ket{\Phi_\beta}\right>\,,
\end{align}
where $\alpha,\beta\in{\rm q.d.}(\alpha_0)$. Its eigenvectors are the components of the $\left.\ket{\Phi}\right>$'s in the basis $\qty{\left.\ket{\Phi_\alpha}\right>|\alpha\in{\rm q.d.}(\alpha_0)}$ and its eigenvalues are the Floquet energies $\hbar\epsilon$ of the $\left.\ket{\Psi}\right>$'s corresponding to these $\left.\ket{\Phi}\right>$'s when turning off the perturbation $\mathbb{V}\rightarrow 0$ [see Eqs.~\eqref{eq:Psi-Phi-G0V} and \eqref{eq:pt_E-E_0}]. Due to the correspondence $\lim\limits_{\mathbb{V}\rightarrow 0}\left.\ket{\Psi}\right>=\left.\ket{\Phi}\right>$, we will refer to the $\left.\ket{\Phi}\right>$ as the 'correct' unperturbed eigenstates.

\subsection{Application to odd harmonic resonances}\label{app:pt_odd}
Now we consider the specific system in Eq.~\eqref{eq:Hamiltonian_Floquet}. We are interested in the correct perturbed Floquet energies at anti-crossing points in Fig.~\ref{fig:4}, i.e., we consider odd harmonic resonances $\Omega_{\rm R}\approx (2p-1)\Omega_{\rm ac}$ with $p=1,2,3..$. Under these conditions the two Floquet states 
\begin{equation}\label{eq:pt_phi_alpha}
    \left.\ket{\Phi_1}\right>=\left.\ket{+,n}\right>\text{ and }\left.\ket{\Phi_2}\right>=\left.\ket{-,n+(2p-1)}\right>
\end{equation}
introduced in Sec.~\ref{sec:floquet_bands} are quasi-degenerate with respect to $\mathbb{H}_0$ with unperturbed energies 
\begin{align}
    \hbar\epsilon_1^{(0)}&=\hbar\Omega_{\rm R}/2+n\hbar\Omega_{\rm ac}\,,\\
    \hbar\epsilon_2^{(0)}&=-\hbar\Omega_{\rm R}/2+(n+2p-1)\hbar\Omega_{\rm ac}\approx \hbar\epsilon_1^{(0)}\,.
\end{align}
All other eigenstates of $\mathbb{H}_0$ have an energy difference to these two states on the order of $\mathcal{O}(\Omega_{\rm ac})$, i.e., they are far away in energy if the resonance condition $|\epsilon_2^{(0)}-\epsilon_1^{(0)}|\ll\Omega_{\rm ac}$ holds sufficiently well, ensuring Eq.~\eqref{eq:not_quasi_degenerate} for all states not in this quasi-degenerate subspace.

To make more progress with Eq.~\eqref{eq:pt_matrix}, we start by noting that $\mathbb{G}_{\alpha_0}$ commutes with $\mathbb{H}_0$, i.e., it does not change the Floquet index $n$ of any given state $\left.\ket{\pm,n}\right>$ when acting on it. The potential $\mathbb{V}$ on the other hand changes the index $n$ to $n\pm 1$, i.e., by an amount of $1$.

We will now distinguish between (A) diagonal and (B) off-diagonal elements in Eq.~\eqref{eq:pt_matrix}:

(A) The diagonal elements $\alpha=\beta$ are associated with states that have the same Floquet index $n$. This implies that any odd amount of operators $\mathbb{V}$ appearing in Eq.~\eqref{eq:pt_matrix} for $\alpha=\beta$ automatically leads to a vanishing of the diagonal elements. 

(B) The off-diagonal elements $\alpha\neq\beta$ are associated with states that differ in Floquet index by the odd amount of $2p-1$. This means that any even number of $\mathbb{V}$ in Eq.~\eqref{eq:pt_matrix} automatically leads to a vanishing of the matrix elements. Also since $\mathbb{V}$ changes the Floquet index by $\pm 1$ when acting on the eigenstates $\left.\ket{\Phi_\alpha}\right>$ of $\mathbb{H}_0$, any odd number $2q-1$ of $\mathbb{V}$ for $q<p$ also leads to a vanishing of the matrix elements with $\alpha\neq \beta$.

With these considerations we can analyze the matrix elements $M_{\alpha\beta}$ in powers of the perturbation $\mathbb{V}$. This leads to
\begin{equation}
    M_{\alpha\beta}\hat{=}\left(\begin{matrix}\hbar\epsilon_1^{(0)}+v_{11}^{(2p-2)}&v_{12}^{(2p-1)}\\v_{21}^{(2p-1)}&\hbar\epsilon_2^{(0)}+v_{22}^{(2p-2)}\end{matrix}\right)+\mathcal{O}\qty(\mathbb{V}^{2p})\,,\label{eq:M_expansion}
\end{equation}
where 
\begin{equation}
    v_{\alpha\beta}^{(j)}=\left<\bra{\Phi_\alpha}\right.\mathbb{V}\sum_{n=0}^{j-1}(\mathbb{G}_{\alpha_0}\mathbb{V})^n\left.\ket{\Phi_\beta}\right>
\end{equation}
are the perturbation-dependent contributions in Eq.~\eqref{eq:pt_matrix} up to including $\mathcal{O}\qty(\mathbb{V}^j)$. 

Assuming for simplicity the resonant case in Eq.~\eqref{eq:M_expansion}, i.e.,
\begin{equation}\label{eq:pt_resonance}
    \hbar\epsilon_1^{(0)}+v_{11}^{(2p-2)}=\hbar\epsilon_2^{(0)}+v_{22}^{(2p-2)}\,,
\end{equation}
the eigenvalues of the matrix in Eq.~\eqref{eq:M_expansion} are given by
\begin{equation}\label{eq:pt_eps_pm}
    \hbar\epsilon_{\pm}=\hbar\epsilon_1^{(0)}+v_{11}^{(2p-2)}\pm |v_{12}^{(2p-1)}|
\end{equation}
with the 'correct' unperturbed eigenstates
\begin{equation}\label{eq:pt_phi_pm}
    \left.\ket{\Phi_\pm}\right>=\frac{1}{\sqrt{2}}\qty(\left.\ket{\Phi_1}\right>\pm \frac{v_{21}^{(2p-1)}}{|v_{12}^{(2p-1)}|}\left.\ket{\Phi_2}\right>)
\end{equation}
being the part of the corresponding eigenstates $\left.\ket{\Psi_{\pm}}\right>$ from Eq.~\eqref{eq:pt_eveq} that lies in the quasi-degenerate subspace spanned by the states in Eq.~\eqref{eq:pt_phi_alpha}. Here we made use of $\qty(v_{12}^{(2p-1)})^*=v_{21}^{(2p-1)}$.

As can be seen in Eq.~\eqref{eq:pt_resonance}, the perturbation $\mathbb{V}$ obviously changes the resonance condition. However, keeping the resonance condition in Eq.~\eqref{eq:pt_resonance} fixed for arbitrary perturbation strength, together with Eq.~\eqref{eq:pt_eps_pm} we obtain 
\begin{equation}\label{eq:pt_eps_pm_V0}
    \eval{\epsilon_{\pm}}_{\mathbb{V}\rightarrow 0}=\epsilon_1^{(0)}=\epsilon_2^{(0)} \Leftrightarrow \Omega_{\rm R}=(2p-1)\Omega_{\rm ac}\,.
\end{equation}
Assuming that the resonance condition in Eq.~\eqref{eq:pt_resonance} holds, thus automatically leads to the correct resonance condition at vanishing perturbation. Note that Eq.~\eqref{eq:pt_eps_pm_V0} together with the discussion (A) from above about diagonal matrix elements implies 
\begin{equation}
    \epsilon_\pm=\epsilon^{(0)}+\mathcal{O}(\mathbb{V}^2)
\end{equation}
with
\begin{equation}
    \epsilon^{(0)}=\Omega_{\rm R}/2+n\Omega_{\rm ac}=\epsilon_1^{(0)}\,.
\end{equation}
This is very helpful for rewriting the Green's function in terms of a perturbative expansion in $\mathbb{V}$
\begin{align}\label{eq:G_0_expand}
    \mathbb{G}_{\alpha_0}&=(\hbar\epsilon-\mathbb{H}_0)^{-1}(1-\mathbb{P}_{\alpha_0})\notag\\
    &=(\hbar\epsilon^{(0)}-\mathbb{H}_0)^{-1}(1-\mathbb{P}_{\alpha_0})+\mathcal{O}\qty(\mathbb{V}^2)\,.
\end{align}
Note that there can be no divergent terms in the denominator here due to the projection with $(1-\mathbb{P}_{\alpha_0})$, removing any contribution from the quasi-degenerate subspace spanned by the states in Eq.~\eqref{eq:pt_phi_alpha}.

Next we calculate the off-diagonal matrix elements $v_{12}^{(2p-1)}$ in Eq.~\eqref{eq:M_expansion}. According to the previous discussion (B) only the $(2p-1)$-th order contributes, since the states in Eq.~\eqref{eq:pt_phi_alpha} have Floquet indices differing by $2p-1$. Thus we obtain
\begin{align}\label{eq:v_21}
    v_{21}^{(2p-1)}&=\left<\bra{\Phi_2}\right.\mathbb{V}(\mathbb{G}_{\alpha_0}\mathbb{V})^{2p-2}\left.\ket{\Phi_1}\right>\notag\\
    &=\left<\bra{\Phi_2}\right.\mathbb{V}\qty[(\hbar\epsilon^{(0)}-\mathbb{H}_0)^{-1}(1-\mathbb{P}_{\alpha_0})\mathbb{V}]^{2p-2}\left.\ket{\Phi_1}\right>\notag\\
    &+\mathcal{O}\qty(\mathbb{V}^{2p+1})\,,
\end{align}
where we used the expansion in Eq.~\eqref{eq:G_0_expand}. For these off-diagonal matrix elements we only need to consider the part of $\mathbb{V}$ in Eq.~\eqref{eq:Hamiltonian_Floquet} that either reduces or increases the Floquet index, since otherwise we cannot connect the states $\left.\ket{\Phi_1}\right>$ and $\left.\ket{\Phi_2}\right>$ from Eq.~\eqref{eq:pt_phi_alpha}, that differ by $2p-1$ in terms of the Floquet index. Using
\begin{equation}
    \mathbb{V}\left.\ket{\sigma,n}\right>=-i\hbar\frac{A_{\rm ac}}{4}\left.\ket{-\sigma,n+1}\right>+ \text{term with }n-1\,,
\end{equation}
we find 
\begin{align}
    &[(\hbar\epsilon^{(0)}-\mathbb{H}_0)^{-1}\mathbb{V}]^{2}\left.\ket{+,n}\right>\notag\\
    &\quad=\frac{\qty(\frac{A_{\rm ac}}{4i})^2\left.\ket{+,n+2}\right>}{\qty[\epsilon^{(0)}+\frac{\Omega_{\rm R}}{2}-(n+1)\Omega_{\rm ac}]\qty[\epsilon^{(0)}-\frac{\Omega_{\rm R}}{2}-(n+2)\Omega_{\rm ac}]}\notag\\
    &\quad+ \text{irrelevant terms}\notag\\
    &\quad=\frac{\qty(\frac{A_{\rm ac}}{4i})^2\left.\ket{+,n+2}\right>}{\qty(\Omega_{\rm ac}-\Omega_{\rm R})\qty(2\Omega_{\rm ac})}\notag\\
    &\quad+ \text{irrelevant terms}\,,
\end{align}
where we used $\epsilon^{(0)}=\Omega_{\rm R}/2+n\Omega_{\rm ac}$. To calculate the matrix element $v_{21}^{(2p-1)}$ using Eq.~\eqref{eq:v_21} we need to act with the operator $[(\hbar\epsilon^{(0)}-\mathbb{H}_0)^{-1}\mathbb{V}]^{2}$ a total of $p-1$ times, each time moving further away from $\epsilon^{(0)}$, i.e., acting again we have
\begin{align}
    &[(\hbar\epsilon^{(0)}-\mathbb{H}_0)^{-1}\mathbb{V}]^{4}\left.\ket{+,n}\right>\notag\\
    &\quad=\frac{\qty(\frac{A_{\rm ac}}{4i})^4\left.\ket{+,n+4}\right>}{\qty(\Omega_{\rm ac}-\Omega_{\rm R})\qty(2\Omega_{\rm ac})\qty(3\Omega_{\rm ac}-\Omega_{\rm R})\qty(4\Omega_{\rm ac})}\notag\\
    &\quad+ \text{irrelevant terms}\,.
\end{align}
Repeating in this fashion, we can generalize to
\begin{align}
    &[(\hbar\epsilon^{(0)}-\mathbb{H}_0)^{-1}\mathbb{V}]^{2(p-1)}\left.\ket{+,n}\right>\notag\\
    &\quad=\frac{\qty(\frac{A_{\rm ac}}{4i})^{2(p-1)}\left.\ket{+,n+2(p-1)}\right>}{\prod_{j=1}^{p-1}\qty[(2j-1)\Omega_{\rm ac}-\Omega_{\rm R}]\qty(2j\Omega_{\rm ac})}\notag\\
    &\quad+ \text{irrelevant terms}\,.
\end{align}
Inserting this result into Eq.~\eqref{eq:v_21} finally yields
\begin{align}
    v_{21}^{(2p-1)}&=\frac{-i\hbar\qty(\frac{A_{\rm ac}}{4})^{2p-1}}{\prod_{j=1}^{p-1}\qty[\Omega_{\rm R}-(2j-1)\Omega_{\rm ac}]\qty(2j\Omega_{\rm ac})}\,,\label{eq:app_v_21}
\end{align}
such that we can finally rewrite Eqs.~\eqref{eq:pt_eps_pm} and \eqref{eq:pt_phi_pm} as
\begin{align}
    \hbar\epsilon_{\pm}=\hbar\tilde{\epsilon}^{(0)}\pm \hbar\frac{\Delta_{\epsilon}}{2}\label{eq:app_E1_eps_pm}
\end{align}
and
\begin{align}
    \left.\ket{\Phi_\pm}\right>=\frac{1}{\sqrt{2}}\qty(\left.\ket{\Phi_1}\right>\mp i\left.\ket{\Phi_2}\right>)\label{eq:app_E1_phi_pm}
\end{align}
with $\hbar\Delta_{\epsilon}=2|v_{21}^{(2p-1)}|$ and $\hbar\tilde{\epsilon}^{(0)}=\hbar{\epsilon}^{(0)}+v_{11}^{(2p-2)}$.

\subsection{Application to even harmonic resonances}\label{app:pt_even}
Considering points of line crossings in Fig.~\ref{fig:4}, i.e., even harmonic resonances $\Omega_{\rm R}\approx 2p\Omega_{\rm ac}$ with $p=1,2,3..$, the two Floquet states 
\begin{equation}
    \left.\ket{\Phi_1}\right>=\left.\ket{+,n}\right>\text{ and }\left.\ket{\Phi_2}\right>=\left.\ket{-,n+(2p)}\right>
\end{equation}
introduced in Sec.~\ref{sec:floquet_bands} are quasi-degenerate with respect to $\mathbb{H}_0$. All other eigenstates of $\mathbb{H}_0$ have an energy difference to these two states on the order of $\mathcal{O}(\Omega_{\rm ac})$, i.e., they are far away in energy if the resonance condition $|\epsilon_2^{(0)}-\epsilon_1^{(0)}|\ll\Omega_{\rm ac}$ holds sufficiently well, ensuring Eq.~\eqref{eq:not_quasi_degenerate} for all states not in this quasi-degenerate subspace.

However these unperturbed quasi-degenerate states have opposite parity [see Eq.~\eqref{eq:parity}]. Since all operators in the matrix elements in Eq.~\eqref{eq:pt_matrix} commute with the parity operator, the matrix $M_{\alpha\beta}$ is already diagonal, such that for even harmonic resonances the states $\left.\ket{\Phi_{\alpha=1,2}}\right>$ are already the 'correct' unperturbed eigenstates with energies $\hbar\epsilon_{\alpha}=\hbar\epsilon^{(0)}_{\alpha}+v_{\alpha\alpha}^{(2p-2)}+\mathcal{O}\qty(\mathbb{V}^{(2p)})$, i.e., there is only renormalization of energy levels leading to change in the resonance condition analog to Eq.~\eqref{eq:pt_resonance}, but no coherent coupling between the states. This means that for even harmonic resonances the approximate eigenstates $\left.\ket{\Phi_\alpha}\right>$ which have opposite parity have degenerate eigenvalues and the lines in Fig.~\ref{fig:4} cross.

\section{Non-unitary perturbation theory: Transition between damping regimes}\label{app:pt_nonunitary}
As in the unitary case presented in Sec.~\ref{sec:floquet_bands}, we can set up a perturbation theory in Liouville space using the periodic Floquet states $\ketL{V_r(t)}$, defined in Eq.~\eqref{eq:def_V_r}. Analog to Eq.~\eqref{eq:floquet_states_eom}, we find from Eq.~\eqref{eq:V_eom} and Eq.~\eqref{eq:def_V_r}
\begin{equation}\label{eq:eom_V_r}
    \dv{t}\ketL{V_r(t)}=(i\delta_r+\gamma_r)\ketL{V_r(t)}+\mathcal{L}(t)\ketL{V_r(t)}
\end{equation}
and we define the scalar product for periodic Liouville space states
\begin{equation}
    \expL{\expL{A|B}}=\frac{1}{T}\int_0^T\dd t\, \expL{A(t)|B(t)}
\end{equation}
analog to Eq.~\eqref{eq:scalar_prod_floquet_unitary}. Considering a basis $\ketL{B_p}$ of Liouville space and its reciprocal basis $\ketL{\overline{B}_p}$ [see Eqs.~\eqref{eq:reciprocal_basis}], we can define a basis and reciprocal basis which are orthonormal with respect to this scalar product as
\begin{equation}
    \left.\ketL{B_p,m}\right)=\ketL{B_p}e^{im\Omega t}\,,\qquad \ketL{\overline{B}_p,m}\big)=\ketL{\overline{B}_p}e^{im\Omega t}
\end{equation}
analog to Eq.~\eqref{eq:floquet_basis}, such that
\begin{subequations}
\begin{equation}
    \expL{\expL{\overline{B}_p,m|B_{q},n}}=\delta_{pq}\delta_{mn}\,
\end{equation}
and
\begin{equation}
    1=\sum_{mp}\left.\ketL{B_p,m}\right)\big(\braL{\overline{B}_p,m}\,.
\end{equation}
\end{subequations}
Calculating the matrix elements of the operators in Eq.~\eqref{eq:eom_V_r} analog to Eq.~\eqref{eq:matrix_elem_unitary}, we obtain
\begin{align}\label{eq:matrix_elem_non_unitary}
    in\Omega \delta_{pq}\delta_{mn}&-(i\delta_r+\gamma_r)\delta_{pq}\delta_{mn}\notag\\
    &=\expL{\expL{\overline{B}_p,m|\mathcal{L}|B_{q},n}}\notag\\
    &=\frac{1}{T}\int\limits_0^T\dd t\, e^{i(n-m)\Omega t}\expL{\overline{B}_p|\mathcal{L}(t)|B_{q}}\,.
\end{align}
From the discussion in App.~\ref{app:floquet_eigen_diss}, we already know that there is a subspace of traceless solutions and a single solution with non-vanishing trace with the eigenvalue $\mu_0=0$. We therefore can fully focus on the subspace of traceless solutions, which is closed under the action of the Lindblad operator [see Eq.~\eqref{eq:lindbladian_matrix_3d}]. We can separate the Lindblad operator, acting on the traceless states, into a time-independent component $\mathcal{L}_{\Tr=0,0}$ and a time-dependent component $\mathcal{L}_{\Tr=0,I}(t)$, such that
\begin{subequations}
\begin{align}
    \mathcal{L}_{\Tr=0}(t)&=\mathcal{L}_{\Tr=0,0}+\mathcal{L}_{\Tr=0,I}(t)\,,\\
    \mathcal{L}_{\Tr=0,0}&\hat{=}\left(\begin{matrix}-\gamma&0&0\\0&-\gamma&-\Omega_{\rm R}\\0&\Omega_{\rm R}&-\gamma_{\rm xd}\end{matrix}\right)\,,\\
    \mathcal{L}_{\Tr=0,I}(t)&\hat{=}\left(\begin{matrix}0&-\Delta(t)&0\\\Delta(t)&0&0\\0&0&0\end{matrix}\right)\,,
\end{align}
\end{subequations}
where we used Eq.~\eqref{eq:lindbladian_matrix_3d} with $h_1(t)=\hbar\Omega_{\rm R}/2$, $h_2(t)=0$ and $h_3(t)=\hbar\Delta(t)/2$. Note that the matrix elements are obtained via the Pauli matrices $\frac{1}{2}\left(\sigma_i\right|{\mathcal{L}}\left|\sigma_j\right)=\frac{1}{2}\Tr\qty(\sigma_i^\dagger\mathcal{L}\qty[\sigma_j])$, $i=1,2,3$. Next we solve for the eigenvectors $\ketL{B_p}$ and eigenvalues $\lambda_p^{(0)}$ of the time-independent Lindblad operator $\mathcal{L}_{\Tr=0,0}\ketL{B_p}=\lambda_p^{(0)}\ketL{B_p}$, yielding
\begin{subequations}\label{eq:lambda_0}
\begin{align}
    \lambda_1^{(0)}&=-\gamma\,,\\
    \lambda_2^{(0)}&=\qty[-\frac{\gamma+\gamma_{\rm xd}}{2}+\sqrt{\frac{(\gamma-\gamma_{\rm xd})^2}{4}-\Omega_{\rm R}^2}]\,,\\
    \lambda_3^{(0)}&=\qty[-\frac{\gamma+\gamma_{\rm xd}}{2}-\sqrt{\frac{(\gamma-\gamma_{\rm xd})^2}{4}-\Omega_{\rm R}^2}]\,,
\end{align}
\end{subequations}
with eigenstates
\begin{subequations}
\begin{align}
    \ketL{B_1}&=\ketL{\sigma_1}\,,\\
    \ketL{B_2}&=\ketL{\sigma_2}-\frac{\lambda_2^{(0)}+\gamma}{\Omega_{\rm R}}\ketL{\sigma_3}\,,\\
    \ketL{B_3}&=\ketL{\sigma_2}-\frac{\lambda_3^{(0)}+\gamma}{\Omega_{\rm R}}\ketL{\sigma_3}\,.
\end{align}
\end{subequations}
Note that the eigenvalues $\lambda_2^{(0)}$ and $\lambda_3^{(0)}$ are degenerate for $\Omega_{\rm R}=|\gamma-\gamma_{\rm xd}|/2$ with the two eigenvectors coalescing in this case $\ketL{B_2}=\ketL{B_3}$, i.e., this corresponds to an exceptional point in the non-unitary dynamics of the system in the absence of acoustic modulation $\Delta(t)=0$~\cite{muller2008exceptional,minganti2019quantum}. When considering the Floquet eigenvalue equation~\eqref{eq:eigen_floquet} with the parametrization in Eq.~\eqref{eq:parametrization} and $\Delta(t)=0$, we can identify $\lambda_r^{(0)}=-i\delta_r-\gamma_r$ in the case of vanishing acoustic modulation. For small Rabi frequencies below the exceptional point $\Omega_{\rm R}<|\gamma-\gamma_{\rm xd}|/2$, all eigenvalues $\lambda_r^{(0)}$ are completely real and we only have a single Floquet frequency $\delta_r=0$. For large Rabi frequencies above the exceptional point $\Omega_{\rm R}>|\gamma-\gamma_{\rm xd}|/2$, we have three Floquet frequencies $\delta_1=0$ and $\delta_{2/3}=\mp \sqrt{\Omega_{\rm R}^2-\frac{(\gamma-\gamma_{\rm xd})^2}{4}}$. The first case corresponds to an overdamped regime, while the second case corresponds to an underdamped regime in which the Mollow triplet emerges ($\delta_{2/3}\approx \mp\Omega_{\rm R}$ for $\Omega_{\rm R}\gg\gamma_{\rm xd,pd}$)~\cite{scully1997quantum}. The exceptional point therefore corresponds to a transition between these two regimes. 

In the following we focus on the underdamped regime, in which the Mollow triplet is visible, i.e., $\Omega_{\rm R}>|\gamma-\gamma_{\rm xd}|/2$, such that the eigenstates $\ketL{B_p}$ form a basis. The eigenvalues $\lambda_{2/3}^{(0)}$ are then complex with $\lambda_{2}^{(0)*}=\lambda_{3}^{(0)}$ and the corresponding reciprocal basis vectors obeying $\expL{\overline{B}_p|B_q}=\delta_{pq}$ are given by
\begin{subequations}
\begin{align}
    \ketL{\overline{B}_1}&=\frac{1}{2}\ketL{\sigma_1}\,,\\
    \ketL{\overline{B}_2}&=\frac{\qty(\lambda_{2}^{(0)}+\gamma)\ketL{\sigma_2}+\Omega_{\rm R}\ketL{\sigma_3}}{4i\Im\qty(\lambda_2^{(0)})}\,,\\
    \ketL{\overline{B}_3}&=\frac{\qty(\lambda_{3}^{(0)}+\gamma)\ketL{\sigma_2}+\Omega_{\rm R}\ketL{\sigma_3}}{4i\Im\qty(\lambda_3^{(0)})}\,.
\end{align}
\end{subequations}
With these eigenstates, unperturbed by the acoustic modulation $\Delta(t)=A_{\rm ac}\sin(\Omega_{\rm ac}t)$ in Eq.~\eqref{eq:detuning_sinus}, i.e., $\Omega=\Omega_{\rm ac}$, we can write the matrix elements in Eq.~\eqref{eq:matrix_elem_non_unitary} as
\begin{align}\label{eq:matrix_elem_non_unitary_L0}
    in\Omega_{\rm ac} \delta_{pq}\delta_{mn}&-(i\delta_r+\gamma_r+\lambda_p^{(0)})\delta_{pq}\delta_{mn}\\
    &=\frac{1}{T}\int\limits_0^T\dd t\, e^{i(n-m)\Omega_{\rm ac} t}\expL{\overline{B}_p|\mathcal{L}_{\Tr=0,I}(t)|B_{q}}\notag\\
    &=\qty(\expL{\overline{B}_p\ketL{\sigma_2}\braL{\sigma_1} B_{q}}-\expL{\overline{B}_p\ketL{\sigma_1}\braL{\sigma_2} B_{q}})\notag\\
    &\qquad\times\frac{1}{2T}\int\limits_0^T\dd t\, e^{i(n-m)\Omega_{\rm ac} t}\Delta(t)\notag\\
    &=\frac{A_{\rm ac}}{4i}(\delta_{n,m-1}-\delta_{n,m+1})\notag\\
    &\times \left[\delta_{q 1}(1-\delta_{p1})\frac{\lambda_{p}^{(0)*}+\gamma}{i \Im\qty(\lambda_{p}^{(0)*})}-2\delta_{p1}(1-\delta_{q1})\right]\notag\,.
\end{align}
As in Eq.~\eqref{eq:Hamiltonian_Floquet}, the acoustic modulation leads to transitions between the optically dressed states $\ketL{B_p}$, while at the same time changing the Floquet index by $\pm 1$. In absence of acoustic modulation we obtain
\begin{equation}\label{eq:floquet_nonunitary_0}
    (i\delta_r+\gamma_r)\delta_{pq}\delta_{mn}=(-\lambda_p^{(0)}+in\Omega_{\rm ac})\delta_{pq}\delta_{mn}\,,
\end{equation}
i.e., the corresponding matrix is diagonal and we can easily read off the Floquet eigenvalues. Considering resonant excitation with $\Omega_{\rm ac}\approx\Omega_{\rm R}$ and weak acoustic modulation in the limit of strong optical driving $\Omega_{\rm R}\gg \gamma_{\rm xd,pd}$, the unperturbed eigenstates $\ketL{B_1,n}\big)$, $\ketL{B_2,n+1}\big)$ and $\ketL{B_3,n-1}\big)$ are quasi-degenerate with unperturbed Floquet frequencies $\delta_1^{(0)}=n\Omega_{\rm ac}$, $\delta_2^{(0)}=-\Omega_{\rm R}+(n+1)\Omega_{\rm ac}$ and $\delta_3^{(0)}=\Omega_{\rm R}+(n-1)\Omega_{\rm ac}$. All other unperturbed eigenstates have Floquet frequencies differing by at least $\Omega_{\rm ac}$ and are coupled via the acoustic modulation amplitude $A_{\rm ac}$. For $\Omega_{\rm ac}\gg A_{\rm ac}$ we may therefore focus on the subspace of these three quasi-degenerate states and rewrite Eq.~\eqref{eq:matrix_elem_non_unitary_L0} into an eigenvalue equation on that subspace. To this aim, we project on this subspace to obtain
\begin{align}\label{eq:eigenvalue_eq_nonunitary}
    \qty(\begin{matrix}-\lambda_{1}^{(0)}&i\frac{A_{\rm ac}}{2}&-i\frac{A_{\rm ac}}{2}\\\frac{A_{\rm ac}\qty(\lambda_{2}^{(0)*}+\gamma)}{4\Im\qty(\lambda_{2}^{(0)*})}&-\lambda_{2}^{(0)}+i\Omega_{\rm ac}&0\\-\frac{A_{\rm ac}\qty(\lambda_{2}^{(0)}+\gamma)}{4\Im\qty(\lambda_{2}^{(0)})}&0&-\lambda_{2}^{(0)*}-i\Omega_{\rm ac}\end{matrix})\vb{v}_r=\qty(i\delta_r+\gamma_r-in\Omega_{\rm ac})\vb{v}_r\,.
\end{align}
Here we used $\lambda_{2}^{(0)*}=\lambda_{3}^{(0)}$ and the eigenvector is given by $\vb{v}_r=\qty(\expL{\expL{\overline{B}_1,n|V_r}},\expL{\expL{\overline{B}_2,n+1|V_r}},\expL{\expL{\overline{B}_3,n-1|V_r}})^T$. In the unitary limit $\gamma_{\rm xd,pd}\rightarrow 0$, the eigenvalues can be easily obtained, since then $\lambda_1^{(0)}=0$ and $\lambda_2^{(0)}=i\Omega_{\rm R}$, such that the eigenvalue equation reads
\begin{align}
    \qty(\begin{matrix}0&i\frac{A_{\rm ac}}{2}&-i\frac{A_{\rm ac}}{2}\\i\frac{A_{\rm ac}}{4}&i(\Omega_{\rm ac}-\Omega_{\rm R})&0\\-i\frac{A_{\rm ac}}{4}&0&-i(\Omega_{\rm ac}-\Omega_{\rm R})\end{matrix})\vb{v}_r=\qty(i\delta_r+\gamma_r-in\Omega_{\rm ac})\vb{v}_r\,,\qquad \gamma_{\rm xd,pd}=0\,.
\end{align}
This yields the eigenvalues $\delta_1=n\Omega_{\rm ac}$, $\delta_{2/3}=n\Omega_{\rm ac}\pm \sqrt{(\Omega_{\rm ac}-\Omega_{\rm R})^2+A_{\rm ac}^2/4}$ as well as $\gamma_r=0$ for the unitary limit with $\gamma_{\rm xd,pd}=0$, consistent with Eqs.~\eqref{eq:app_v_21} and \eqref{eq:app_E1_eps_pm} for $p=1$, i.e., this describes the anti-crossing pattern at resonance $\Omega_{\rm ac}\approx\Omega_{\rm R}$. The full eigenvalue equation with dissipation, i.e., Eq.~\eqref{eq:eigenvalue_eq_nonunitary}, is not as easily solvable. However, if we focus only on the exact resonance $\Im\qty(\lambda_2^{(0)})=\sqrt{\Omega_{\rm R}^2-\frac{(\gamma-\gamma_{\rm xd})^2}{4}}=\Omega_{\rm ac}$, we obtain the simpler eigenvalue equation using Eq.~\eqref{eq:lambda_0}
\begin{align}\label{eq:eigenvalue_eq_nonunitary_resonance}
    \qty(\begin{matrix}\gamma&i\frac{A_{\rm ac}}{2}&-i\frac{A_{\rm ac}}{2}\\-\frac{A_{\rm ac}\qty(\frac{\gamma-\gamma_{\rm xd}}{2}-i\Omega_{\rm ac})}{4\Omega_{\rm ac}}&\frac{\gamma_{\rm xd}+\gamma}{2}&0\\-\frac{A_{\rm ac}\qty(\frac{\gamma-\gamma_{\rm xd}}{2}+i\Omega_{\rm ac})}{4\Omega_{\rm ac}}&0&\frac{\gamma_{\rm xd}+\gamma}{2}\end{matrix})\vb{v}_r=\qty(i\delta_r+\gamma_r-in\Omega_{\rm ac})\vb{v}_r\,,\quad \sqrt{\Omega_{\rm R}^2-\frac{(\gamma-\gamma_{\rm xd})^2}{4}}=\Omega_{\rm ac}\,.
\end{align}
From this we obtain the Floquet eigenvalues
\begin{subequations}
\begin{align}
    i\delta_1+\gamma_1&=in\Omega_{\rm ac}+\frac{\gamma_{\rm xd}+\gamma}{2}\,,\\
    i\delta_{2/3}+\gamma_{2/3}&=in\Omega_{\rm ac}+\frac{\gamma_{\rm xd}+3\gamma}{4}\notag\\
    &\pm\frac{1}{8}\sqrt{(\gamma_{\rm xd}-\gamma_{\rm pd})^2-16 A_{\rm ac}^2}
\end{align}
\end{subequations}
at resonance, where we used $\gamma=(\gamma_{\rm xd}+\gamma_{\rm pd})/2$. The modulated system thus shows another overdamped-to-underdamped transition, apart from the standard weak-to-strong driving transition leading to the appearance of the Mollow triplet~\cite{scully1997quantum}. Here we find the transition from an overdamped regime for $|\gamma_{\rm xd}-\gamma_{\rm pd}|>4A_{\rm ac}$ to an underdamped regime for $|\gamma_{\rm xd}-\gamma_{\rm pd}|<4A_{\rm ac}$. If the acoustic driving is not sufficiently strong compared to the damping, i.e., in the overdamped regime, all Floquet frequencies are equal $\delta_r=n\Omega_{\rm ac}$. The anti-crossing pattern is only visible for sufficiently strong acoustic modulation in the underdamped regime. In the extreme case of $A_{\rm ac}\gg \gamma_{\rm xd,pd}$ we recover the result from the unitary calculation at resonance $\delta_{2/3}=n\Omega_{\rm ac}\pm A_{\rm ac}/2$ [see Eq.~\eqref{eq:line_distance_resonant}]. To ensure that we are in the underdamped regime at resonance, we can set $4A_{\rm ac}>\max(\gamma_{\rm xd},\gamma_{\rm pd})$ which is automatically $\geq|\gamma_{\rm xd}-\gamma_{\rm pd}|$.

\end{appendices}
\section*{References}
\bibliographystyle{iopart-num}
\bibliography{all_refs}

\end{document}